\documentclass[preprint,11pt,3p,times,authoryear]{elsarticle}

\usepackage{amssymb}
\usepackage{amsmath}

\usepackage{pifont,graphicx,amssymb,color,setspace,lineno,comment,
	booktabs,dcolumn,tabularx,pdflscape,longtable,float,lscape,subcaption}
\usepackage{caption}
\usepackage[table,xcdraw]{xcolor}
\usepackage{xspace}

\journal{ICARUS}

\usepackage{hyperref}   
\begin{document}
\newcommand{\maps}{Meteoritics \& Planetary Science}

\begin{frontmatter}

\title{The Enceladian crater production function}

\author[UniGe,ELSI]{E. W. Wong\corref{cor}}
\author[UiO]{S. C. Werner}
\author[SW]{M. R. Kirchoff}
\author[CSFK,UiO]{R. Brasser}

\affiliation[UniGe]{organization={Geneva Observatory, University of Geneva},
    addressline={Chemin Pegasi 51},
    city={Versoix},
    postcode={CH-1290},
    country={Switzerland}}

\affiliation[ELSI]{organization={Earth Life Science Institute, Tokyo Institute of Technology},
    addressline={Meguro-ku},
    city={Tokyo},
    postcode={152-8550},
    country={Japan}}

\affiliation[UiO]{organization={Centre for Planetary Habitability, Department of Geosciences, University of Oslo},
    postcode={N-0315},
    city={Oslo},
    country={Norway}}

\affiliation[SW]{organization={Southwest Research Institute},
    addressline={1301 Walnut St, Suite 400},
    city={Boulder},
    state={CO},
    country={USA}}

\affiliation[CSFK]{organization={Konkoly Observatory, HUN-REN CSFK, MTA Centre of Excellence},
    addressline={15-17 Konkoly Thege Miklos St.},
    postcode={H-1121},
    city={Budapest},
    country={Hungary}}

\cortext[cor]{Corresponding author}

%
\begin{abstract}
Interpreting Enceladus’s past and present surface history and interior state remains challenging, owing to uncertain prescription of its impact bombardment history and limited interpretation of its crater statistics. Further progress in understanding its evolutionary history can be achieved with an improved crater chronology model and a thorough assessment of Enceladus’s surface.
Here we present the first step in the form of a comprehensive, global crater catalogue with geomorphology survey for Enceladus. From our dataset we build the crater production function (CPF), which is the underlying, unmodified crater size-frequency distribution of the satellite surface, assuming no subsequent modification.
We obtained the CPF with a data-driven approach;  therefore it makes no assumptions about the impactor source population or planet evolution models or the timing of impact.
We fit the CPF with a high-order polynomial as is customary for the Moon and Mars, capturing the slope variations across different crater diameter ranges.
The Enceladian CPF generally has a steeper cumulative slope than that of the Moon and Mars for small crater diameters $D_{\rm cr}<10$~km , as well as that of the size-frequency distribution of trans-Neptunian objects. This CPF serves as a critical observational input for an Enceladian crater chronology model, enabling the conversion of crater densities into absolute surface ages. Extending the crater cataloguing and CPF derivation of this study to other Saturnian satellites will determine whether the Enceladian CPF is unique; a shared CPF would indicate a common impactor population, providing  observational constraints on the size-frequency distribution of small bodies in the outer Solar System.\\
\end{abstract}

\begin{highlights}
\item New global crater catalogue of Enceladus down to 800 m diameter
\item First Enceladus-specific crater production function is derived
\item CPF is constructed directly from Enceladus' crater record
\item Provides a framework for future age dating of icy satellites
\end{highlights}


\end{frontmatter}


\section{Introduction}
\label{intro}

Enceladus exhibits ongoing cryovolcanic activity, most prominently expressed by plume emissions at its south pole observed by Cassini \citep{Porco2004,Spencer2006}. Multiple lines of evidence, including active degassing and physical libration, point to the presence of a subsurface ocean \citep{Thomas2016,Nimmo2023}. However, the internal heat required to sustain this activity exceeds expectations from tidal dissipation and radiogenic heating alone \citep{Meyer2007,Roberts2008}. Together with the coexistence of heavily cratered and smooth sparsely cratered terrains, this indicates a complex and evolving geological history. Constraining the timing and duration of this activity through robust surface age estimates is therefore essential.\\

All key ingredients conventionally considered necessary for life have been identified on Enceladus. Its subsurface ocean contains organic molecules \citep{Peter2024} and the essential elements carbon, hydrogen, nitrogen, oxygen, phosphorus, and sulfur \citep{Hsu2015,Sekine2015,Bouquet2015,Choblet2017,Nimmo2018}, along with inferred energy sources such as tidal heating \citep{Thomas2016,Nimmo2023}, hydrothermal activity and water-rock chemical reactions \citep{Waite2017,Deamer2017,Choblet2017}. A critical, yet overlooked and poorly constrained, factor is the duration over which Enceladus has maintained these potentially habitable conditions. On Earth, microbial life emerged within ~0.5–0.7 Gyr after stable oceans formed 4.54 billion years ago \citep{Wilde2001,Nutman2016,Bell2015}, highlighting the importance of timescales for habitability. Determining the longevity of Enceladus’ internal activity therefore remains a central question.\\

In the absence of sample-return data for radiometric dating, crater counting remains the primary method for reconstructing the geological history and surface evolution of icy satellites like Enceladus. By analysing the size-frequency distribution of impact craters, researchers can estimate relative and absolute surface ages, constraining the timing and duration of resurfacing events. However, interpreting crater records is challenging due to limited feasibility of observations of impactor population.
In the Saturnian system, it remains uncertain whether the dominant impactors are heliocentric comets or system-internal debris. While localised features may reflect planetocentric populations \citep[e.g.][]{Cuk2016,Ferguson2022}, the prevailing consensus is that heliocentric impactors, especially those from the scattered disc, are the primary source of cratering across the outer solar system for craters above a few kilometres in diameter \citep{Shoemaker1982,Chapman1986,Zahnle1998,Hirata2016,Nesvorny2019,Wong2023,Nesvorny2023}. Planetocentric debris may dominate the production of smaller craters, though the exact diameter threshold is debated and likely depends on target properties such as surface material and average impact velocity \citep{Bell2020}.\\

Against this background, our study avoids assumptions about impactor populations or scaling from lunar chronology and small body size–frequency distributions. Here, we present key dataset and complementary tools: first, a global crater catalogue for Enceladus down to 800~m crater diameter, based on a unified classification of terrain units and crater morphologies; and second, a new data-driven crater production function derived directly from Enceladus’ observed crater record, using a comprehensive and minimally eroded crater catalogue, independent of external impactor models. \\

These results advance data-driven approaches for icy satellite and outer solar system small bodies studies. The \textit{crater production function} (CPF) -- the expected size–frequency distribution of impact craters, often hypothesised to be time-invariant, though it may have varied during the earliest epochs (e.g., the first $\sim$40~myr, see \citet{Bottke2023,Bottke2024})
-- forms the basis for computing surface ages on the solid icy crust of the giant planets' satellites, including Enceladus.

\subsection{Early mapping of Enceladus}
Enceladus, an active body with diverse terrain, has been recognised as such since the Voyager era, beginning with the first images from Voyager 2 at a typical resolution of 2 to 5 km/pixel, with the best images of few limited areas reaching $\sim$1~km/pixel resolution at the closest approach. Early Voyager-era studies examined crater distributions and geological histories, providing initial insights into surface evolution and internal activity \citep{Smith1982, Plescia1983, Schenk1989, Pozio1990, Kargel1996}. \citet{Smith1982} used crater abundance and landform occurrence to broadly categorize the trailing hemisphere (30$^\circ$W to 120$^\circ$E) into old, heavily cratered terrains, sparsely cratered plains, ridge plains, and the youngest smooth plains. \\

During the Cassini mission \citet{Spencer2009} produced the first global map, broadly dividing Enceladus into four geological provinces: cratered terrain, leading hemisphere fractured plains, trailing hemisphere fractured plains, and south polar terrain. Further subdivisions within each province, based on surface morphology and structural features, were established by \citet{Helfenstein2010} and they refine the relative chronology of geological events. In this work, we largely reference the global geomorphology and unit definitions of \citet{CrowWillard2015}, who retained the four provinces of \citet{Spencer2009} but refined the subunits based on morphological and structural criteria.\\

Early investigations of Enceladus, based on its orbital properties, predicted that tidal heating could drive internal activity, as was also proposed for Io and Europa of the Galilean satellites \citep{Cassen1979, Peale1979}. Direct evidence for such activity came with Cassini observations, which revealed active south polar plumes and confirmed substantial internal heating \citep{Porco2006, Spencer2006}. Cassini imaging further revealed a geologically diverse and heterogeneous surface, including tectonic-like deformation, ridge plains with linear features, smooth plains with few or no craters, heavily cratered terrains, and large troughs crossing multiple units \citep{Helfenstein2010}. Previous crater mapping and counting studies have primarily targeted local regions with similar surface morphologies, especially the heavily cratered plains \citep{Kirchoff2009, Kinczyk2024, Blanco-Rojas2024}, providing valuable but spatially limited insights into surface ages and geological processes \citep{Kirchoff2010, Bland2012, Martin2017}.

\subsection{Previous crater counting on Enceladus}
\citeauthor{Kirchoff2009} (\citeyear{Kirchoff2009}, \citeyear{Kirchoff2010}) conducted detailed, region-specific crater counts and surface unit mapping on Enceladus’ trailing hemisphere using Cassini images, analysing crater distributions on cratered plains and ridged plains. \citet{Kirchoff2009} defined manual crater identification criteria requiring at least one quarter of the crater rim to be visible. Diameters were measured rim-to-rim using one or two orthogonal measurements, while degraded craters were assessed from inferred centres and rim traces enhanced by contrast stretching. Only craters with diameters at least ten times the image resolution (e.g., $>$10 pixels) were included, minimising artificial roll-off in the size-frequency distribution. Non-impact circular features, such as collapse pits, were explicitly excluded.\\

Crater counts from \citeauthor{Kirchoff2009} (\citeyear{Kirchoff2009}, \citeyear{Kirchoff2010}) were used to construct size-frequency distributions and spatial density maps for the trailing hemisphere. Within the Enceladian cratered plains, \citet{Kirchoff2009} subdivided the study area into two mid-latitude ($\sim$30$^{\circ}$N to $\sim$60$^{\circ}$N) and two equatorial ($\sim$50$^{\circ}$S to 40$^{\circ}$N) patches based on distinctive tectonic-like features to investigate spatial variation in crater density. The resulting dataset demonstrates strong heterogeneity in crater density among different regions, likely reflecting variations in resurfacing or geological history. Mid-latitude cratered plains contain approximately three times higher crater densities than equatorial plains, consistent with latitude-dependent resurfacing such as viscous relaxation or deposition from the south polar plume and E-ring. The study explicitly concludes that Enceladus’ crater distribution is modified and spatially heterogeneous. 
In this study, the mapping of the trailing hemisphere units follows Fig~.4 of \citet{Kirchoff2009}, particularly the ridged plains, and further subdivides the four cratered plains from \citet{Kirchoff2009} into smaller units.\\

In a later study, \citet{CrowWillard2015} present a global structural-geological map of Enceladus aimed at characterising tectonic organisation and deformation history instead of  cataloguing impact craters. In this framework, craters serve primarily as qualitative and semi-quantitative indicators, used alongside cross-cutting relationships of large-scale surface features to establish the relative ages of major tectonised terrains -- namely the trailing hemisphere, leading hemisphere, and south polar terrains -- as opposed to conducting comprehensive crater analysis. The crater density values cited in \citet{CrowWillard2015} largely derive from earlier studies, particularly \citet{Kirchoff2009}. While cratered terrains, covering $\sim$43\% of the surface, are identified, they were not subjected to size-frequency analysis. Nevertheless, the study provides valuable insights into how resurfacing across different terrains may have modified crater morphology and counts, and it serves as a key reference for global mapping, classification, and nomenclature based on surface morphology and resurfacing or formation history. For instance, the mapping of the leading hemisphere and south polar terrains in the present work follows \citet{CrowWillard2015}.\\

Furthermore, \citet{Kinczyk2024} conducted a regional geological and crater analysis of Enceladus’ cratered terrains, integrating geological mapping, structural characterisation, and regional crater size-frequency distributions to interpret long-term surface evolution. The study focused on seven discrete regions within the cratered terrains, selected based on qualitative similarities in surface morphology and crater preservation. These regions span a ranges of longitudes -- both far-side and near-side to Saturn -- as well as latitudes, including equatorial, mid-latitude, and northern polar areas. Craters larger than 500~m in diameter were mapped to ensure completeness for craters exceeding 1~km, and size-frequency distributions were constructed using cumulative, differential, and relative (R-)plot methods. Crater depth was estimated from interior shadow lengths to assess relative depth-to-diameter ratios and compare degradation states, with analyses restricted to areas of similar illumination to minimise systematic bias.\\

Their crater counts are regionally focused, aimed at reconstructing the relative geological timeline in contrast to quantifying global impact flux. Comparison among the seven regions reveals that crater density, preservation, and size-frequency distributions vary with latitude and hemisphere, even within the cratered plains. \citet{Kinczyk2024} interpret these variations -- including the paucity of craters in equatorial plains, first noted by \cite{Kirchoff2009} -- as unlikely to be fully explained by viscous relaxation and plume deposition, since these mechanisms alone cannot erase craters over the age of the Solar System \citep{Bland2012}. Additionally, \citet{Kinczyk2024} suggest that tectonic resurfacing and the formation of terrains at different geological times are key mechanisms constraining crater preservation. This finding directly motivates the present study's approach to further (sub-)define and analyse distinct surface units globally, avoiding the mixing of crater populations.\\

\citet{Blanco-Rojas2024} introduced a machine-learning-based crater detection method, a convolutional neural network (CNN) trained to identify crater highlight–shadow pairs for deriving depth–diameter ratios. Their catalogue includes 5,240 craters, extending to small diameters ($<$1–2~km), with crater sizes measured using minimum bounding circles and depths estimated from shadow lengths under strict illumination constraints ($>60^{\circ}$ solar incidence and $>26^{\circ}$ camera phase angles). The study reports a missing rate of 17\%–26\%, particularly for subtle or degraded craters that are more readily identified manually. Although the CNN was applied globally, effective detection is strongly latitude dependent, with most reliably identified craters between $\sim70^{\circ}$N and $\sim40^{\circ}$S. For depth–diameter analysis, the dataset was further restricted to $0$–$30^{\circ}$N, where shadow visibility is maximised.\\

While the crater density was evaluated globally, \citet{Blanco-Rojas2024} did not perform region-based counts or distinguish between resurfaced and smooth terrains. Their catalogue aggregates craters across surface units spanning multiple ages and formation histories. Compared to earlier manual catalogues of \citet{Kirchoff2009}, the CNN-based study identifies more craters, primarily due to the inclusion of smaller diameters and the use of the improved global basemap from \citet{Bland2018}, which is adopted as the base map in this study. Previous studies explicitly targeted cratered plains and incorporated geological context (e.g., \citealt{CrowWillard2015}), whereas the CNN-based catalogue prioritises detection completeness over stratigraphic control. Although \citet{Blanco-Rojas2024} provides the most extensive crater dataset to date and therefore a valuable resource, it is not suitable for constructing a CPF because counts are not restricted to geologically homogeneous units, detection completeness varies spatially with illumination and projection, and resurfacing effects and crater retention states are not explicitly controlled.\\

While recent studies have mapped surface units and performed crater counting or size–frequency distribution analyses for selected regions -- primarily within the cratered plains -- no previous work has produced a global, unit-consistent crater dataset using uniform criteria across all terrain types. Consequently, existing crater catalogues are spatially limited and often combine craters from geologically distinct units, complicating efforts to derive an Enceladus-specific CPF or to estimate surface ages at the global scale.\\

Here, we address these limitations by presenting the first global, unit-specific crater catalogue for Enceladus, constructed using a unified geomorphological map and consistent crater identification criteria applied across all surface units. This approach enables spatially resolved surface age estimates and supports the development of a data-driven CPF, representing a substantial advance over previous regionally focused or morphologically mixed studies.

\subsection{Crater size-frequency measurement and distribution}
\label{subsec:sfd}
Impact craters form on the solid crust of the satellites. The term ``{\it size-frequency distribution}" (SFD) refers to a model of a statistical relationship between the diameters of craters (in km) and their frequency of occurrence, which is expressed as per unit area (km$^{-2}$). Smaller craters are relatively more abundant than larger craters. 
The cumulative size-frequency distribution is commonly expressed as a power law:
\begin{equation}
  n_{\rm >D} \propto (D_0/D)^\alpha, 
  \label{eq:sfd}
\end{equation}
where $D_0$ is a scaling diameter in km, and $\alpha$ is the cumulative slope for that given diameter range. For a constant impactor population and bombardment rate, an older surface exhibits a higher crater density, manifested as an up-shift in the SFD. \\

It is important to distinguish between two related but distinct terms used to describe crater populations. \textit{Size–frequency measurements} refer to the raw crater counts derived from image surveys of a specific surface unit and plotted as discrete data points. These measurements are subsequently fitted and summarised as a \textit{size–frequency distribution}, which represents an idealised description of the observed crater population. \\

The CPF, described in detail in Section~2, represents the expected initial SFD for a planetary surface unmodified by the geological process. It is derived from physical considerations of the impactor population and cratering mechanics, assumed to be globally applicable with time-invariant impactor SFD shape, and makes no assumption regarding the impact flux.
The distinction between the SFD and the CPF is that the SFD describes the observed crater population of a specific geological unit and is therefore surface-age dependent, whereas the CPF is (assumed to be) time-independent and represents the underlying impactor population. 
The CPF can be applied to different surface units --- and, in principle, across planetary bodies sharing the same projectile population --- to predict the expected SFD for a given surface age, while observed SFDs provide the empirical basis for constraining the CPF.\\

Throughout this study, crater populations are presented as \textit{cumulative size-frequency distributions}, in which the number of craters larger than a given diameter is plotted as a function of diameter, resulting in a monotonically decreasing curve. 
We adopt cumulative representations to maintain consistency with previous studies of crater chronology and crater production functions, particularly those developed for the Moon and other planetary bodies \citep[e.g.,][]{Neukum1983,Neukum2001}, thereby facilitating comparison with existing chronology frameworks. 
In addition, the observed crater populations on Enceladus exhibit complex variations in slope across different diameter ranges that cannot be adequately represented by a single, or even simple piecewise, power-law form. 
A cumulative representation therefore provides a convenient framework for describing the Enceladian CPF, which is later parameterised using a polynomial formulation.
Relative (R-)plots, which normalise crater frequencies to a $D^{-3}$ reference slope, remain useful for highlighting deviations from simple power-law behaviour. 
However, for consistency with established chronology methodology and previously derived production functions, the Enceladian CPF is represented here using a cumulative distribution.\\

\section{The Crater Production Function}
\label{subsec:intro_cpf}
The \textit{crater production function} (CPF) represents the underlying, unmodified crater size–frequency distribution on a planetary surface \citep{Neukum1975}. The CPF is presumed to remain unchanged over time \citep{Neukum1975,Neukum1983,Werner2014,Werner2023}, forming the basis for crater-based age estimates. Deviations between the actually observed crater SFD and the CPF yield insight into the satellite's geological history, in particular when specific areas were (partially) resurfaced (see Appendices B and C in \citet{Werner2009} for a detailed description).
Although the assumption of a time-invariant CPF could be seen as a limitation, recent studies suggest it is broadly valid: \citet{Bottke2024} found that the SFD of the heliocentric projectiles in the outer solar system stabilised within $\sim$40 Myr of planetesimal formation and has remained largely unchanged for the past 4.3 Ga.\\

While several ``crater production functions'' have been proposed for outer Solar System satellites based on impactor models or limited crater datasets (e.g., \citealt{Zahnle2003, Kirchoff2009, Singer2019, Wong2023}), a fully observed, body-specific crater production function for Enceladus (or any icy satellite) has not previously been constructed. Here, we present an Enceladian crater production function by quantifying the spatial distribution and size–frequency characteristics of craters across all major surface units, yielding a functional description of the unaltered crater population that is internally consistent at the global scale. \\

The CPF is commonly expressed as a polynomial function relating the logarithmic of the cumulative crater size-frequency ($N_{D}$) to the logarithm of the crater diameter ($D$) in km:
\begin{equation}
  \log_{10}(N_{D}) = \sum_{i=0}^{j} a_{i} \times [\log_{10}(D)]^i
  \label{eq:cpf}
\end{equation}
For the lunar CPF in the inner Solar System, the Neukum-style formulation uses $j = 11$, resulting in an 11$^{th}$-degree polynomial \citep{Neukum2001}. The polynomial form was adopted because the lunar crater SFD exhibits genuine slope variations across diameter ranges that a simple power law cannot capture \citep{Neukum1983}; the same reasoning applies to Mars \citep{Ivanov2001} and, as we demonstrate here, to Enceladus. Other production functions have also been defined for the Moon (e.g., \citealt{Marchi2009}).\\

\subsection{Crater and source population size-frequency distribution}
\begin{figure}
    \centering
    \includegraphics[width=0.6\linewidth]{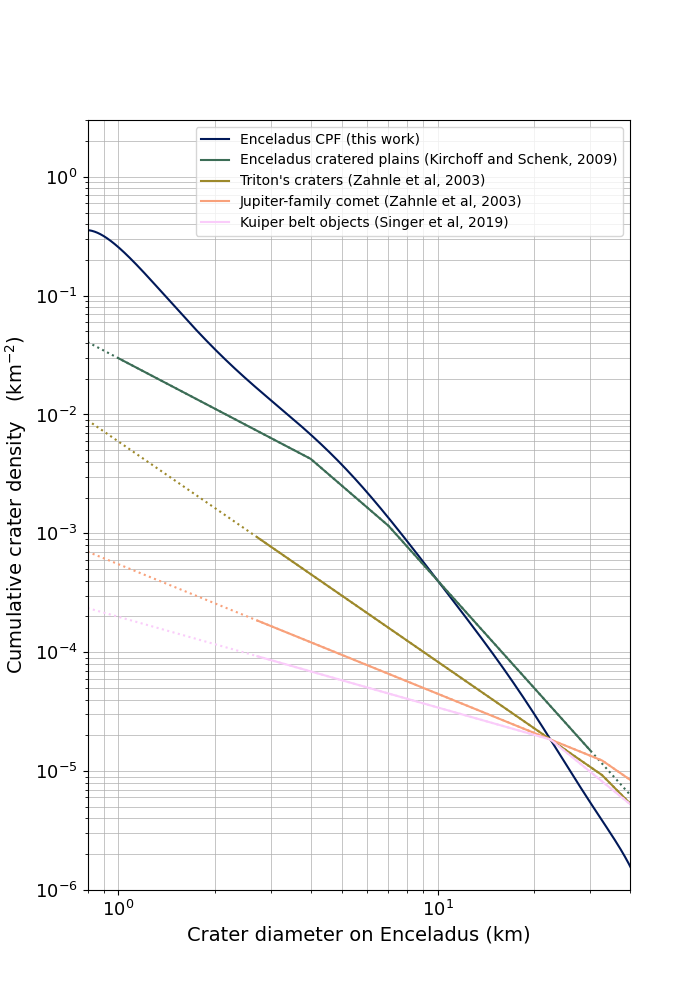}
    \caption{Enceladus crater production function and size-frequency distributions of craters and impactors in the outer Solar System. The displayed distributions include the Enceladus crater production function derived in this study (in blue), the crater size-frequency distribution fitted from the trailing hemisphere mid-latitude crater plains (\citet{Kirchoff2009}; in pine green), the impactor size-frequency distribution resembling Jupiter-family comets (case A from \citet{Zahnle2003}; in orange) and craters on Triton of the Neptunian satellite (case B from \citet{Zahnle2003}; in olive yellow), and Kuiper belt objects (\citet{Singer2019}; in pink). Impactor size-frequency distributions are converted to crater diameters using the \citet{Zahnle2003} crater scaling law for icy targets; adopting a different scaling law would horizontally shift the converted impactor distributions, but would not affect the observed crater-based production function. The dashed lines indicate diameter ranges where the distribution was not covered in the reference data but was extended with the cumulative slope of the adjacent diameter ranges.}
    \label{fig:cpf_compare}
\end{figure}

Previous crater studies of Enceladus and the outer solar system icy satellites focused on specific terrains, particularly cratered plains, and inferred projectile SFDs from three main sources:
a) Jupiter-family comets (yellow line in Fig.~\ref{fig:cpf_compare}), b) craters on Triton 
and c) trans-Neptunian objects (pink line in Fig.~\ref{fig:cpf_compare}) \citep{Zahnle2003, Singer2019, Bottke2024}.
Although these may all be viable sources, the comparisons rely on SFDs derived from different satellite systems (i.e., Jovian, Saturnian, Neptunian, and Pluto-Charon, respectively), where variations in planetary encounter velocities and therefore average impact velocities with the satellites, as well as material of the target surfaces, affect crater formation, potentially limiting their direct applicability to Enceladus and the other Saturnian satellites \citep{Neukum1983,Zahnle2003,DiSisto2007,DiSisto2011,DiSisto2013,DiSisto2016,Rossignoli2019,DiSisto2020,Rossignoli2022}.\\

Moreover, these projectile SFDs mainly describe kilometre-scale objects, with the best constraints for bodies with diameter $>$10~km. Yet, most observable craters on Enceladus range from a few hundred metres to 34~km in diameter. Given Enceladus’ low mass and the high impact velocities of heliocentric impactors (relative to some of the other mid-sized Saturnian satellites; \citealt{Zahnle2003, Wong2021}), most visible craters on Enceladus were formed by sub-kilometre impactors. Using the impactor-to-crater scaling law of \citet{Zahnle2003} for icy targets --- which depends various parameters such as surface gravity and mean impact velocity --- a 1~km impactor produces a crater of $\sim$22.6~km on Enceladus, a diameter exceeded by only seven observed craters on the surface. This sub-kilometre size range is precisely where the projectile population remains largely unknown, making it unreliable for constructing a CPF. \\

Previously, \citet{Wong2023} assumed an outer Solar System-wide CPF derived from the Pluto-Charon craters \citep[pink line in Fig~\ref{fig:cpf_compare}]{Singer2019}. However, such a distribution does not match that of Enceladus or the other Saturnian satellites of previous studies \citep[pine green line]{Kirchoff2022}. Similarly, a CPF constructed from craters on other satellites in different systems, such as Ganymede, Callisto, and Europa of the Jovian system \citep[orange line]{Zahnle2003}, exhibit similar discrepancies \citep{Wong2023}. Consequently, no existing reference provides a CPF derived solely from Enceladus data, prompting the need to construct one. The CPF resulting from the work presented here is shown in black in Fig~\ref{fig:cpf_compare}. \\

\subsection{Challenges in deriving crater production functions from crater counts}
Even when the CPF is inferred directly from crater counts on Enceladus, an accurate and unbiased representation is not guaranteed. Although extensive crater counts improve statistical precision, crater abundance alone is insufficient. Robust CPF modelling requires not only large datasets but also explicit consideration of how crater SFDs are modified by surface processes. 
On Enceladus, heavily cratered plains preserve abundant craters over a wide diameter range and therefore commonly serve as reference regions for crater-based age analyses. A prominent example located at 90$^\circ$E and 150$^\circ$W, and 30$^\circ$N and 80$^\circ$N, where the Far-Mid-CP1 to Far-Mid-CP6 units (``Far'' denoting the far side, ``Mid'' the mid-latitudes, and ``CP'' cratered plains) are located (Fig.~\ref{fig:division}). 
\citet{Wong2023} adopted the broken power-law crater SFD derived by \citet{Kirchoff2009} from part of these plains to estimate surface ages on Enceladus. While this approach is suitable for similarly cratered regions and adjacent terrains, systematic deviations emerge when the same SFD is applied to less densely cratered surfaces or to regions further from the reference plains.\\

These deviations do not necessarily indicate changes in the impactor population, but can instead arise from surface modification processes that preferentially affect smaller craters. Crater SFDs evolve over time, with the earliest and most pronounced effects at small diameters. Small craters are abundant, prone to overlap, and are selectively removed by obliteration from subsequent large impacts, erosion or burial by geological activity (such as plume fallout from cryovolcanism at Enceladus’ south pole), and observational incompleteness near image resolution limits. The primary limitation is image resolution (typically $\sim$100~m/pixel for high-resolution Cassini images of Enceladus), which hinders the detection of small craters (diameter $\lesssim$1 km) and causes a flattening of the distribution near the resolution threshold. Previous crater count studies (e.g., \citealt{Kirchoff2009, Kinczyk2024}) often identified craters below the resolution limit, but applied a cut-off above this limit to ensure completeness. \\

Furthermore, crater saturation happened in densely cratered units, where new impacts overwrite existing craters, preferentially erasing smaller ones. Crater destruction and production eventually reach equilibrium, beginning with small diameters; this state is characterised by a SFD with a cumulative power-law slope of approximately $-2$ \citep{Gault1970, Richardson2009}. On Enceladus, this equilibrium is more evident in cratered plains with higher overall crater densities than in sparsely cratered ridged plains. Consequently, small craters are more likely to be undercounted due to limited resolution, more readily erased, and more strongly affected by crater saturation, producing a characteristic roll-off at the small-diameter end of crater SFDs. Nevertheless, small craters remain essential for constraining young surface ages, where larger craters are rare or absent. \\

\begin{figure}
  \label{fig:resurface_exp}
	\centering
	\begin{subfigure}{1.0\textwidth}
		\centering
		\captionsetup{width=.9\linewidth}
		\resizebox{\hsize}{!}{\includegraphics{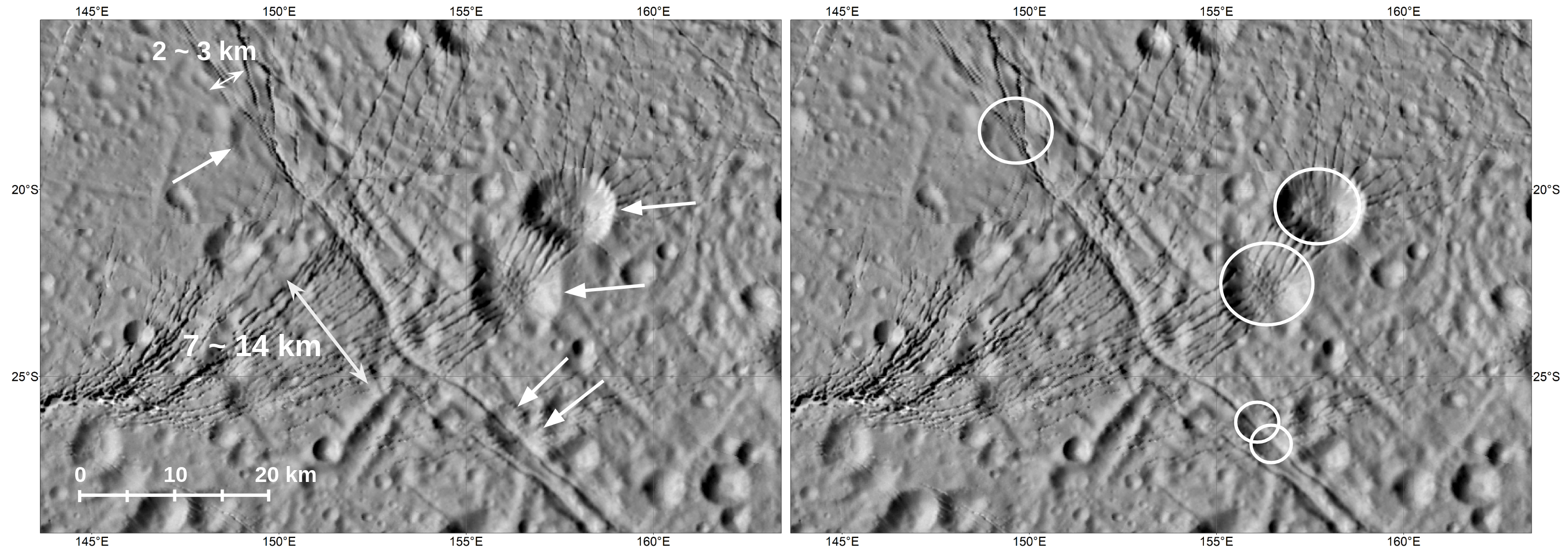}}
    \phantomcaption          
		\label{fig:cp_exp}
    \caption*{Figure~\ref{fig:cp_exp}: Located within the far-side equatorial cratered plain (Far-Eq-CP8). On the {\it left}, arrows indicate the craters partially eroded by two types of trough structures: (1) slightly wider trough, each about 2 to 3 km in width, occurring individually or in small groups, transects the craters from the NW-SE across the images, and (2) a cluster of narrower troughs that together span a width of around 7 to 14 km in the SW-NE-trending. This group of finer troughs has a large area coverage, and seems to diverge into separated branchs in its path and transect more craters. On the {\it right}, the same satellite image with circles highlighting the partly eroded craters ranging from 4 km to 10 km.\\}
	\end{subfigure}
	\newline
	\begin{subfigure}{.48\textwidth}
		\centering
		\captionsetup{width=.9\linewidth}
		\resizebox{\hsize}{!}{\includegraphics{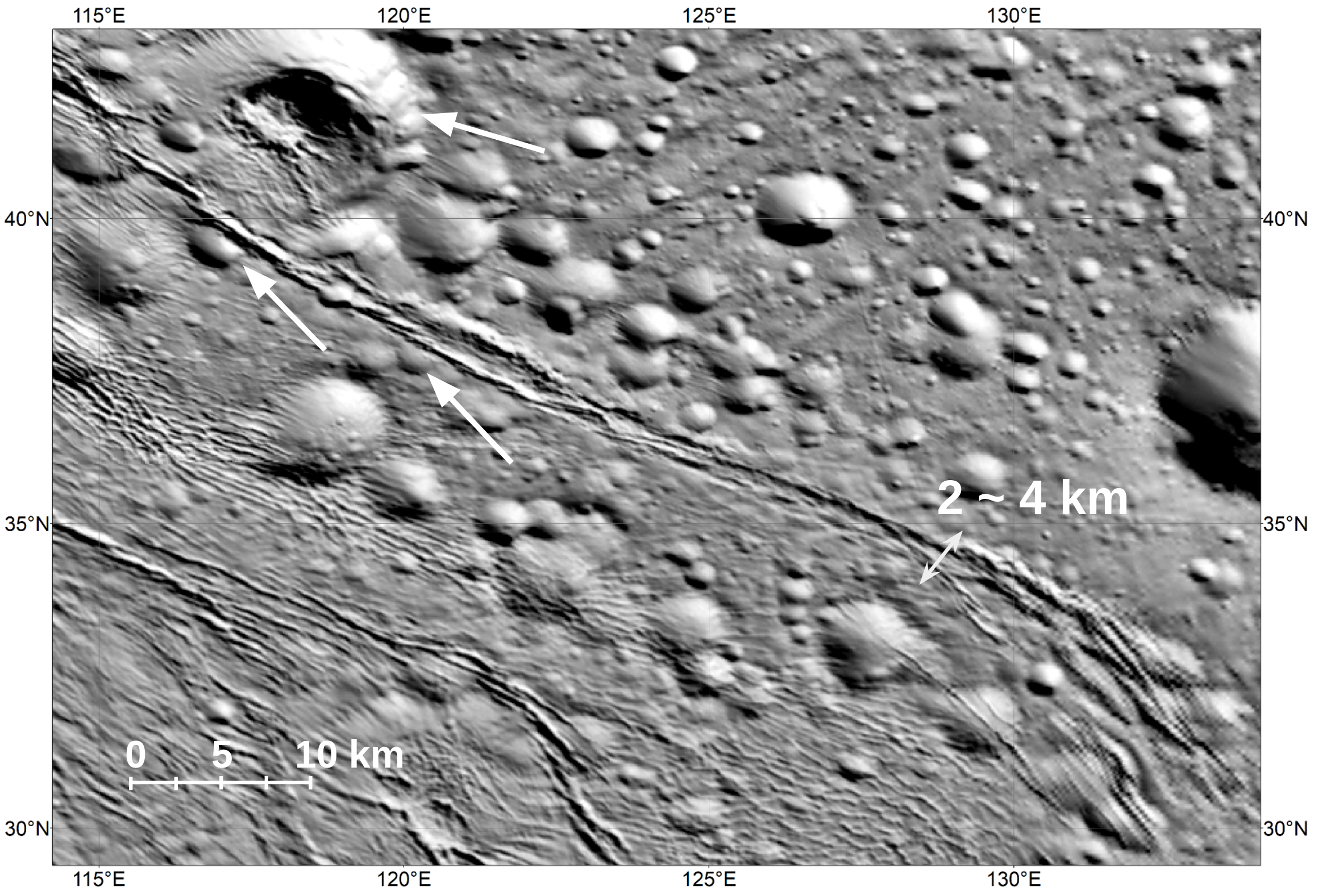}}
		\phantomcaption          
    \label{fig:cp_exp2}
    \caption*{Figure~\ref{fig:cp_exp2}: Near the boundary of the far-side northern mid-latitude cratered plain (Far-Mid-CP4) and the right-arms of the trailing hemisphere terrains (CW-T-curvilinear-right), arrows indicate craters partially eroded by trough formations with varying widths from 2 to 4 km.\\}
	\end{subfigure}
	\begin{subfigure}{.48\textwidth}
		\centering
		\captionsetup{width=.9\linewidth}
		\resizebox{\hsize}{!}{\includegraphics{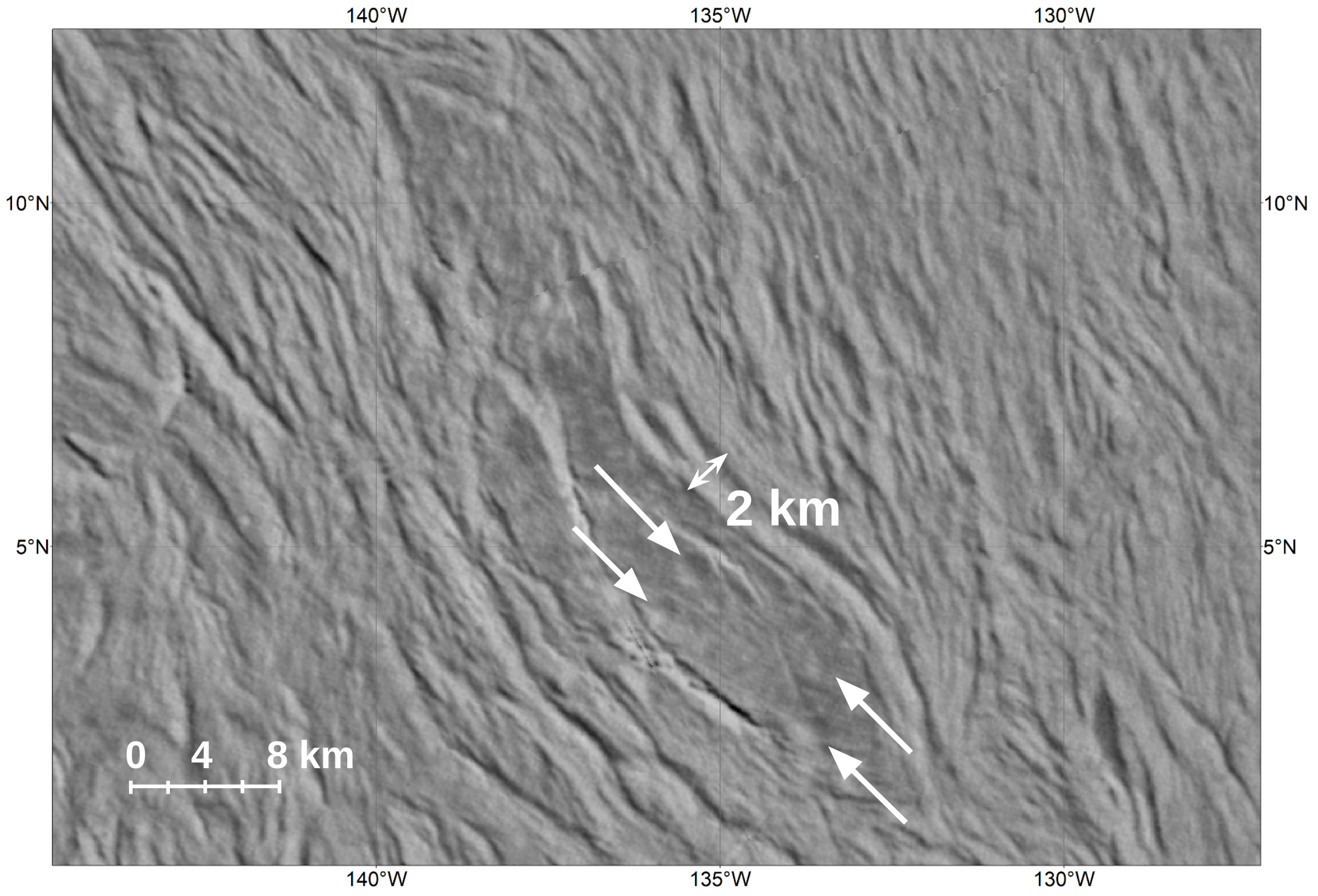}}
    \phantomcaption          
		\label{fig:rp_exp}
    \caption*{Figure~\ref{fig:rp_exp}: In the west region of the leading hemisphere curvilinear terrains (CW-L-curvilinear), there is a small patch of the less-altered area showing some subdued craters, ranging from 1.5 to 2 km.\\  \\}
	\end{subfigure}
\end{figure}

On Enceladus, the formation of geological structures and the deposition of plume material can modify crater records, particularly at small diameters, partially or entirely resetting crater counts. Surface morphologies -- including troughs, ridges, scarps, and fractures pit chains, listed here in decreasing scale and occurrence frequency \citep{CrowWillard2015, Martin2023} -- may form either before or after crater emplacement. When craters predate these structures, their preservation depends on the scale of deformation, such as the width of individual structures and the spacing between closely grouped features.\\

Cassini's images show geological structures intersecting and partially erasing craters in both heavily cratered plains (Fig.~\ref{fig:cp_exp} and Fig.~\ref{fig:cp_exp2}) and sparsely cratered ridge plains (Fig.~\ref{fig:rp_exp}). Figure~\ref{fig:cp_exp} shows the far-side equatorial cratered plains (Far-Eq-CP8), where craters smaller than, or comparable in size to, the intruding troughs ($\sim$ 2–10~km wide, depending on the structure) are susceptible to complete removal or severe degradation, effectively excluding them from crater counts. In contrast, only craters larger than the structural scale are likely to survive or remain detectable, as illustrated in Fig.~\ref{fig:rp_exp}. In addition, small craters with initially shallow depths, depending on their depth-to-diameter ratios, may be more rapidly infilled by plume deposition \citep{Kirchoff2009}.\\

If the surface undergoes complete renewal, all pre-existing craters are erased, and a new crater record begins at a different surface age. In this case, the resulting SFD is expected to follow the CPF until limited by image resolution or crater saturation, allowing age fitting to proceed. However, when resurfacing is partial or selectively affects craters within specific diameter ranges, pre- and post-resurfacing craters coexist. These populations may be distinguished based on crater morphology: craters that predate resurfacing commonly appear subdued, distorted, or intersected by tectonic features, whereas post-resurfacing craters tend to be more pristine, with sharper rims and deeper floors. Even in terrains affected by multiple resurfacing events and complex geomorphology, both surface solidification ages and resurfacing ages can be inferred from crater size-frequency measurements, as distinct drop or deviations from the CPF indicate resurfacing episodes \citep{Michael2010}. 
Some mechanisms of selective crater removal can further distort the SFD. As proposed by \citet{Kirchoff2009}, a relative depletion of larger craters may result from the combined effects of viscous relaxation and plume burial, which can preferentially degrade or obscure larger craters on Enceladus. 
Additional processes, such as impact gardening \citep[e.g.][]{Costello2021}, may further modify crater preservation states through progressively reworks the uppermost surface and can contribute to degradation and loss of small craters over time.\\

In summary, assuming a unique impactor population, variations in crater size-frequency measurements within the same terrain or planetary body can be attributed to: i) differences in image resolution and crater detectability between surface units; ii) crater saturation, where older and more heavily cratered surfaces exhibit roll-over at larger diameters; and iii) selective removal of craters below a threshold diameter, typically associated with partial resurfacing. 
The relative contribution of these effects depends on the surface unit and the diameter range considered. Image resolution primarily affects the small-diameter end of the SFD, crater saturation dominates in cratered plains and can even influence larger diameters in heavily cratered terrains, and selective crater removal introduces more complex distortions. However, most of these modifications are easily assessed as documented in Figure~\ref{fig:cp_exp} to \ref{fig:rp_exp}.\\

\subsection{Extracting the Enceladian crater production function} 
Previous surface age estimates for Enceladus have relied on imported or theoretical CPFs, either scaled from the lunar chronology \citep{Spencer2006} or derived from modelled outer solar system impactor populations \citep{Kirchoff2009,Wong2023,Bottke2024}. None of these approaches have been directly validated using Enceladus' own crater record. In contrast, the CPF developed here is explicitly anchored to the observed global crater population of Enceladus, providing a reliable, body-specific framework that minimises assumptions inherited from other planetary environments and physical parameters.\\

The key to extracting the underlying CPF is recognising that each surface unit preserves a reliable crater record only over a restricted diameter range that remains largely unaffected by geological modification. Diameter ranges affected by crater saturation, observational incompleteness, resurfacing, or other modification processes are excluded from the fitting procedure. The CPF shape is then recovered by combining the remaining valid crater populations from multiple surface units of different ages and morphologies through overlap normalisation, following the approach of \citet{Neukum1983}. The consistent size--frequency distribution that emerges from these overlapping crater populations is interpreted as the underlying Enceladian CPF. The detailed implementation, representative unit selection criteria, and reliable crater diameter ranges are described in Section~\ref{subsec:cpf}.\\

Importantly, our CPF is not constructed solely from heavily cratered terrains, which may under-represent smaller craters owing to erosion, resurfacing, or potential crater saturation effects. Instead, guided by the established principles of lunar crater chronology \citep{Neukum1983}, we derive the Enceladian CPF from a global analysis spanning multiple surface units with different crater densities and geomorphological characteristics. This approach minimises biases against small craters by incorporating terrains where smaller crater remain comparatively well preserved, while simultaneously allowing assessment of regional variations in crater size-frequency distributions that may reflect differences in resurfacing history or, potentially, variations in impactor populations. By leveraging global coverage and unit-to-unit comparisons, the resulting CPF provides a more robust and internally consistent framework for future age-dating applications on Enceladus and other icy satellites. We note, however, that crater relaxation driven by viscous ice flow and crater floor infill by E-ring material cannot be fully quantified, as Enceladus' historical heat flux and long-term south polar plume deposition rates remain poorly constrained. These processes therefore represent
recognised limitations of the present analysis.\\


\section{Data and Methods}
\label{sec:data}

\subsection{Global crater survey on Enceladus}
\label{subsec:survey}
Crater counting is a subjective process whose outcomes vary among researchers \citep{Robbins2014}. To ensure reliability, this study independently conducted crater counts on Enceladus by E. Wong (EW) and M. Kirchoff (MK). EW surveyed craters from 50$^\circ$S to 90$^\circ$N, while MK covered from 90$^\circ$S to 60$^\circ$N, with an overlap at 50$^\circ$S to 60$^\circ$N (approximately 82\% of Enceladus' global area). Counts, where overlapping, were cross-checked for statistical robustness. EW’s and MK’s set had no specific lower limit to identify craters; a minimum cutoff at 800 m for EW and 1 km for MK is applied after surveying and recording all candidate craters. A consensus cutoff is introduced for the entire survey of 16,958 craters. 
The global mosaic image of \citet{Bland2018} was adopted for their unified coordinate system and comprehensive surface coverage. As variations in emission angles, phase angles, and image resolution persist among the constituent images across the mosaic, we restrict our study to craters with diameters larger than 800~m; these imaging and instrumental effects are discussed further in Section~\ref{subsec:instrumental}.\\

EW’s survey utilised the geodetically controlled image sets released by \citet{Bland2018}: a global mosaic of 108 high-resolution images (55–419 m/pixel) in equirectangular projection, covering 96\% of the surface at better than 200 m/pixel and 18\% at better than 100 m/pixel, as well as north and south polar stereographic mosaics. Despite the high coverage, variations in image resolution and lighting conditions affect the minimum resolvable crater diameter. Sub-kilometre craters are identifiable in mid-latitude and equatorial regions but are more difficult to discern near the north pole due to lower resolution, even though similar crater abundances are expected based on crater count at large diameters ranges and the region's proximity. Furthermore, solar incidence angle influences crater identification \citep{Ostrach2011, Robbins2025}, especially on the central leading hemisphere, where low incidence angles reduce shadow length and crater visibility. For crater counting and geomorphological analysis in the polar regions—particularly the heavily cratered north pole—EW used polar stereographic projections to accurately identify craters and surface features. At high latitudes, craters appear increasingly elliptical and distorted in equirectangular projection, often as straight horizontal lines, so polar stereographic images from \citet{Bland2018} were used for crater identification and size measurement in polar regions.\\

MK's counts have been gathered for over a decade (2007–current) and have used the greyscale mosaics generated by Paul Schenk throughout that time frame \citep{Schenk2011,Schenk2018} with the most recent being published in \citet{Schenk2024}. Measurements were updated and added in both location and diameter as cartography and imaging improved \citet{Kirchoff2016,Kirchoff2018}. As the images used in the mosaic did not vastly differ from those used by \citet{Bland2018}, similar issues affected MK's counts as EW's counts discussed above.\\

EW employed ArcGIS with the CraterTools add-in \citep{Kneissl2011} and QGIS with the OpenCraterTool add-in \citep{Heyer2023} to locate and document the diameter and coordinates of candidate craters. MK employed both USGS ISIS (for only the earliest counts; \citealt{Kirchoff2009}) and JMARS (http://jmars.asu.edu/; \citealt{Christensen2009}) to measure crater diameters and locations.  In JMARS both the 3-point crater counting and ellipse shape tools were used.

\subsection{Assessing the global crater record in subdivided units}
\label{subsec:units}
Previous mapping efforts have often grouped extensive regions as single units. For example, \citet{CrowWillard2015} classified 320,000~km$^{2}$ (41\% of Enceladus’ surface) as a single cratered plain. Although \citet{Kirchoff2009} separated the anti-Saturn side cratered plains into mid-latitude and equatorial units, these still encompass broad areas with apparent internal differences. Combining surfaces with potentially different solidification times and geological histories -- especially when younger, lower-crater-density surfaces are included -- dilutes crater statistics by mixing distinct crater populations. This mixing appears as drops in incremental frequencies or shallowing in cumulative frequencies, since smaller craters are more numerous, resulting in turnover or a reduced slope in the SFD at small diameters.\\

\begin{figure}
    \centering
    \includegraphics[width=1.0\textwidth]{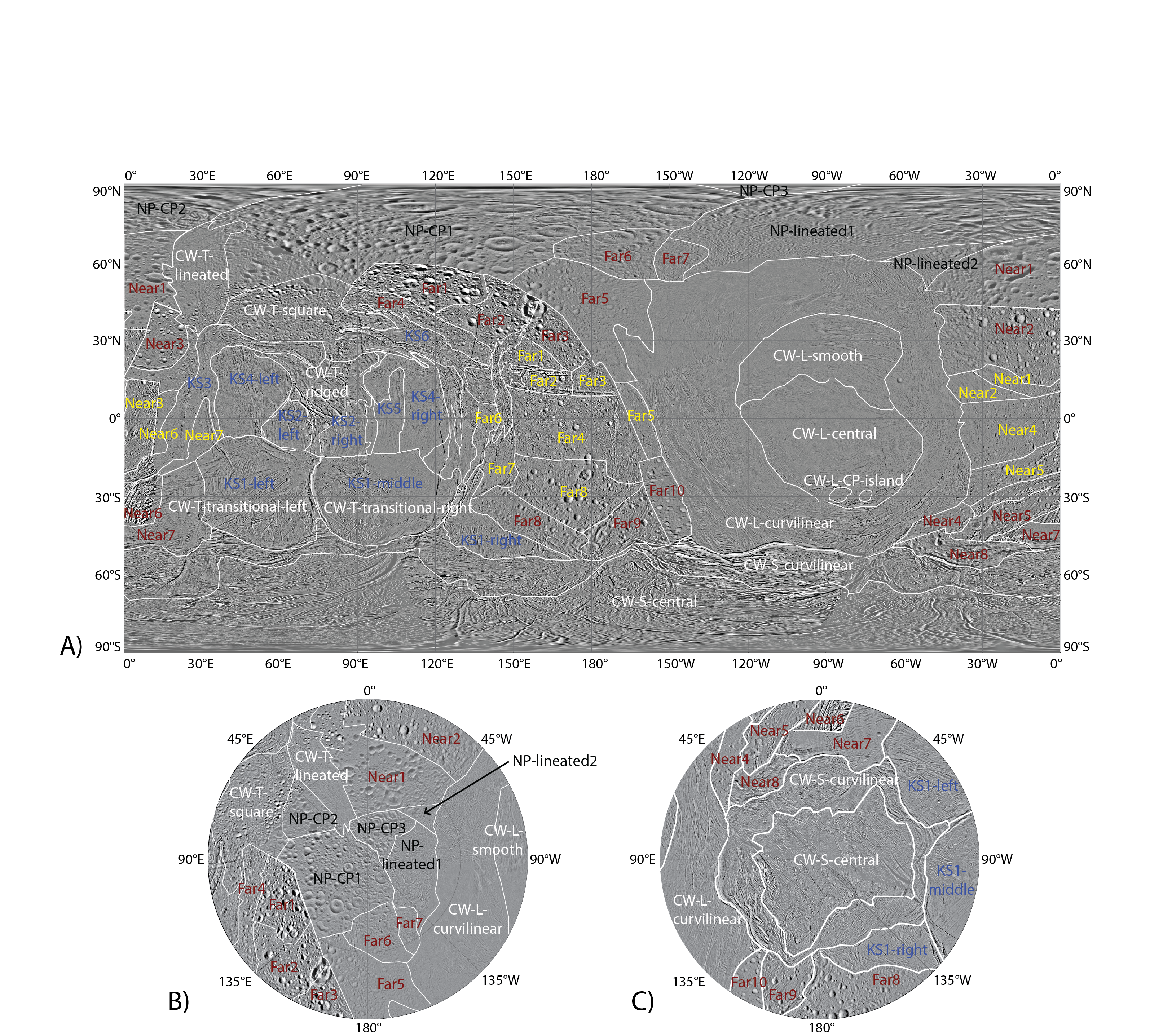}
    \caption{Division of our units on Enceladus. \textbf{(A)} in equirectangular projection from 90°N to  90°S, and \textbf{(B)} and \textbf{(C)}  in polar stereographic from pole to 30° latitude in the north and south poles, respectively. Unit names follow the convention: ``Far'' or ``Near'' denotes the far or near side, ``Mid'' or ``Eq'' the mid-latitude or equatorial region, and ``CP'' the cratered plain. Due to space constraints, not all units are labelled with full names. The prefix “RP” for ridged plains from \citet{Kirchoff2009} is omitted. Colour variations (dark red and yellow) distinguish between ``Mid'' (for mid-latitude) and ``Eq''' (for equator) without repeated labelling. Units in the north pole defined in this work are labelled in black. Units referencing \citet{Kirchoff2009} are in blue, while those referencing \citet{CrowWillard2015} are in white.}
    \label{fig:division}
\end{figure}

To ensure representative crater statistics and size-frequency measurements, both of which are essential for interpreting surface ages and regional geological history, we subdivided Enceladus’ surface into 62 analytical units (Fig.~\ref{fig:division}) based on three simultaneously considered criteria: 
(i) large-scale geomorphological structures --- ridges, troughs, and surface texture transitions (Section~\ref{subsec:geomorphology}); 
(ii) spatial variations in crater density, reflecting differences in resurfacing history and solidification age (Section~\ref{subsec:crater_distribution}); 
and (iii) imaging properties, particularly image resolution and solar incidence angle, to prevent instrumental artefacts from being misinterpreted as geological signals (Section~\ref{subsec:instrumental}).
Consequently, some adjacent units may represent a single geological province and could be merged in future work with improved imaging. 
15 of the units are based on global geomorphological mapping \citep{CrowWillard2015}, 10 are adapted from previous crater distribution studies \citep{Kirchoff2009}, and the remainder are newly defined here, with particular focus on finer subdivision of the heavily cratered terrains, which cover $\sim$43\% of the surface \citep{CrowWillard2015}.\\

The names and crater information for all 62 units are listed in Table~\ref{tab:analytical_units} and their spatial distribution is shown in Fig.~\ref{fig:division}. Units previously defined by \citet{CrowWillard2015} are prefixed with ``CW'', those by \citet{Kirchoff2009} with ``KS'', and are further labelled ``T'' for the trailing hemisphere or ``L'' for the leading hemisphere. Newly defined units, primarily within the cratered plains, 
are prefixed with ``Near'' for units on the sub-Saturn (Saturn-facing) hemisphere  and `` Far'' for units on the anti-Saturn hemisphere,
followed by ``Mid'' for mid-latitude or ``Eq'' for equatorial regions. North polar units are labelled ``NP'', followed by ``CP'' for crater plains or ``lineated'' for lineated plains. For the south polar regions, we adopt the nomenclature of \citet{CrowWillard2015}, including the south polar curvilinear plains (CW-S-curvilinear) and central plain (CW-S-central).\\

Of the 62 units, 13 were selected as representative units for CPF derivation following the criteria of \citet{Neukum1983}: units must exhibit spatially homogeneous crater densities, show no evidence of significant partial resurfacing or preferential small-crater loss, and contain sufficient craters for statistically robust size-frequency measurements. The minimum reliable crater diameter was assessed individually per unit based on surface morphology and imaging conditions, typically ranging from 1 to 5~km. More explanations is described in next section, full documentation and justification for each representative unit is provided in Supplementary Section~S1.\\

Units adopted from \citet{CrowWillard2015} and \citet{Kirchoff2009} retain their original boundaries as published and are not further subdivided, with two exceptions: (i) the heavily cratered plains, which are further subdivided in this study, and (ii) the paired units on the trailing hemisphere --- CW's striated and transitional plains and KS's ridged plain (rp-)1, rp2, and rp4 --- which each comprise a left and right part (see Fig.~2 of \citealt{CrowWillard2015} and Fig.~4 of \citealt{Kirchoff2009}). For these paired units, crater counts and size-frequency measurements were conducted separately for each part, indicated by the ``-left'' 
and ``-right'' labels. \\

\begin{longtable}{p{4.0cm} p{3.0cm} p{2.0cm} p{3.0cm}}
\caption{List of the 62 units, among them 13 representative units, with $*$ superscript, are used to derive the crater production function. From left to right are the name codes, rough coordinates of the centre of the unit, the number of craters counted within the unit, and the maximum crater diameter from the unit.
Alphabetical subscripts identify seven units from \citet{CrowWillard2015} and nine units from \citet{Kirchoff2009} that cover overlapping regions on the ridged plains of the trailing hemisphere; units sharing the same subscript spatially overlap, where one CW unit may correspond to one or two KS units. Their boundaries differ in extent and do not coincide exactly.}
\label{tab:analytical_units}\\
\hline
  \textbf{Analytical unit} & \textbf{Lat., Long. (°)} & \textbf{No. Craters} & \textbf{Max. D$_{Cr}$ (km)} \\
\hline
\endfirsthead  

\multicolumn{4}{l}{\textit{Table~\ref{tab:analytical_units} continued from previous page}}\\
\hline
  \textbf{Analytical unit} & \textbf{Lat., Long. (°)} & \textbf{No. Craters} & \textbf{Max. D$_{Cr}$ (km)} \\
\hline
\endhead       

\hline
\multicolumn{4}{r}{\textit{Continued on next page}}\\
\endfoot       

\hline
\endlastfoot   

CW-L-central$^{*}$ & 90°W, 10°S & 34 & 3.1 \\
CW-L-CP-island & 86°W, 27°S  & 37 & 4.2 \\
CW-L-curvilinear &  140°W, 10°N & 7 &  13.1 \\
CW-L-smooth &  90°W, 25°N & 274 &  2.0 \\
CW-T-curvilinear-left$^{A}$ & 25°E, 5°N & 93 &  15.3\\
CW-T-curvilinear-right$^{B}$ & 125°E, 20°S & 143 &  15.8\\
CW-T-north-lineated & 30°E, 55°N & 187 &  16.3\\
CW-T-ridged$^{C}$ & 80°E, 5°N & 134 &  11.4 \\
CW-T-square & 70°E, 40°N & 653 &  21.4 \\
CW-T-striated-left$^{*,D}$ & 50°E, 10°N & 86 & 4.6 \\
CW-T-striated-right$^{E}$ & 110°E, 5°N & 120 & 4.4 \\
CW-T-transitional-left$^{*,F}$ & 35°E, 30°S & 125 & 7.2 \\
CW-T-transitional-right$^{G}$ & 155°E, 30°S & 132 &  4.3 \\
CW-S-curvilinear & 0°, 60°S & 39 &  3.9 \\
CW-S-central & 0°, 90°S & 32 &  2.8 \\
Far-Eq-CP1 & 155°E, 25°N & 199 &  6.9 \\
Far-Eq-CP2 & 157°E, 15°S & 210 &  15.1 \\
Far-Eq-CP3 & 180°, 15°N & 149 &  10.2\\
Far-Eq-CP4$^{*}$ & 175°E, 5°S & 1576 & 13.8 \\
Far-Eq-CP5 & 162°W, 0° & 213 &  4.7 \\
Far-Eq-CP6$^{*}$ & 137°E, 10°S & 125 & 5.6 \\
Far-Eq-CP7 & 144°E, 15°S & 219 &  8.8 \\
Far-Eq-CP8$^{*}$ & 175°E, 30°S & 857 & 20.8\\
Far-Mid-CP1$^{*}$ & 115°W, 50°N & 638 & 15.4 \\
Far-Mid-CP2$^{*}$ & 142°W, 42°N & 579 & 31.2 \\
Far-Mid-CP3$^{*}$ & 166°W, 30°N & 436 & 31.2\\
Far-Mid-CP4 & 112°E, 40°N & 243 &  12.1\\
Far-Mid-CP5$^{*}$ & 180°W, 40°N & 813 & 31.2\\
Far-Mid-CP6 & 170°W, 63°N & 164 &  22.5\\
Far-Mid-CP7 & 147°W, 62°N & 49 &  16.8\\
Far-Mid-CP8 & 155°E, 40°S & 369 &  9.2\\
Far-Mid-CP9 & 165°W, 40°S & 315 &  22.4\\
Far-Mid-CP10 & 152°W, 30°S & 338 & 13.1 \\
Near-Eq-CP1 & 20°W, 15°N & 180 &  7.3\\
Near-Eq-CP2 & 30°W, 10°N & 164 &  8.1\\
Near-Eq-CP3 & 20°E, 20°N & 172 &  10.7\\
Near-Eq-CP4 & 20°W, 5°S & 622 &  10.3\\
Near-Eq-CP5 & 2°E, 15°S & 609 &  15.3\\
Near-Eq-CP6 & 15°E, 2°S & 43 &  6.7\\
Near-Eq-CP7 & 30°E, 10°S & 105 &  6.5\\
Near-Mid-CP1$^{*}$ & 15°W, 55°N & 1138 & 34.0 \\
Near-Mid-CP2$^{*}$ & 20°W, 30°N & 1450 & 15.0 \\
Near-Mid-CP3 & 15°E, 30°N & 448 &  10.7\\
Near-Mid-CP4 & 45°W, 42°S & 150 &  8.5\\
Near-Mid-CP5 & 20°W, 37°S & 128 &  21.8\\
Near-Mid-CP6 & 5°W, 32°S & 20 &  2.6\\
Near-Mid-CP7 & 20°E, 40°S & 44 &  6.8\\
NP-CP1 & 120°E, 75°N & 899 &  25\\
NP-CP2 & 55°E, 70°N & 192 &  16.4\\
NP-CP3 & 30°W, 80°N & 96 &  23.3\\
NP-lineated1 & 105°W, 75°N & 42 &  30.9\\
NP-lineated2 & 55°W, 70°N & 10 &  16.7\\
KS-RP1-left$^{F}$ & 50°E, 25°S & 80 &  7.2\\
KS-RP1-middle$^{G}$ & 100°E, 30°S & 113 &  4.3\\
KS-RP1-right$^{G}$ & 85°E, 5°S & 24 &  2.9\\
KS-RP2-left$^{C}$ & 64°E, 3°S & 23 &  4.3\\
KS-RP2-right$^{C}$ & 140°E, 45°S & 41 &  3.6\\
KS-RP3$^{A}$ & 35°E, 30°N & 157 &  9.2\\
KS-RP4-left$^{D}$ & 45°E, 12°S & 67 & 4.1 \\
KS-RP4-right$^{*,E}$ & 120°E, 5°N & 97 & 5.2 \\
KS-RP5 & 100°E, 1°S & 47 &  5.0 \\
KS-RP6$^{B}$ & 140°E, 15°N & 209 & 15.8 \\
\end{longtable}

While a unit typically denotes a region with similar geological history and geomorphology, our definition is also influenced by variations in image resolution and imaging conditions. 
A geologically coherent region that was imaged by Cassini as separate images of different resolution would exhibit an artificial flattening in the SFD at small diameters in the lower-resolution portion, which is indistinguishable from geological resurfacing unless accounted for; such regions are therefore subdivided into separate units to ensure reliable crater statistics within each unit.
Where adjacent units are separated for instrumental in addition to geomorphological and crater distribution reasons, their SFDs are expected to be mutually consistent well above the reliable diameter limits --- set by the poorer-resolution constituent image. In a comprehensive geological map developed in future work, instrumentally separated areas would likely be combined and represented by a single geologic unit.\\ 


\subsection{Preparing crater counts to derive the Enceladus-specific crater production function}
\label{subsec:cpf}
Tectonic activity and other erosive processes are hypothesised to preferentially remove smaller craters while leaving larger ones relatively intact \citep{Michael2010}. Consequently, small and potentially removed craters are under-represented in the observed crater size-frequency measurements, causing the cumulative distributions to gradually, and sometimes irregularly, roll off towards smaller diameters. Directly fitting such modified crater populations may therefore lead to a misinterpretation of the CPF.\\

In heavily cratered terrains, subsequent impacts preferentially erase smaller craters, flattening the small-diameter end of the observed crater SFD through crater saturation and equilibrium processes \citep{DiSisto2016,Minton2019}. In tectonically active regions, ridges, troughs, and resurfacing fronts may similarly obscure or remove craters of comparable or smaller size. Conversely, smoother terrains with lower crater densities may preserve smaller craters more effectively, with some units retaining crater populations down to diameters of $\sim$1~km. These effects motivate the identification of a threshold diameter above which crater populations are considered minimally modified and therefore suitable for CPF derivation.\\

To mitigate these effects, we analysed the global crater database in conjunction with surface morphology to identify, for each unit, a minimum threshold diameter below which crater counts are no longer considered reliable for a given unit. Specifically, we examined craters with diameters comparable to tectonic features traversing the same terrain, or those potentially susceptible to removal by partial resurfacing, often extending from adjacent (younger) terrains.
The threshold diameter was determined using three complementary assessments: (i) a morphological assessment, in which the sizes of tectonic features --- including ridges, troughs, and resurfacing fronts --- were measured, as craters of comparable or smaller size are likely to have been obscured or erased; 
(ii) an image-resolution assessment, in which the diameter below which craters become increasingly difficult to resolve was estimated from the image pixel scale and illumination conditions of each unit; 
and (iii) an SFD assessment, in which the diameter at which the cumulative crater size-frequency distribution begins to flatten or roll off was identified, indicating the onset of preferential crater removal or observational incompleteness, as corroborated by criteria (i) and (ii). For each unit, the adopted threshold was taken as the largest of the three estimates and typically ranges from 1 to 5~km.\\

Given these considerations, not all mapped units are suitable for reconstructing undisturbed crater production distributions. We therefore selected a list of units that best preserve presumed primary crater populations, using established photogeological criteria for reliable crater statistics as recommended by \citet{Neukum1983} for constructing the lunar crater production function.
Table~\ref{tab:representative_units} lists the selected representative units, their central coordinates, number of craters within the unit, and the crater diameter ranges used to construct the CPF.\\

Unit selection followed the rigorous requirements outlined by \citet{Neukum1983}, ensuring that crater populations reflect a homogeneous surface age. We prioritised geologically coherent units, that is, units sharing a similar formation history and bounded by mappable structural or morphological boundaries --- with uniform crater densities (avoiding those exhibiting gradual spatial variation). We evaluated whether observed craters formed during the exposure of the mapped surface or represented remnants of older, partially eroded terrains. Units exhibiting significant surface modification, such as partial resurfacing or preferential loss of small craters, were excluded. 
Secondary craters were not considered, as their expected sizes fall below the effective resolution of available Enceladus imagery. The maximum diameter of secondary craters is typically $\sim$5\% of the primary crater diameter \citep{Schultz1980, Bierhaus2018} and decreases with distance from the source crater \citep{McEwen2006}. 
Based on the largest crater on Enceladus ($\sim$35~km), the maximum expected secondary diameter is $\sim$1.7~km; because Enceladus hosts only a limited number of large craters (approximately a dozen craters that exceed 20~km), the resulting secondaries are expected to be sub-kilometre in size.
At the image resolutions used in this study (typically $\ge$200 m pixel$^{-1}$), such small craters would fall near or below the reliable detection limit and are therefore unlikely to be represented in the counted crater population. Documentation and justification for choosing each of the 13 representative units is provided in Supplementary Section~S1.\\

To ensure statistical robustness across the full diameter range, we normalised and 
joined the crater size-frequency measurements from the representative units spanning distinct crater-size regimes through an overlap normalisation procedure: adjacent units contributing overlapping diameter ranges 
are scaled to a common reference density by averaging the crater frequency ratios within shared diameter bins, following \citet{Neukum1983}.
Crater diameters were measured consistently at the rim crest, and unit surface areas were precisely defined to ensure accurate normalisation of crater densities.\\

Small craters ($<3$~km) occur widely across Enceladus. However, to derive an unbiased CPF, small-crater counts were obtained exclusively from sparsely cratered smooth plains selected as representative units of the small-crater population (e.g., CW-L-central, CW-T-transitional-left, CW-T-striated-left, and KS-RP4-right; marked from navy to green in Fig.~\ref{fig:age}). Their low crater densities minimise potential smoothing of the crater SFD due to crater saturation, while the absence of nearby large impacts reduces possible removal by ejecta emplacement, local resurfacing, and contamination by secondary craters.\\

In contrast, large craters ($>10$~km) are intrinsically rare, such that crater counts from individual units suffer from substantial Poisson uncertainties. To improve counting statistics, crater populations from neighbouring heavily cratered mid-latitude units exhibiting comparable geological histories and similar crater SFDs were combined: Near-Mid-CP1 and Near-Mid-CP2 on the near side (marked as pink crosses in Fig.~\ref{fig:age}B--D), and Far-Mid-CP1, Far-Mid-CP2, and Far-Mid-CP3 on the far side (marked as pink stars in Fig.~\ref{fig:age}B--D). 
Combining these adjacent units substantially reduces statistical scatter at large diameters, yielding ten sampling datasets derived from the 13 representative units. 
Together — smooth plains for the small-diameter range, and combined heavily cratered terrains for the large-diameter range — these datasets enabled us to derive the Enceladian CPF over crater diameters from 800~m to 35~km, encompassing the minimum resolvable to the maximum observed diameter.

\subsection{Scaling and fitting the crater production function}
\begin{table}[t]
\centering
    \begin{tabular}{p{4.0cm} p{3.0cm} p{2.0cm} p{3.0cm}}
\hline
\textbf{Representative unit} & \textbf{Lat., Long. (°)} & \textbf{No. Craters} & \textbf{Fitting D$_{cr}$ ranges (km)} \\
\hline
CW-L-central & 90°W, 10°S & 34 & 0.8 $\sim$ 2.5 \\
CW-T-transitional-left & 35°E, 30°S & 125 & 0.9 $\sim$ 3.8 \\
CW-T-striated-left & 50°E, 10°N & 86 & 1.0 $\sim$ 2.0 \\
KS-RP4-right & 120°E, 5°N & 97 & 1.0 $\sim$ 4.4 \\
Far-Eq-CP4 & 175°E, 5°S & 1576 & 2.0 $\sim$ 10.2\\
Far-Eq-CP6 & 137°E, 10°S & 125 & 1.6 $\sim$ 5.0 \\
Far-Eq-CP8 & 175°E, 30°S & 857 & 6.0 $\sim$ 14.9 \\
Near-Mid-CP1 & 15°W, 55°N & 1138 & 8.0 $\sim$ 34.0 \\
Near-Mid-CP2 & 20°W, 30°N & 1450 & 8.0 $\sim$ 15 \\
Far-Mid-CP1 & 115°W, 50°N & 638 & 9.0 $\sim$ 15.4 \\
Far-Mid-CP2 & 142°W, 42°N & 579 & 9.0 $\sim$ 19.5 \\
Far-Mid-CP3 & 166°W, 30°N & 436 & 9.0 $\sim$ 31.2 \\
Far-Mid-CP5 & 180°W, 40°N & 813 & 4.0 $\sim$ 13.1 \\
\hline
\end{tabular}
\caption{List of the representative units used to derive the crater production function. From left to right are the name codes, rough coordinates of the centre of the unit, number of crater counted within the unit, and the crater diameter ranges (in km) used to fit and scale for the crater production function.}
\label{tab:representative_units}
\end{table}

CPF construction methodology established a clear distinction between the time-independent shape of the CPF and the time-dependent absolute crater density. \citet{Neukum2001} and \citet{Ivanov2001} demonstrated that crater size–frequency measurements from surfaces of different ages can be directly compared by vertical normalisation (shifting of crater densities or count in logarithmic scale), provided the distributions overlap in diameter ranges unaffected by saturation, resurfacing, or resolution limits. In this framework, the production function shape is considered invariant, while differences in crater density reflect variations in exposure time. Importantly, because no single surface samples the full range of crater diameters or densities without bias, or fully records all resurfacing events, the CPF must be assembled from multiple surfaces with different crater densities.\\

Some units contain abundant craters spanning a broad diameter range, while others --particularly smaller ridged or smooth units -- may only have limited crater counts at small diameters. Ideally, each diameter range would be sampled by multiple measurements from different units; however, practical constraints such as image resolution, resurfacing, and post-impact modification restrict the usable size range for each region. Maximising overlap among measurements minimises statistical uncertainties and prevents gaps in the CPF.\\

Practical guidance, such as \citet{Neukum1983,Xiao2015,Robbins2018}, emphasised selecting overlapping diameter intervals, excluding bins affected by crater saturation, geological modification or incomplete detection, and deriving normalisation factors by averaging crater frequency ratios within these overlap intervals in contrast to fitting entire distributions. These studies demonstrate that overlapping diameter ranges are the statistically robust foundation for constructing CPFs from distinct surface with heterogeneous crater count datasets.\\

Following the established ``envelope-building'' approach, originally developed for the lunar CPF \citep{Neukum1983}, we apply the same method to construct the Enceladian CPF from selected representative crater size–frequency measurements across geological units. 
In this approach, crater SFD measurements from multiple geological units of different ages and crater densities are each treated as time-scaled realisations of the same underlying production function. The measurements are normalised vertically by scaling each unit's crater density to a common reference level — achieved by aligning overlapping diameter intervals between adjacent units — so that the composite of all normalised measurements traces the time-independent shape of the production function.
This normalisation procedure follows established CPF construction methods developed for the Moon \citep{Neukum1975, Neukum1983, Werner2005}. 
\citet{Neukum1983} formalised this method for the Moon across units spanning ${\sim}$0.1 to more than 4~billion years and diameters from 10~m to 300~km, demonstrating that the composite SFD envelope is time-invariant and deviates significantly from a simple power law. The distribution was instead described by an 11th-degree polynomial in $\log$--$\log$ space (Eq.~\ref{eq:cpf}). We adopt the same strategy here for Enceladus.\\

The 13 representative geological units listed in Table~\ref{tab:representative_units} are marked on the global mosaic images of Enceladus in Fig.~\ref{fig:age}A. We illustrate the observed crater size-frequency measurements for these units in Fig.~\ref{fig:age}B.
Crater diameters are binned logarithmically with a bin width of 0.05 in $\log D$, covering diameters from 0.8 to 34 km. 
This fine binning preserves statistical resolution, particularly in the units where crater counts are low (e.g., CW-L-central), reducing fluctuations that would otherwise arise with unbinned data. It also facilitates consistent comparison across the crater counts from different units with different crater densities and areal extents.
Conventional bin widths (e.g., $D$ to $D \times \sqrt{2}$, or $\sim0.15$ in log scale, as used in Hartmann's incremental plots) would yield only 12 bins, potentially obscuring features, whereas, 0.05 intervals provide 34 bins, allowing more detailed analysis and robust surface age-fitting in the future.\\

Individual measurements sample different diameter ranges depending on unit size, image resolution, surface morphology that's affected by post-impact modification and resurfacing events. Consequently, no single unit provides a complete and statistically robust crater SFD over the full diameter range on Enceladus. To construct a CPF, we therefore combine multiple measurements that partially overlap in diameter space. Only diameter ranges judged to be reliably preserved and well resolved are retained for each unit.\\

Figure~\ref{fig:age}C illustrates the normalisation procedure and how the crater SFDs from different units are vertically scaled and aligned through their overlapping diameter intervals to form a continuous composite envelope. For each unit, a normalisation factor is calculated as the average ratio of crater frequencies within diameter ranges that overlap with the reference unit, so that the SFD from different units with different crater densities (in the vertical dimension) can be linked together and exhibit consistent SFD slopes. \\

We adopt the Far-Mid-CP5 terrain as the reference unit because it spans a broad diameter range and exhibits an intermediate crater density, providing stable overlap with both more sparsely and more densely cratered terrains. 
Heavily cratered units are more likely to approach the crater saturation limit, where crater destruction begins to balance production and the SFD shape becomes unreliable as a proxy for the production function; units with lower crater densities have accumulated too few intermediate and large craters to provide statistically robust counts at those diameter ranges, resulting in large Poisson uncertainties and poor count resolution. Far-Mid-CP5 sits between these extremes, making it the most robustly constrained anchor for the normalisation.
In Fig.~\ref{fig:age}D, the same measurements are shown after applying vertical scaling factors to align crater densities across overlapping diameter intervals. After normalisation, the standardised crater size–frequency measurements from all units are merged to form a composite envelope that traces the time-independent shape of the crater production function (black line in Fig.~\ref{fig:age}D). 
%
\begin{figure}[H]
    \centering

    \begin{subfigure}{\textwidth}
        \centering
        \includegraphics[width=\linewidth]{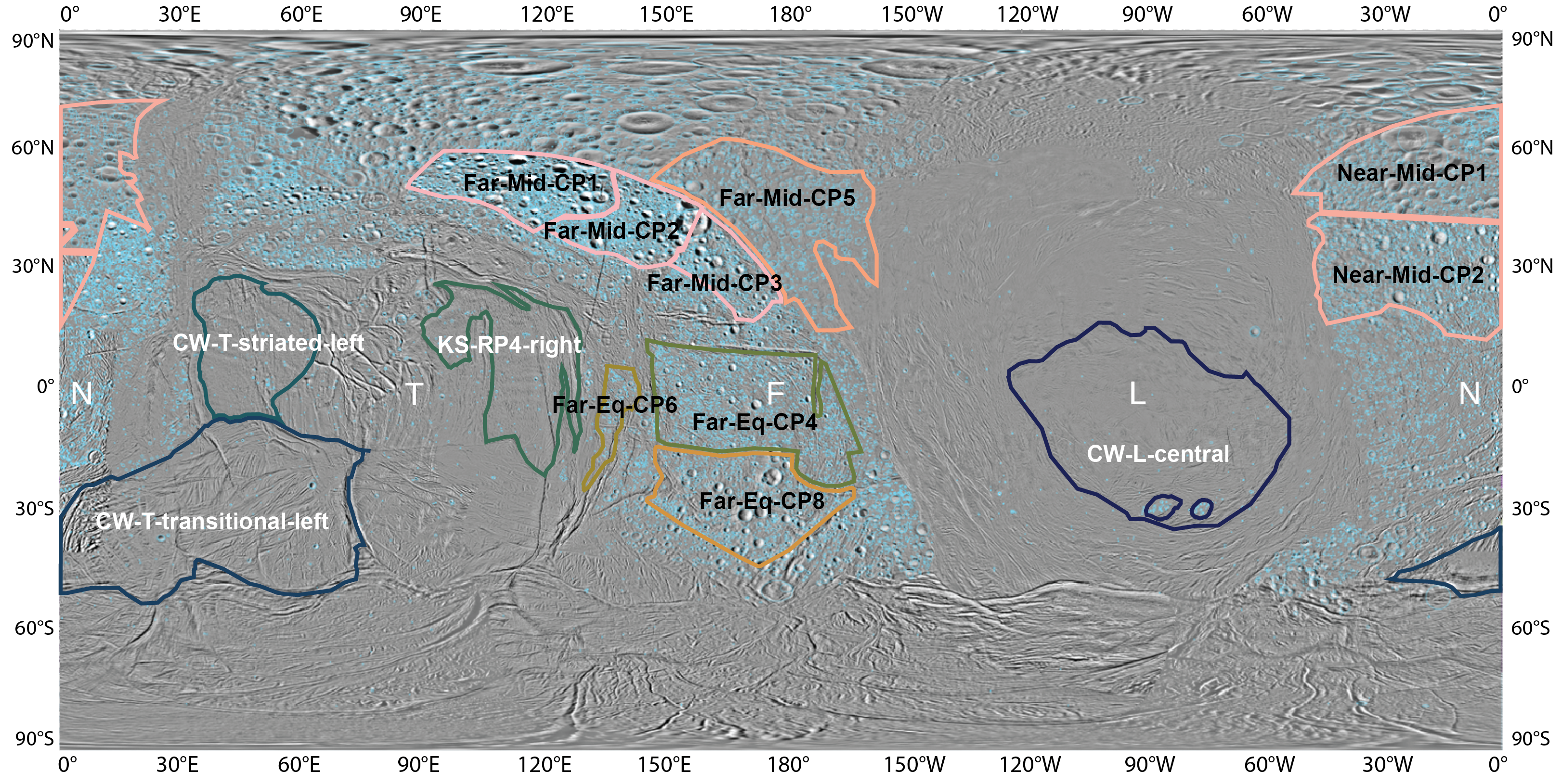}
        \caption*{\raggedright\textbf{A)}}
    \end{subfigure}


    \begin{subfigure}{0.5\textwidth}
        \centering
        \includegraphics[width=\linewidth]{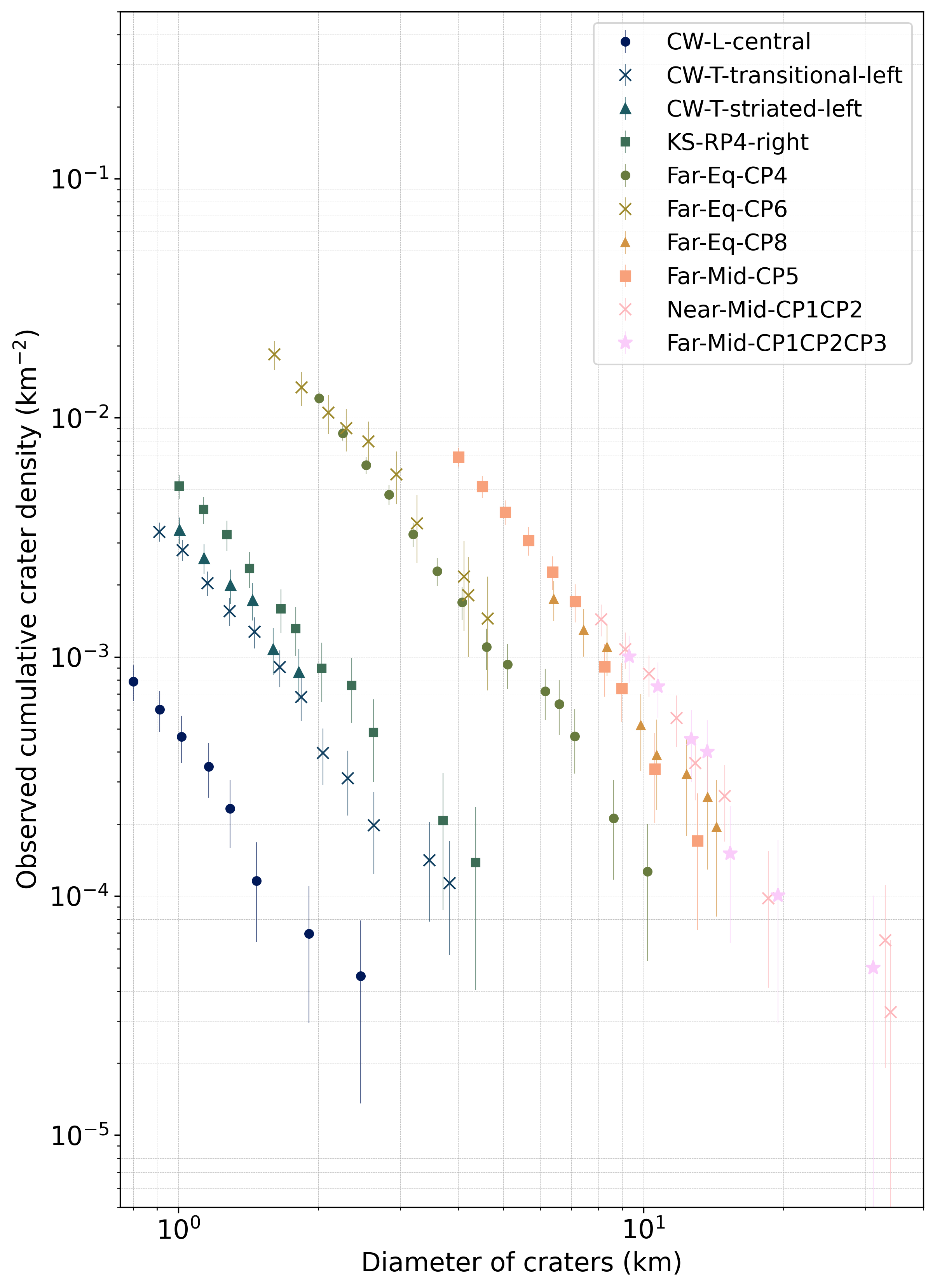}
        \caption*{\raggedright\textbf{B)}}
    \end{subfigure}

    \caption{Construction of the crater production function.
        \textbf{(A)} Representative units overlaid on global mosaic images of Enceladus. White names denote units referenced from \citet{Kirchoff2009} and \citet{CrowWillard2015}, while black names denote units defined in this study.
        \textbf{(B)} Crater size-frequency measurements for ten sampling datasets with logarithmic binning at 0.05 intervals. To improve large-crater statistics, neighbouring heavily cratered mid-latitude units with comparable size-frequency distributions are combined: Near-Mid-CP1\,\&\,CP2 (near side) and Far-Mid-CP1, CP2,\,\&\,CP3 (far side), yielding ten datasets from the 13 representative units. Transitioning colours (navy to pink) indicate increasing crater density and larger typical diameters. The error bars represent Poisson errors.
        (Continued on next page.)
        }
    \label{fig:age}
\end{figure}

\clearpage   

\begin{figure}[H]
    \ContinuedFloat          
    \centering

    \begin{subfigure}{0.5\textwidth}
        \centering
        \includegraphics[width=\linewidth]{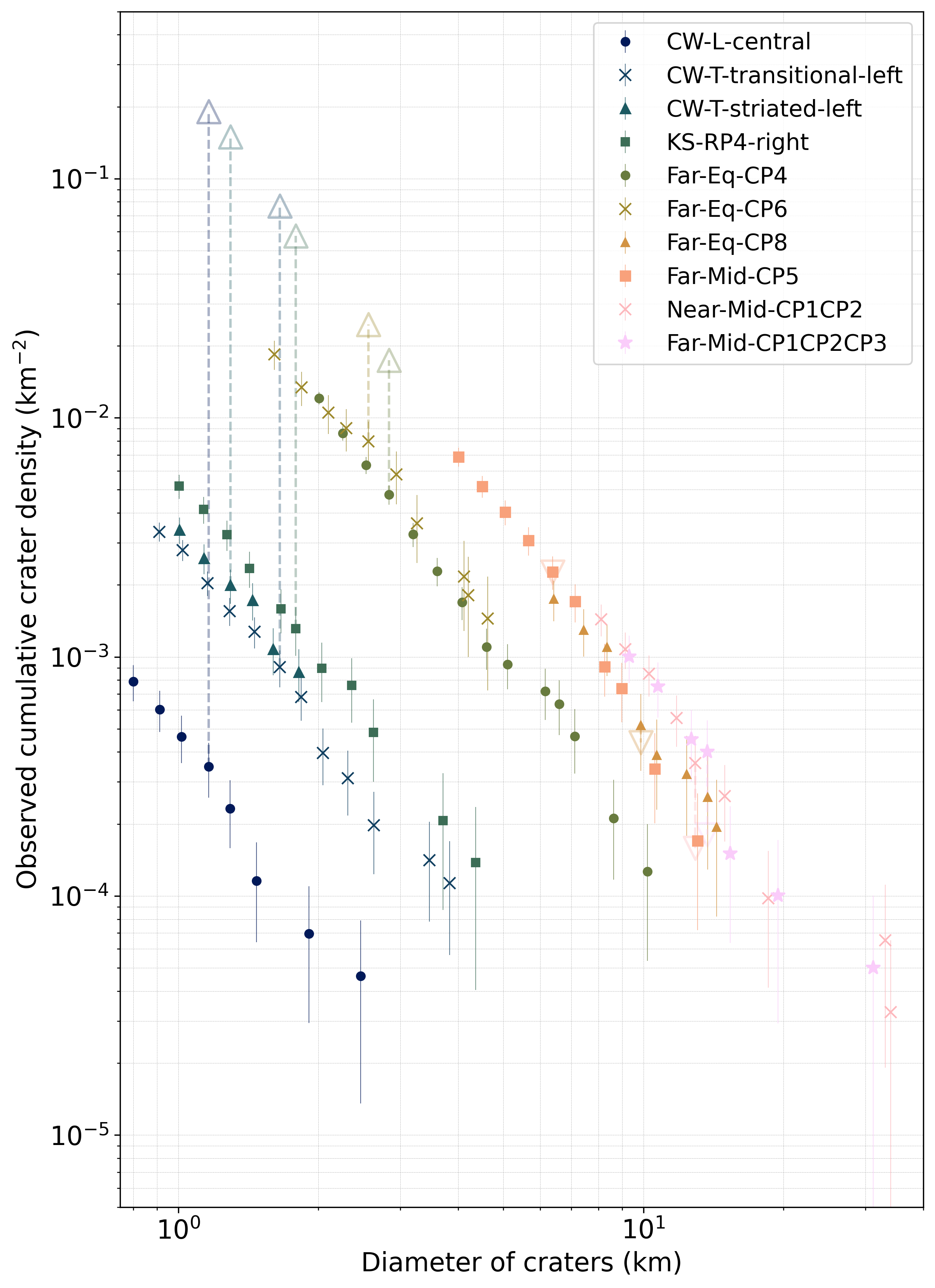}
        \caption*{\raggedright\textbf{C)}}
    \end{subfigure}%
    \hfill
    \begin{subfigure}{0.5\textwidth}
        \centering
        \includegraphics[width=\linewidth]{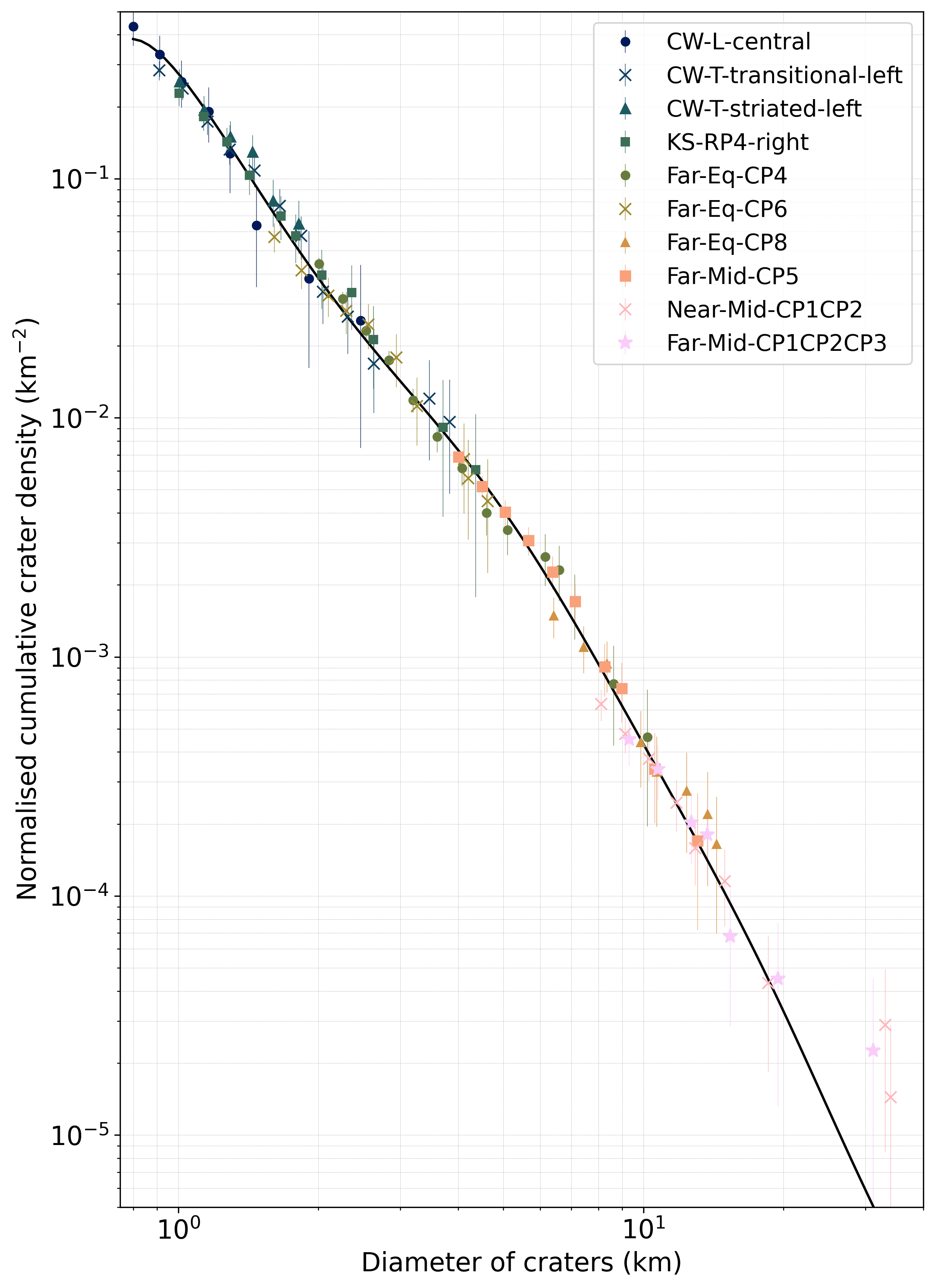}
        \caption*{\raggedright\textbf{D)}}
    \end{subfigure}

    \caption[]{(Continued.)
        \textbf{(C)} Visualisation of the vertical scaling applied to align crater size-frequency measurements to the reference crater density of Far-Mid-CP5. Arrows indicate the direction and relative magnitude of each unit's scaling: upward-pointing arrows denote units with lower crater densities that are scaled up, while downward-pointing arrows denote more densely cratered units scaled down; the latter are less visually prominent due to the logarithmic scale.
        \textbf{(D)} Normalised size-frequency measurements relative to the crater density of Far-Mid-CP5 (selected as reference owing to its broad diameter coverage and intermediate crater density, providing stable overlap with both sparser and denser units), with the black line showing a 10$^{\rm th}$-degree polynomial fit for the crater production function.
        }
    \label{fig:age_cont}   
\end{figure}


\section{Results}

\subsection{Global crater distribution}
\label{subsec:crater_dist}

\begin{figure}
    \centering
    \begin{subfigure}[t]{0.9\linewidth}
      \includegraphics[width=\linewidth]{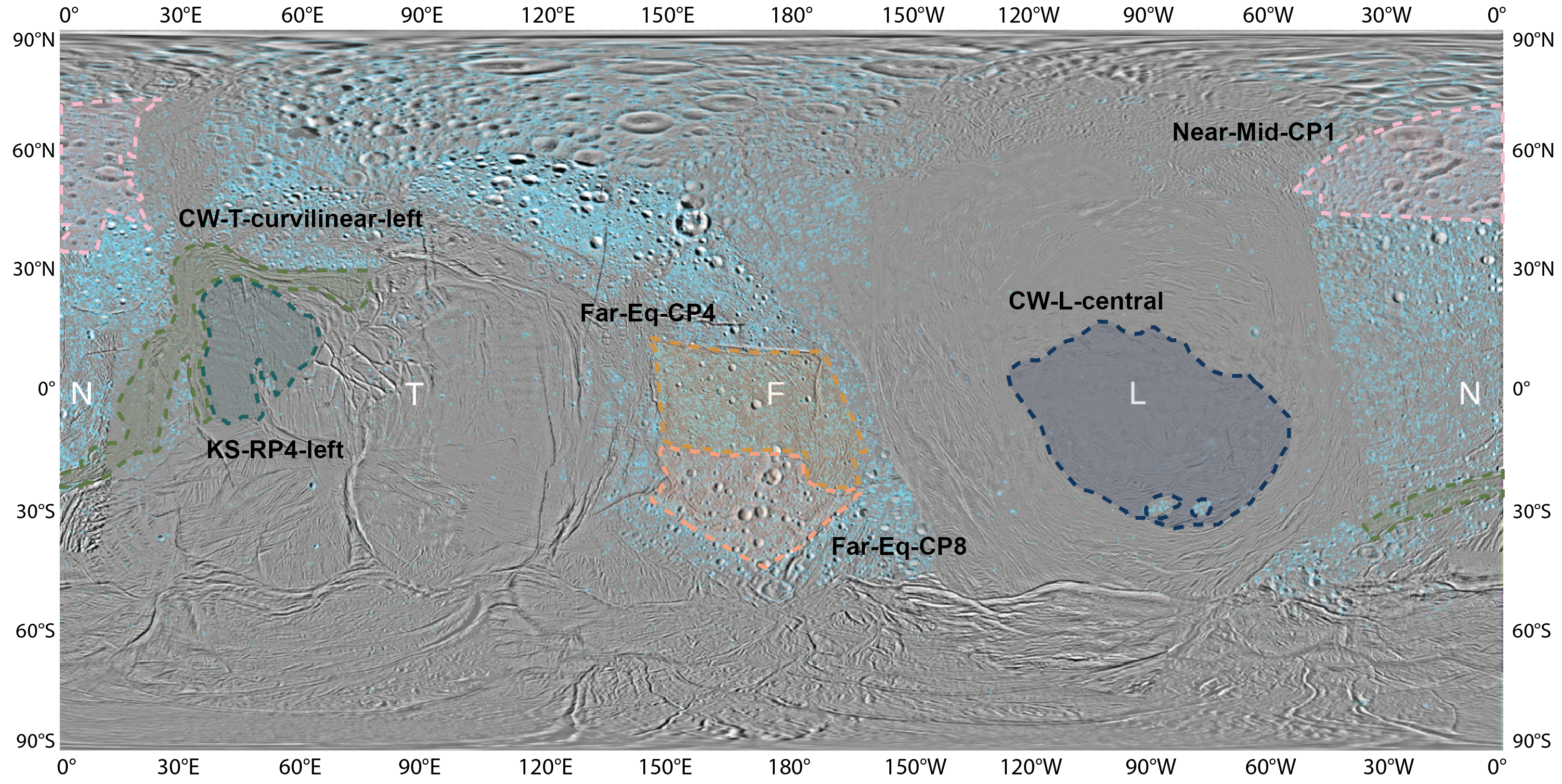}
      \caption*{\raggedright\textbf{A)}}
    \end{subfigure}
    \begin{subfigure}[t]{0.6\linewidth}
      \vspace{-1.0em}
      \includegraphics[width=\linewidth]{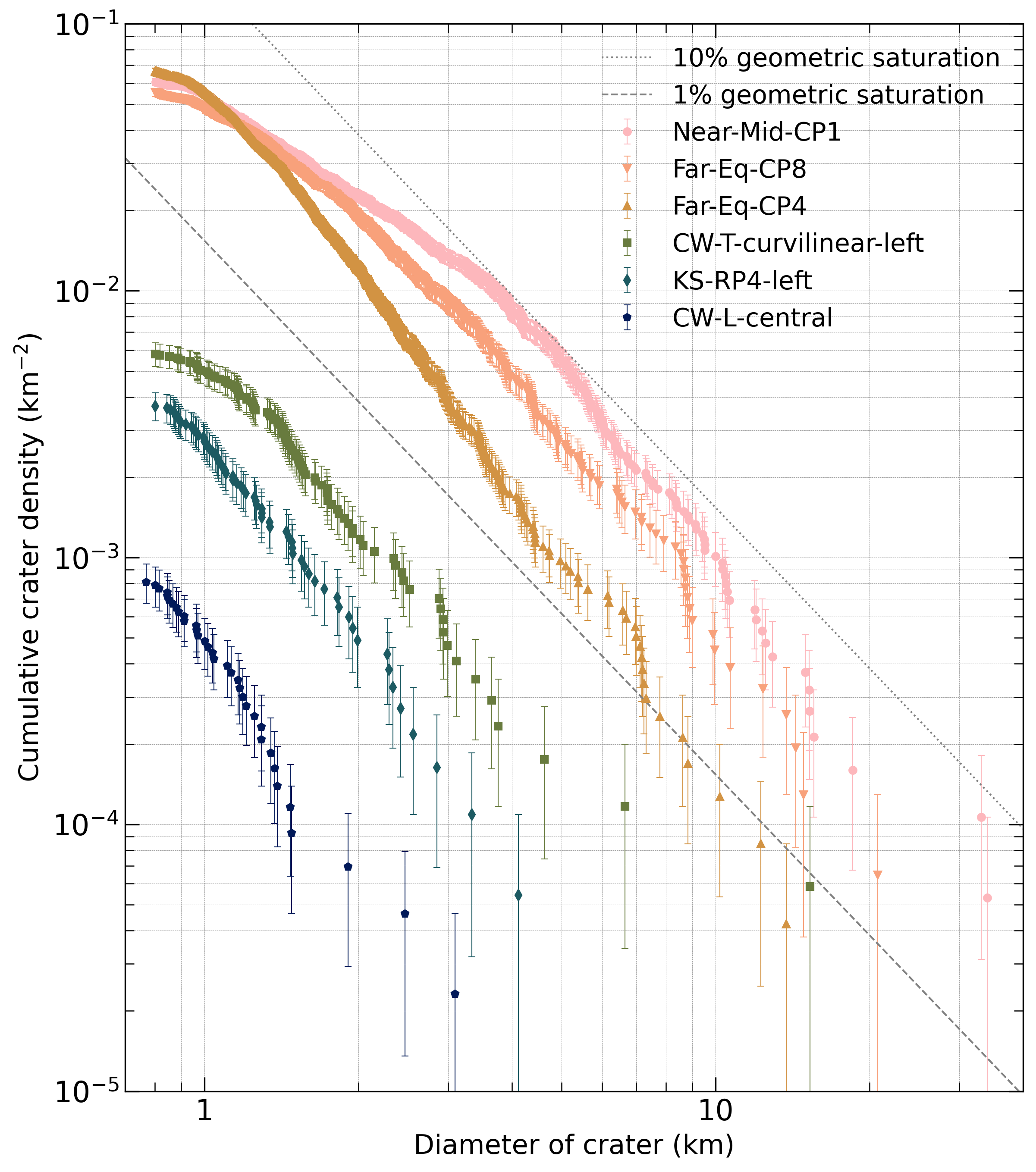}
      \caption*{\raggedright\textbf{B)}}
    \end{subfigure}
    \caption{Demonstration of different crater populations across six units.
        \textbf{(A)} Cyan elliptical outlines highlight the uneven distribution of craters across Enceladus. Thin dashed lines bounding the semi-transparent areas mark the six distinctive units, each labelled with its name. Capital letters along the equator (N, T, F, L) represent the Near-side, Trailing, Far-side, and Leading hemispheres, aiding in unit location reference.
       \textbf{(B)} Raw and unbinned crater size-frequency measurements in cumulative distribution for the six units, with colours matching those in (A). Each unit was chosen to represent a distinct terrain type on Enceladus, selected to minimise partial resurfacing effects that would otherwise obscure slopes, illustrating the global heterogeneity in crater densities and size-frequency distributions. Differences in slope reflect variations in geological setting and degree of resurfacing. Error bars represent Poisson errors. 
        The dashed grey line references the 1\% geometrically saturated crater density, and the dot-dashed grey line references the 10\% geometrically saturated crater density.
      }
    \label{fig:crater}
\end{figure}

Enceladus' craters are heterogeneously distributed across its icy surface. The spatial variation in crater density is illustrated in Figure~\ref{fig:crater}A, where cyan elliptical outlines and varying cyan colour intensity overplotted on the surface images of Enceladus. As shown in Fig~\ref{fig:crater}A, the northern high latitude region (beyond 40$^\circ$N), as well as narrow longitudinal bands along the sub-Saturn (0$^\circ$) and anti-Saturn (180$^\circ$) sides, display particularly dense concentrations of craters. Additionally, clusters of craters, such as those around 80$^\circ$W, 30$^\circ$S, and 30$^\circ$E, 0$^\circ$, are observed within the more expansive yet scarcely cratered leading and trailing hemispheres. The heterogeneity in the crater distribution is attributed to regional resurfacing of the icy crust, which has erased previous impact records, as opposed to being caused by preferential impact at specific locations on the satellite, i.e. the apex-antapex impact asymmetry. \\

Figure~\ref{fig:crater}B compares the observed crater SFDs across different analytical units on Enceladus. These measurements reveal varying crater densities and span different diameter ranges. Due to differences in morphology of the units, offsets in the crater record, and limitations imposed by image resolution, only selected portions of these measurements can be used for deriving the Enceladian CPF. 
Notably, the presumed young, low-density units (greenish and bluish curves) exhibit relatively steep size-frequency distributions at small crater diameters below ${\sim}$3~km, while their slopes at intermediate diameters are broadly comparable to those of other units; large craters exceeding ${\sim}$10~km are absent or extremely rare in these low-density units. The steeper small-diameter slopes may reflect a lower degree of erasure by subsequent large impacts, or could partly arise from crater saturation effects; the precise cause remains uncertain and is discussed further in Section~\ref{subsec:saturation}.
In contrast, the densely cratered (presumed old) units show apparently variable slopes (yellowish to pinkish curves), and converge at smaller crater diameter ranges, suggesting variable resurfacing activity and a strong dependence on image resolution.
In contrast, the presumed old, densely cratered units (yellowish to pinkish curves) show apparently variable slopes that converge at smaller diameter ranges, suggesting variable resurfacing activity and a strong dependence on image resolution.

\subsection{Fitting of the crater production function}
\label{subsec:fitting}
The crater production function (CPF) for Mercury \citep{Neukum2001}, Mars \citep{Ivanov2001, Hartmann2005}, and asteroids Vesta, Ida \citep{Schmedemann2014}, and Ceres \citep{Hiesinger2016}, have been derived by fitting a mathematical function to the normalised crater SFDs, typically using polynomials or piecewise power laws. The former has traditionally been used for the Lunar \citep{Neukum2001}, and Martian CPFs \citep{Ivanov2001}, while the latter is more commonly used to describe impactor distributions in the outer Solar System \citep{Zahnle2003}. These empirical functions enable the translation of crater densities across different size ranges. \\

For Enceladus, we derived a cumulative CPF by fitting the normalised and combined SFD (coloured points in Fig.~\ref{fig:age}D), we found that a 10$^{\rm th}$-degree polynomial (black line) best represents the CPF over crater diameters between 800~m and 35~km, while preserving a continuous downward slope. 
To avoid unphysical absolute values and pre-assumption of any impactor source, and dependence on impactor-to-crater diameter scaling relations, we express the polynomial fit of the CPF in a \emph{normalised form}, characterised by a \emph{dimensionless shape function} $\widetilde{S}(D_{\rm cr})$. It is important to note that $\widetilde{S}(D_{\rm cr})$ alone does not represent a physical crater density or cumulative crater count; it provides only a relative scaling with diameter, capturing the \emph{shape} of the Enceladian production function. Physical crater densities are recovered by anchoring $\widetilde{S}$ to a directly observed crater density on Enceladus at a chosen reference diameter.
The CPF is scaled such that the cumulative crater density at 10~km diameter matches the directly observed value from the Far-Mid-CP5 reference unit, corresponding to $\widetilde{S}(D_{\rm cr}) = 1$ at reference diamater ($D_{\rm ref}) = 10$~km. The choice of $D_{\rm ref}$ is arbitrary; the shape of the production function is invariant to this choice. By anchoring the normalisation directly to an observed Enceladus crater density rather than to an impactor flux model or impactor-to-crater diameter conversion, the CPF remains self-consistently grounded in the Enceladian crater record and independent of crater scaling relations.
\begin{equation}
    \log \widetilde{S}(D_{\rm cr}) = \sum_{i=0}^{10} a_i \, \bigl[\log D_{\rm cr}\bigr]^i,
    \label{eq:cpf_scale}
\end{equation}

\begin{table}[H]
\centering
\begin{tabular}{cc}
\hline
  \textbf{Polynomial degree ($n$)} & \textbf{Coefficient ($a_n$)} \\
\hline
0 & -0.5932  \\
1 & -2.3503 \\
2 & -5.5573 \\
3 & 24.951  \\
4 & -77.268 \\
5 & 212.99  \\
6 & -406.68 \\
7 & 464.58  \\
8 & -303.37 \\
9 & 104.74  \\
10 & -14.850 \\
\hline
\end{tabular}
\caption{Coefficients of the polynomial representation of Enceladus’ cumulative crater production function. The polynomial describes the relative shape of the cumulative distribution; absolute normalisation is defined separately.}
\label{tab:cpf}
\end{table}

Table~\ref{tab:cpf} lists the polynomial coefficients $a_{0}$–$a_{10}$. The coefficient $a_{0}$ encodes the absolute crater density of a given terrain unit and serves as a relative age measure; the remaining coefficients $a_{1}$–$a_{10}$ define the shape of the production function, and their ratios remain invariant across units.\\

To ensure reproducibility and facilitate use of our CPF, we provide a documented Python code\footnote{Available at \url{https://github.com/EmiWW/enceladus_crater_production_function}.} that (i) implements the polynomial functions of the CPF, (ii) includes routines to compute crater densities for arbitrary crater diameters as scaled to the crater densities of 10~km diameter craters on the Far-Mid-CP5 reference unit, and (iii) allows comparisons with published impactor and crater SFDs, including those from \citet{Kirchoff2009}, \citet{Zahnle2003}, and \citet{Singer2019}; for comparisons with impactor SFDs, the impactor-to-crater diameter conversion is performed using the $\pi$-scaling law of \citet{Zahnle2003}, accounting for Enceladus-specific gravity, impact velocity, and impact probability.\\

To facilitate comparison to a piecewise linear power-law fitting, where different crater diameter ranges follow distinct power-law slopes, we note that the corresponding cumulative power-law slope of this polynomial varies from -2.6 to -4.2.
The slope is shallower below $D_{\rm cr}\!\sim\!3$~km and steepens beyond $D_{\rm cr}\!\sim\!7$~km, although this transition is not pronounced enough to indicate a clear break in the SFD at 7--8~km. Such steepening at the largest diameters likely reflects edge effects due to the rarity of large craters.
For comparison, \citet{Kirchoff2009} reported cumulative slopes of $-1.4$ to $-2.9$
for smaller craters and $-3.0$ to $-4.3$ for craters larger than 7~km (see their Table~2). The slopes derived from our CPF fall comfortably within this previously reported range.\\

We also note a subtle bend in the CPF near $D_{\rm cr}\!\sim\!1.2$~km. This diameter falls well above our 800~m completeness limit — corresponding to 5--20 pixels across a crater at the image resolutions of the contributing units — making image resolution and data completeness unlikely explanations. The polynomial fit faithfully traces the normalised composite data, so the bend is not an artefact of the fitting. It most plausibly reflects genuine slope variation in the crater SFD at small diameters, possibly related to sub-km complexity in the impactor population \citep{Bottke2024}, though we acknowledge it may partly be coincidental given its proximity to the minimum fitting diameter.\\

We end with the following statement: the CPF we present here does not make assumptions about projectile origin, whether heliocentric or planetocentric, or their relative contributions. The CPF is derived directly from crater counts instead of model-based predictions. The piecewise overlap for a large diameter range supports the underlying assumption that the projectile population is similar through time.


\section{Discussion}
\label{sec:discussion}
In this study, we construct a crater production function (CPF) for Enceladus based on the global crater record, and subdivide the surface into units defined by consistent image resolution, geomorphology, and evidence for resurfacing.
Through the envelope-building approach — retaining only reliably preserved diameter ranges from each representative unit and normalising their crater densities to a common reference — we minimise the influence of observational conditions and surface modification on the shape of the resulting production function.
This framework enables us to use deviations from the CPF as a diagnostic of resurfacing, rather than relying on absolute impact rates or timescales, which remain uncertain at the time of this study.\\

In the following, we examine how variations in image properties and surface morphology affect the measured crater SFDs, and how systematic differences between the CPF and observed distributions across Enceladus motivate the subdivision into distinct terrain units. We then explore the potential origin of slope variations, and assess the extent to which processes such as partial resurfacing, crater saturation and crater relaxation may contribute to the observed trends. 

\subsection{Criteria and justification for unit boundary delineation}

\subsubsection{Known effects of image properties}
\setcounter{section}{5}
\setcounter{subsection}{1}
\setcounter{subsubsection}{1}
\label{subsec:instrumental}
The global mosaic comprises 108 images acquired by the Cassini Imaging Science Subsystem Narrow Angle Camera \citep[ISS/NAC;][]{Porco2004} in an equirectangular projection, each with distinct resolution, solar incidence angle, and emission angle (see Fig.~12–14 in \citet{Bland2018}). These variations affect crater counts and geomorphological identification as described in Section~\ref{subsec:units}. Here we focus on their practical implications for unit delineation.\\
Image resolution constrains the smallest detectable crater diameters and can artificially flatten SFDs as diameters approach the resolution limit. If a delineated unit erroneously spans images of differing resolutions, the reduced detection of smaller craters in lower-resolution areas produces an apparent flattening in the SFD, which could be mistaken as resurfacing activity. In our analysis, we carefully separated units with respect to image resolution, to mitigate this effect.\\

Illumination conditions, particularly solar incidence angle—the angle between incoming sunlight and the surface normal— affect shadow length and feature visibility. 
Higher incidence angles produce longer shadows that enhance the visibility of crater walls and other subtle topographic features, whereas lower incidence angles (e.g., near local noon) yield more uniform illumination that can obscure subtle topography, such as relaxed craters and low-relief ridges or troughs.
Viewing geometry further constrains interpretation. The emission angle, defined as the angle between Cassini's camera viewing direction and the surface normal, affects feature appearance and measurement accuracy. High emission angles can distort features and reduce the accuracy of size measurements (e.g., ridge width or crater rim-to-rim diameter). See Fig.~2 in \citet{Bland2018} for visual examples of how incidence and emission angles alter surface appearance of Enceladus.
Recent assessments of crater detectability indicate that solar incidence angles of approximately $74^{\circ}$--$82^{\circ}$ are optimal for crater identification, while reliable crater populations can generally still be obtained for phase angles of $\sim20^{\circ}$--$75^{\circ}$ and emission angles up to $\sim60^{\circ}$ \citep{Robbins2025}. Similarly, \citet{Blanco-Rojas2024} imposed a minimum incidence angle of $\gtrsim60^{\circ}$ for their machine-learning-based crater detection, implying that crater identification becomes increasingly reliable above this threshold, although image resolution remains an additional limiting factor.\\

Thus, variations in imaging conditions inherently produce differences in observed crater densities and size-frequency measurements. Combining crater counts from images with differing conditions may yield measurements unrepresentative of the actual surface ages and geological history. Fortunately, image properties are known and can be considered when assessing the shape of the observed crater SFD, and discontinuities in image resolution and solar incidence angle are often clearly visible in the mosaic as abrupt changes in surface texture or shadow length, and were used to guide unit boundaries.
Adjacent areas imaged under significantly different conditions were therefore treated as separate units to prevent imaging artefacts from being conflated with genuine geological signals. 
Figure~\ref{fig:unit_subdivide_example} shows an example where Far-Mid-CP9 and Far-Mid-CP10, two neighbouring units of otherwise similar geological character, are separated based on a visible discontinuity in image resolution and shadow length.

\subsubsection{Observed effects of surface morphology}
\label{subsec:geomorphology}
Craters, being circular or elliptical, are generally distinguishable from most tectonic features -- such as ridges, troughs, and chasmata -- which are often linear and parallel. However, distinguishing craters from tectonic features of similar scale can be challenging, particularly when the width of curvilinear troughs in transitional units (e.g., CW-T-transitional-left/right; \citet{CrowWillard2015}) resembles that of small craters (1–3.5~km wide). Cross-cutting troughs may produce quadrilateral or circular-like features that mimic small craters. In cratered plains, heavily eroded ``squarish'' craters (see Fig.~\ref{fig:app_erosion1}) and craters intersected by pit chains (Fig.~\ref{fig:app_erosion3}) are often difficult to identify and measure. The size and shape of these ambiguous craters vary across units due to differences in background topography, necessitating separate consideration of crater counts for each unit. All of which results in a possible underdetection of craters at scales similar to the tectonic surface texture. \\

Additionally, Enceladus’ surface is characterised by abundant ridges and troughs that often form large, nearly parallel groups with similar widths, dominating regions such as the far-side southern mid-latitude cratered plain (e.g., Far-Eq-CP3) and the trailing hemisphere transitional plain (e.g., KS-RP1-middle). When a more extensive region displays the same group of ridges or troughs, this pattern indicates partial or complete resurfacing by a single event, which can affect surface age and crater density. However, regions with similar morphology may form at different times through similar processes. Therefore, considering the superposed crater population is essential for distinguishing units or subunits that appear morphologically similar but have distinct formation ages. Thus, we can infer several episodes of tectonic overprinting has modified Enceladus' surface (related to crater density per diameter bin), with the resulting SFDs varying across diameter ranges according to the scale and extent of the underlying tectonic textures. \\

Unit boundaries were accordingly drawn along large-scale geomorphological structures --- ridges, troughs, and surface texture transitions --- which are visually identifiable in the global mosacis, to ensure each unit captures a geologically coherent region. 
Figure~\ref{fig:unit_subdivide_example} shows an example where the eastern boundary of Far-Mid-CP10 is defined by a prominent ridge system — a lineated band of closely-spaced scarps in CW-L-curvilinear — separating the two units.\\

\subsubsection{Spatial crater density variations across Enceladus}
\label{subsec:crater_distribution}
Crater density varies across Enceladus, even within the cratered plains and subdued cratered plains defined by \citet{CrowWillard2015}. Because surface age estimates depend on crater density -- older surfaces accumulate more craters -- combining crater counts from areas that solidified at different times can skew average crater densities and misrepresent solidification ages. Accurate age calculations therefore necessitate bounding units to areas with similar crater densities.\\

However, Enceladus presents additional challenges due to its complex and potentially multiple resurfacing history. Not all surfaces experienced complete resurfacing, which would erase all pre-existing features and fully reset ages; instead, many areas were only partially resurfaced. Some cratered plains show gradual decreases in crater density, for example from high to low latitudes (e.g., Far-Eq-CP8 to Far-Eq-CP4) or near boundaries with other terrains or tectonic features (e.g., Far-Eq-CP4 to Far-Eq-CP7). These variations in crater density reflect differences in surface age and resurfacing intensity, and necessitate the classification of distinct units. 
Regions exhibiting gradual crater density gradients, visible by eye, particular the gradual decrease from high to low latitudes in the crater terrains along the far-side and near-side axis, were therefore subdivided into separate units, as such gradients are indicative of partial resurfacing and represent distinct surface ages. Where such gradients exist, the boundary between units is anchored to the nearest large-scale geomorphological structure to avoid an arbitrary placement. 
Figure~\ref{fig:unit_subdivide_example} illustrates such a density transition, with crater density increasing gradually southward from Far-Eq-CP5 into Far-Mid-CP10, and the corresponding unit boundary between them.\\

\subsection{Observed slope variations}
\label{subsec:rolloff}
From Fig.~\ref{fig:crater}B, we observe that the two most heavily cratered terrains (orange and pink data) exhibit a gradual decrease (roll-off) in crater frequency at smaller diameters ($\sim$ 3--4 km). A similar roll-off also appears in the size-frequency measurements of other heavily cratered units, i.e., CW-T-north-lineated, CW-T-square, Far-Eq-CP8, Far-Mid-CP1, and all units in the north pole region. However, this roll-off results in different slopes for the observed crater SFD. This roll-off is not due to observational bias, as it occurs above the resolution limit of the images used for crater identification. Instead, it likely results from geological processes that have modified the original crater population.\\

One could argue that this rolling off results from a change in the cratering population with time. Yet, geomorphological analysis points towards partial resurfacing as a plausible and more straightforward explanation, selectively removing small craters first, and most prominently for the heavily cratered plains \citep{Michael2010}. Heavily cratered regions have accumulated high crater densities, implying they comprise old surfaces. These regions have since been disturbed by tectonic-like structures that extend from neighbouring terrains. Such tectonic activity has modified the crater record, particularly for craters with diameter ranges comparable to the scale of the tectonic features. Cassini's images reveal ample evidence of partially eroded or traversed craters within the ancient cratered plains, demonstrating the effect of tectonic structures extending from neighbouring, less-cratered younger surfaces. Additional partial resurfacing mechanisms — including resurfacing by basin-forming impacts and local crustal melting consistent with the diachronous crustal solidification argued here — are also plausible contributors.\\

Regarding disturbance by tectonic-like features, for example, in the trailing hemisphere, the relatively young CW-S-curvilinear unit -- identified by its well-preserved curvilinear ridges and scarcity of craters -- has traversed and eroded small craters along the southern boundary of the heavily cratered Far-Mid-CP9 unit (Fig.~\ref{fig:app_erosion3}). Similarly, in the leading hemisphere, the formation of the smooth and sparsely cratered terrain, CW-L-Curvilinear, apparently degraded and obscured craters in the nearby cratered plains, i.e., Near-Mid-CP1 and Near-Mid-CP2, producing ridged and lineated structures that cut across pre-existing craters and preferentially removed smaller ones (Fig.~\ref{fig:app_erosion2}). Additional examples of partial resurfacing, crater erosion, and tectonic-like modification are provided in Supplementary Section~S2.\\

By contrast, scarcely cratered regions (e.g., Fig.~\ref{fig:crater}B green and blue, units around the central leading and trailing hemisphere) appear smoother and morphologically more homogeneous, often dominated by single-directional lineated bands of ridges and troughs. Their low crater density suggests they formed more recently, while their structural uniformity implies resurfacing occurred during a single, coherent geological event. Having undergone fewer episodes of tectonic-like reworking and crater erosion, their crater populations are less modified. 
Consequently, the SFDs of these less-cratered and presumably more recently solidified terrains exhibit little or no roll-off; the resulting steep crater SFDs suggest that the recent impactor population is characterised by a steep size distribution, which cannot be explained by secondary cratering processes associated with impacts on Enceladus itself.\normalcolor\\

\subsection{Crater saturation}
\label{subsec:saturation}
Another possible explanation for the roll-off is crater saturation. While our study does not advocate this as the primary cause, it is worth considering its potential effects. 
In Fig.~\ref{fig:crater}B, the dashed and dot-dashed lines mark the 1\% and 10\% geometrically saturated crater density (N$_{\rm gs}$) respectively, thresholds where crater destruction and production begin to balance, starting with smaller craters \citep{Gault1970}.
On more heavily cratered terrains (pink and orange data), the SFD rolls off at larger diameters, reflecting their longer exposure to impacts and approach toward higher saturation levels. Spatial statistical analyses also suggest that Enceladus’s cratered plains may approach saturation at above 6 km in diameter \citep{Kirchoff2018b}. Moderately cratered terrains (e.g., yellow data) may have reached only minor saturation, with roll-offs beginning at smaller diameters as parts of their distributions fall within the 1\% N$_{\rm gs}$ regime. Despite these insights, \citet{Xiao2015} cautions that conventional evaluation of equilibrium crater densities may be unreliable: saturation may occur at even lower levels than 1\% N$_{\rm gs}$, and that local geology, initial impactor SFD, and tectonic activity (prominent on Enceladus) play more dominant roles in shaping the crater populations.

\subsection{Crater degradation due to viscous relaxation}
\label{subsec:relax}
Crater degradation processes may themselves modify the observed crater size--frequency distribution and therefore potentially influence the derived crater production function. Studies of lunar crater degradation suggest that crater detectability decreases with time and that smaller craters generally have shorter observable lifetimes than larger ones \citep{Fassett2014,Fassett2022}. In contrast, models for Enceladus indicate that viscous relaxation may preferentially affect larger craters owing to their greater depths and the temperature dependence of ice rheology \citep{Bland2012}. Because these studies concern different degradation mechanisms operating on fundamentally different surface materials, their quantitative results cannot be directly compared or transferred to Enceladus.\\

At present, the influence of crater relaxation on Enceladus' crater SFD remains poorly constrained. In this study, we do not explicitly quantify relaxation effects, as measurements of crater depth-to-diameter ratios and rim degradation are beyond the scope of the present work. If relaxation operates globally and preferentially modifies craters within particular diameter ranges, it could locally produce deficits or excesses in specific crater sizes and thereby distort the observed CPF. Alternatively, if relaxation is spatially heterogeneous and linked to local thermal conditions or plume deposition, its effects may primarily contribute to regional slope variations and local departures from the CPF, with some of these variations potentially averaging out when constructing a global production function. Furthermore, if crater degradation timescales are substantially longer than the ages of younger terrains, crater SFD measurements on such surfaces are expected to be less affected by viscous relaxation.\\

Recently \citet{Bland2012,Martin2023} indicate that Enceladus’ low gravity and nominal cold surface temperature of $\sim$70 K, even with a high constant heat flux of 150 mW m$^{-2}$ still make complete crater burial through viscous relaxation and burial by south polar plumes and E-ring material alone unlikely. However, the observed crater relaxation (e.g., \citealt{Bland2012}) and scarcity of large craters near the equator likely required a short period of excessive heat flux ($>$150 mW m$^{-2}$) and warmer subsurface temperatures ($\sim$150 K). Such conditions were possibly enhanced by insulation from regolith or plume material.\\

Under a constant heat flux of 150 mW m$^{-2}$ on a 150 K surface, craters 15–20 km in diameter with depth-to-diameter ratios of $\sim$0.2 could relax by over 80\% . Specifically, at a current maximum equatorial deposition rate of $\sim$10$^{-3}$ mm yr$^{-1}$, these craters could be buried within a few hundred million years \citep{Bland2012}. Although a constant heat flux is unlikely, this relaxation timescale could still be achieved with transient periods of extreme heat flux and higher deposition, e.g., due to a shift in the locus of plume activity \citep{Bland2012}. Ongoing work continues to refine models of historical and current heat flux and deposition.

\section{Conclusions}
\label{sect:conclusion}
In this work, we have derived the first published CPF of an outer Solar System icy satellite. The CPF is the underlying, unmodified crater SFD of the satellite surface; that is, it is the pristine SFD that the craters would have, assuming no subsequent modification. The CPF is therefore essential for computing surface ages when combined with a crater chronology function. \\

For this work EW and MK counted a combined 16,958 craters on the Enceladian surface with diameters $\ge$800~m to compute the CPF, which was fit with a 10$^{\rm th}$ degree polynomial as is customary for the Moon and Mars \citep{Neukum1975, Neukum2001,Ivanov2001}.
The resulting crater size–frequency distribution is overall steeper than implied by previous assessments of the projectile population. The steep slope persists across Enceladian heavily-cratered and lightly-cratered terrains,indicating that the steepness is not primarily the product of extensive secondary cratering, and further implies that the crater-forming projectile population has not undergone any recognisable change through time.\\

Future work will indicate whether the Enceladian CPF can be applied to the other Saturnian satellites, or whether it is unique. Either result would yield greater insight into the source populations responsible for satellite bombardment in the Saturnian system.
With an appropriate description of the projectile flux, it will also be possible to constrain the timing and rates of the processes that modified Enceladus’ surface. In this contribution, however, we focus on describing the nature and sequence of these processes. Importantly, Enceladus’ surface evolution reflects the combined influence of endogenic activity—driven primarily by tidal heating—and exogenic modification through impact cratering. Together, these mechanisms rejuvenate the surface, obscuring or erasing parts of its cratering record while preserving clues to the moon’s complex (and potentially ancient) geological history. A corresponding improved model crater chronology for Enceladus will be presented in a forthcoming paper.

\section*{Acknowledgments}
To prevent visual distortion of the data and exclusion of readers with colour-vision deficiencies, the Scientific colour map batlow from \citet{Crameri2020} is used in all the figures in this work. EW would like to thank the Japan Society for the Promotion of Science (JSPS) for the support provided through the JSPS Research Fellow grant No. 22KJ12886 during which this work originated. SCW and RB acknowledge the Research Council of Norway for funding through its Centres of Excellence funding scheme, project No. 332523 (PHAB). MRK appreciates the support from the NASA Cassini Data Analysis Program, grant No. 80NSSC22K0632. Our QGIS map with complete crater counts and units are available by email request from the corresponding author.

\bibliographystyle{elsarticle-harv}
\bibliography{enceladus_crater}{}
\clearpage

\section*{Supplementary Section}
\setcounter{section}{0}
\setcounter{figure}{0}
\renewcommand{\thesection}{S\arabic{section}}

\section{Documentation and justification of the selected representative units}
\label{subsec:suppB}
\renewcommand{\thefigure}{S.1.\arabic{figure}}

Here, we document and justify the selection of each of the 13 representative units and their associated diameter ranges. The lower diameter cutoff is chosen to exclude features likely influenced by tectonic activity, which dominates at smaller scales. For units previously defined by \citet{CrowWillard2015} (CW15), we reference their documentation and adopt their characteristic properties as justification for selection.

\subsection{Leading hemisphere central plains (CW-L-central)}
\begin{figure}[H]
    \includegraphics[width=1.0\linewidth]{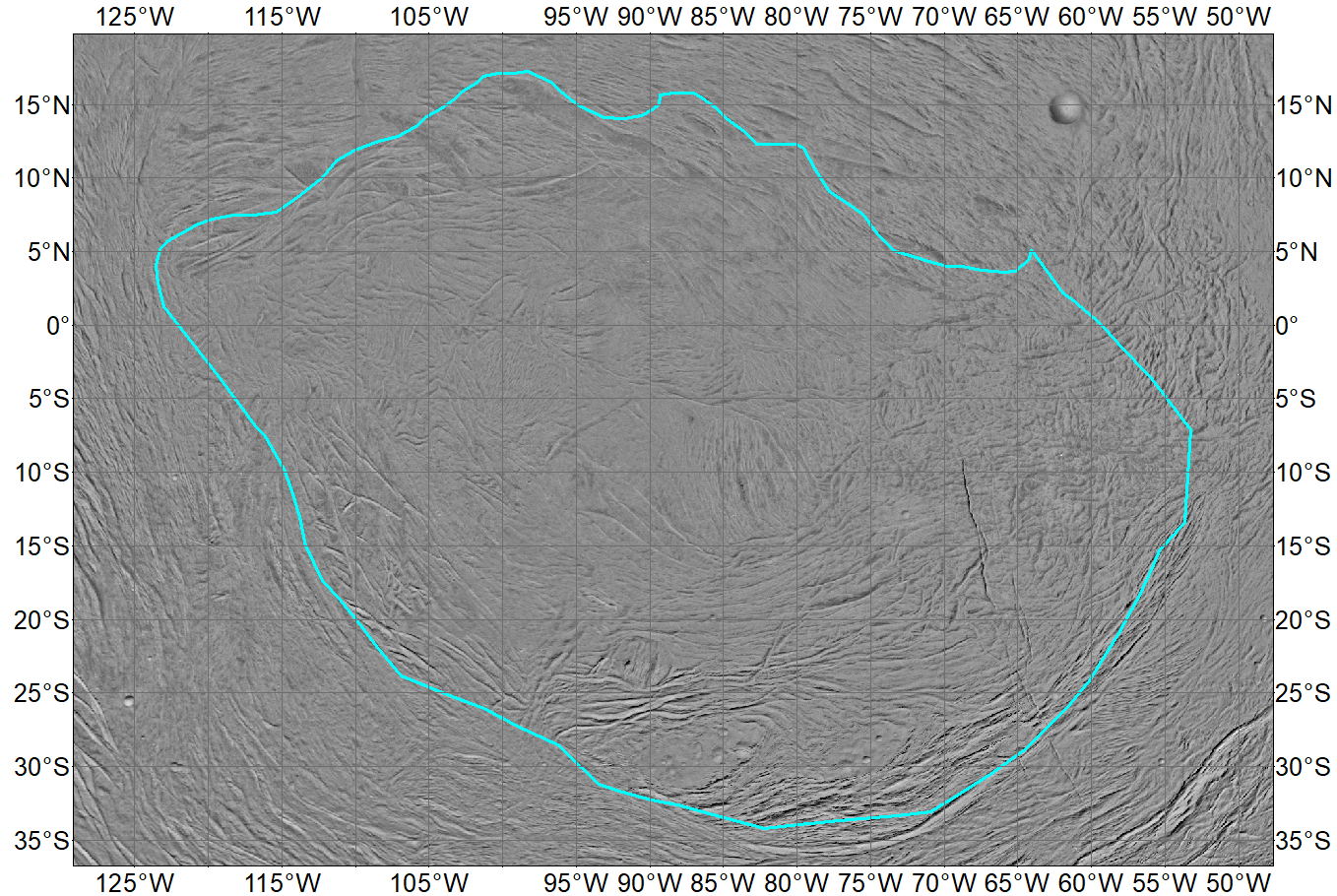}
    \caption{}
    \label{fig:cw-L-central}
\end{figure}
As defined by CW15, CW-L-central is among the smoothest and most homogeneous surface units on Enceladus. It forms a laterally continuous plains unit characterised by a consistent ridge–trough morphology. While tectonic-like features such as ropy and interlaced ridges are present (CW15), they do not subdivide the unit into distinct resurfacing episodes. This morphological uniformity indicates a single dominant surface age, minimising the risk of mixing crater populations from different periods.\\

The surface is notably smooth and shows no evidence of burial by thick mantling deposits. Small craters (0.8–1.5 km in diameter) appear fresh, with well-preserved rims and floors, and are not deformed by the ridge–trough fabric, indicating they postdate the resurfacing event. Within this diameter range -- where small crater depletion often complicates analysis in other terrains -- the crater size–frequency distribution is smooth and lacks inflections or roll-overs, suggesting minimal erosion or partial resurfacing. This implies that pre-existing craters were removed during the formation of CW-L-central, and the current crater population records a single, recent resurfacing event.\\

The only evidence for partial resurfacing is two small, isolated exposures of older cratered plains (CW-L-CP-island) embedded within the unit. These islands are morphologically distinct and spatially well bounded, allowing their crater populations and areas to be clearly identified and excluded from analysis (see Supp.~\ref{fig:app_erosion4}). Collectively, these characteristics satisfy the principal criteria for reliable crater size–frequency analysis \citep{Neukum1983}: geological homogeneity, limited resurfacing, preservation of small craters, and minimal contamination from large impacts. As a result, the crater counts in the 0.8–1.5 km diameter range on CW-L-central provide a robust basis for modelling the small-diameter portion of the Enceladian CPF, offering a well-preserved and representative record of recent impact flux.

\subsection{Trailing hemisphere transitional plains, left (CW-T-transitional-left)}
\begin{figure}[H]
    \includegraphics[width=1.0\linewidth]{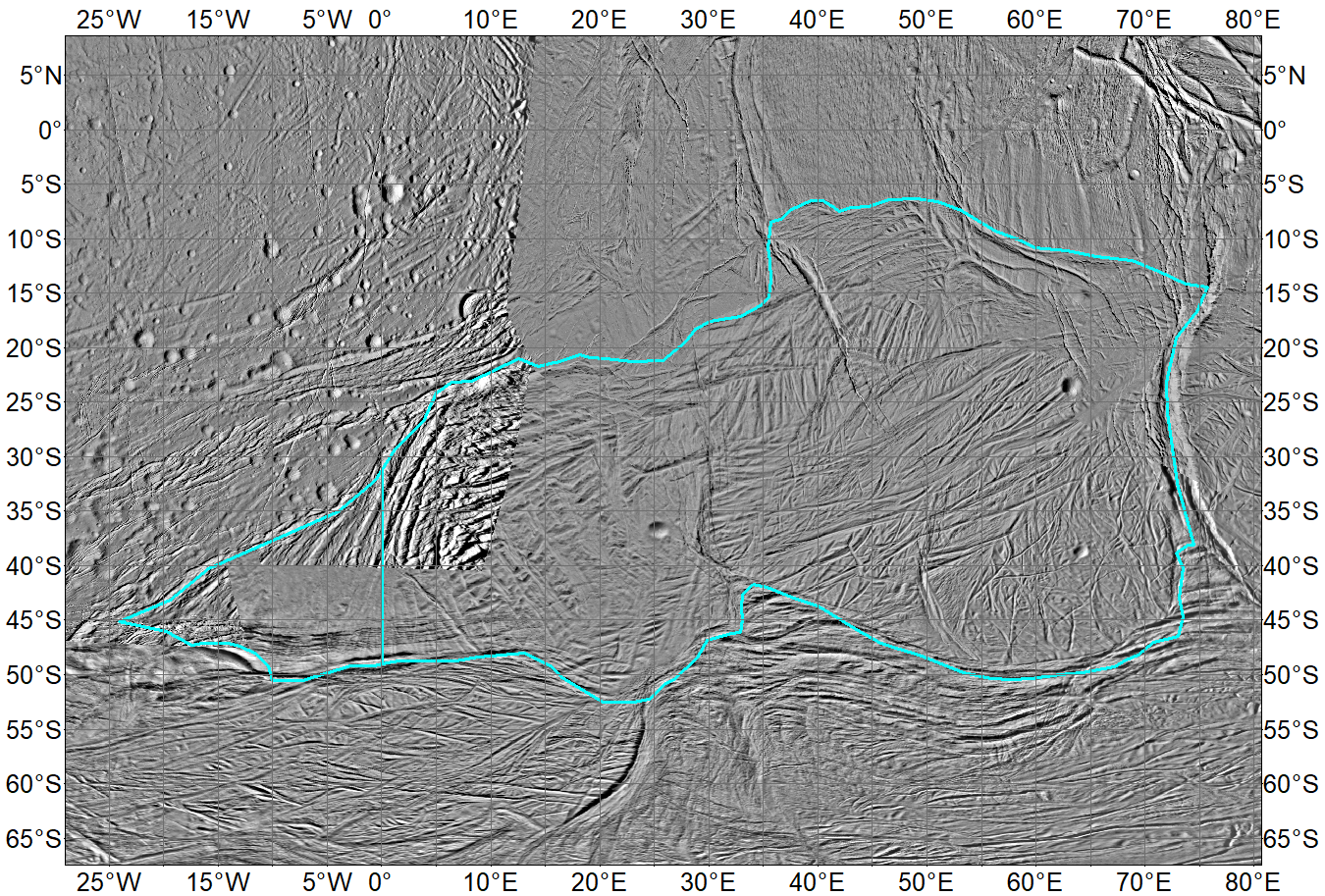}
    \caption{}
    \label{fig:cw-T-transitional-left}
\end{figure}
This unit comprises the left portion of the transitional plains on Enceladus’s trailing hemisphere, as defined by CW15, and overlaps with the western part of Ridge Plain 1 from \citet{Kirchoff2009} (KS-RP1-left). It forms a transition zone between the trailing hemisphere terrains and the south polar region, with surface morphologies partly resembling those of the south polar central unit. The eastern boundary is marked by Labtayt Sulci, a $\sim$1 km–deep canyon-like scarp that sharply separates the left and right transitional plains. The surface is dominated by extensive, NE–SW–trending subparallel troughs (1–3.5 km wide, up to 200 km long), with subordinate cross-cutting structures reflecting localised tectonic modification associated with south polar resurfacing.\\

Despite this structural complexity, the unit preserves a coherent crater population. Several fresh, mid-sized (5–7 km) simple craters superpose the trough fabric, indicating that cratering postdates the main tectonic deformation. Small craters (<2 km) are distributed across the unit; although pit-chain–like troughs occur locally, they are spatially limited and do not systematically disrupt the crater population at the diameters analysed.\\

Crater counts from KS-RP1-left and two subregions (Near-Mid-CP6 and Near-Mid-CP7), which were subdivided due to differences in image resolution and mapping conventions \citep{Kirchoff2009,CrowWillard2015}, exhibit nearly identical crater size–frequency profiles. This consistency indicates a shared cratering history, justifying their combination into a single representative unit to improve statistical robustness, particularly at diameters below 2~km. The resulting size–frequency distribution is smooth at small diameters, with only minor flattening above $\sim$2.5 km. Consequently, craters in the 0.9–2.5 km diameter range from this unit provide a reliable and undisturbed record of recent impact flux, making them well suited for constructing the CPF.

\subsection{Trailing hemisphere striated plains, left (CW-T-striated-left)}
\begin{figure}[H]
    \includegraphics[width=1.0\linewidth]{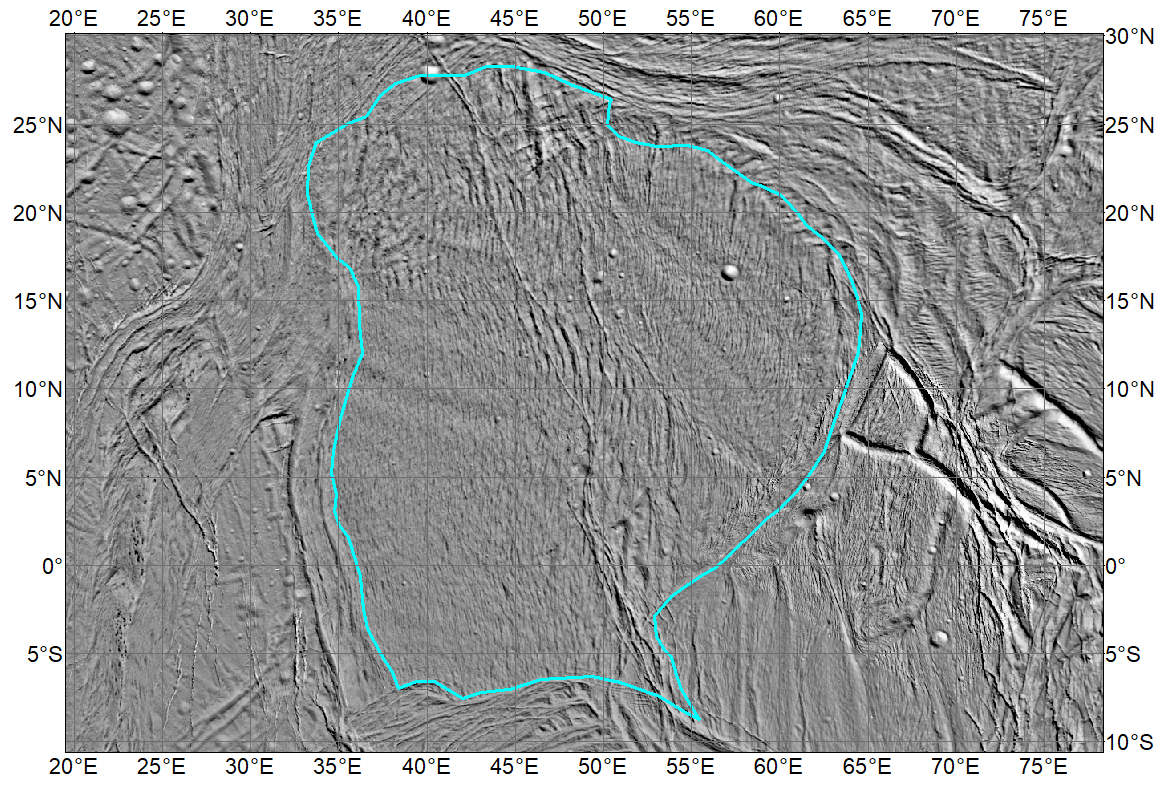}
    \caption{}
    \label{fig:cw-T-striated-left}
\end{figure}
The left portion of the striated plain, as defined by CW15 and also known as Sarandib Planitia, is characterised by numerous 0.6–2 km wide north–south trending ridges and troughs, with groups of small, curved, sigmoidal features arranged in parallel, staggered, en echelon patterns. These morphologies indicate zones of localised shear or distributed tectonic deformation (see Fig. 7 of CW15). A prominent set of wider (2–4 km) NW–ES-trending troughs cross-cuts the unit, locally transecting the finer sigmoidal features. Some craters are superimposed on these troughs, with well-defined floors and rims, demonstrating that the troughs predate the craters.\\

The crater size–frequency distribution is generally smooth across the recorded diameters (0.9–4.1 km), and the regular pattern of sigmoidal features allows small craters to be easily distinguished. However, the wider troughs may obscure some craters larger than 2 km, potentially affecting completeness at the upper end of the size range. To ensure robust and unbiased results, we select craters in the 0.9–2.5 km diameter range for fitting the CPF. This range avoids areas of possible incompleteness and provides a well-preserved, representative record of recent impact flux.

\subsection{Ridged plain-4, right portion (KS-RP4-right)}
\begin{figure}[H]
\includegraphics[width=1.0\linewidth]{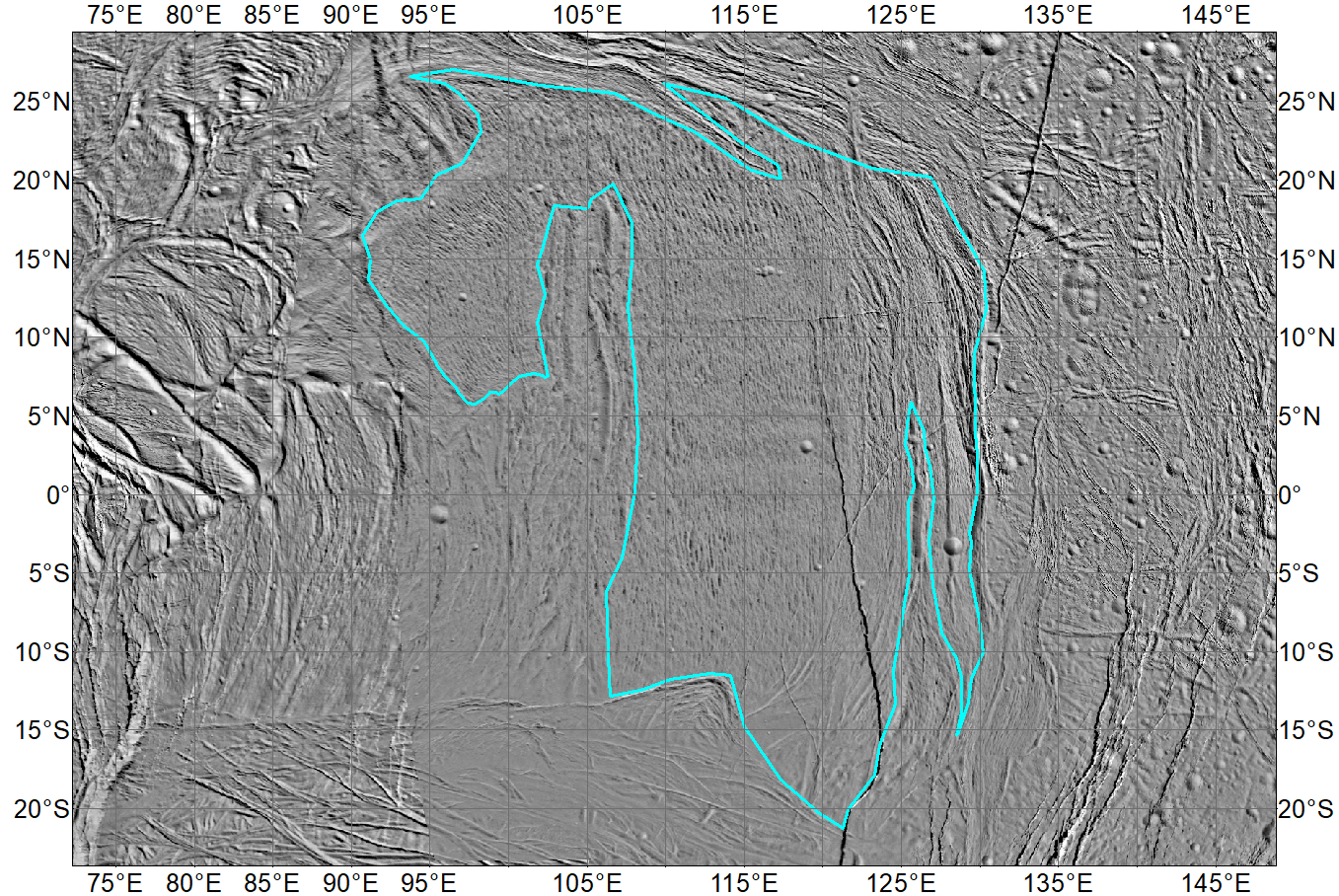}
\caption{}
\label{fig:ks-rp4-right}
\end{figure}
The right portion of the ridged plain-4, as defined by \citet{Kirchoff2009}, occupies the northern and eastern part of the east-side striated plains on Enceladus’ trailing hemisphere and, together with KS-RP5, forms the CW-T-striated-right unit of CW15 (Diyar Planitia). These two subunits (KS-RP4-right and KS-RP5) were separated by \citet{Kirchoff2009} based  on differences in crater densities and surface morphology.\\

KS-RP4-right is distinguished by tightly packed, narrow (0.6–2 km wide), north–south trending ridges and troughs that create a laterally continuous, morphologically uniform background, suggesting a single resurfacing episode. In some areas, notably near 17$^\circ$N, groups of sigmoidal features arranged in en echelon patterns indicate localised shear and tectonic deformation associated with the most recent resurfacing. In contrast, the adjacent KS-RP5 unit exhibits wider ($\sim$2~km), more widely spaced ridges and troughs with variable orientations, indicating a different geological history, thus warrant to be separated.\\

KS-RP4-right also contains two narrow, deep troughs that cut across all ridges and lineated bands, but their limited width means they have minimal impact on the overall crater population. One trough follows the eastern boundary, while the other extends over 300 km (from 11◦ N, 119◦ E to 45◦ S, 110◦ E), intersecting and partially removing the rim of a crater at 6°S, 123°E. The crater size–frequency distribution in KS-RP4-right is smooth and consistent, supporting its treatment as a single surface population. While craters larger than 2.5 km may be reliable, the presence of wider troughs in KS-RP5, covering nearly one-third of the unit, complicates identification of larger craters in that region. Therefore, for KS-RP4-right, we focus on craters in the 0.9–2.5 km diameter range, which are well preserved and unaffected by local tectonic modification. This ensures that the selected crater counts provide a robust and representative basis for modelling the CPF.

\subsection{Far-side equatorial cratered plain 4 (Far-Eq-CP4)}
\begin{figure}[H]
    \includegraphics[width=1.0\linewidth]{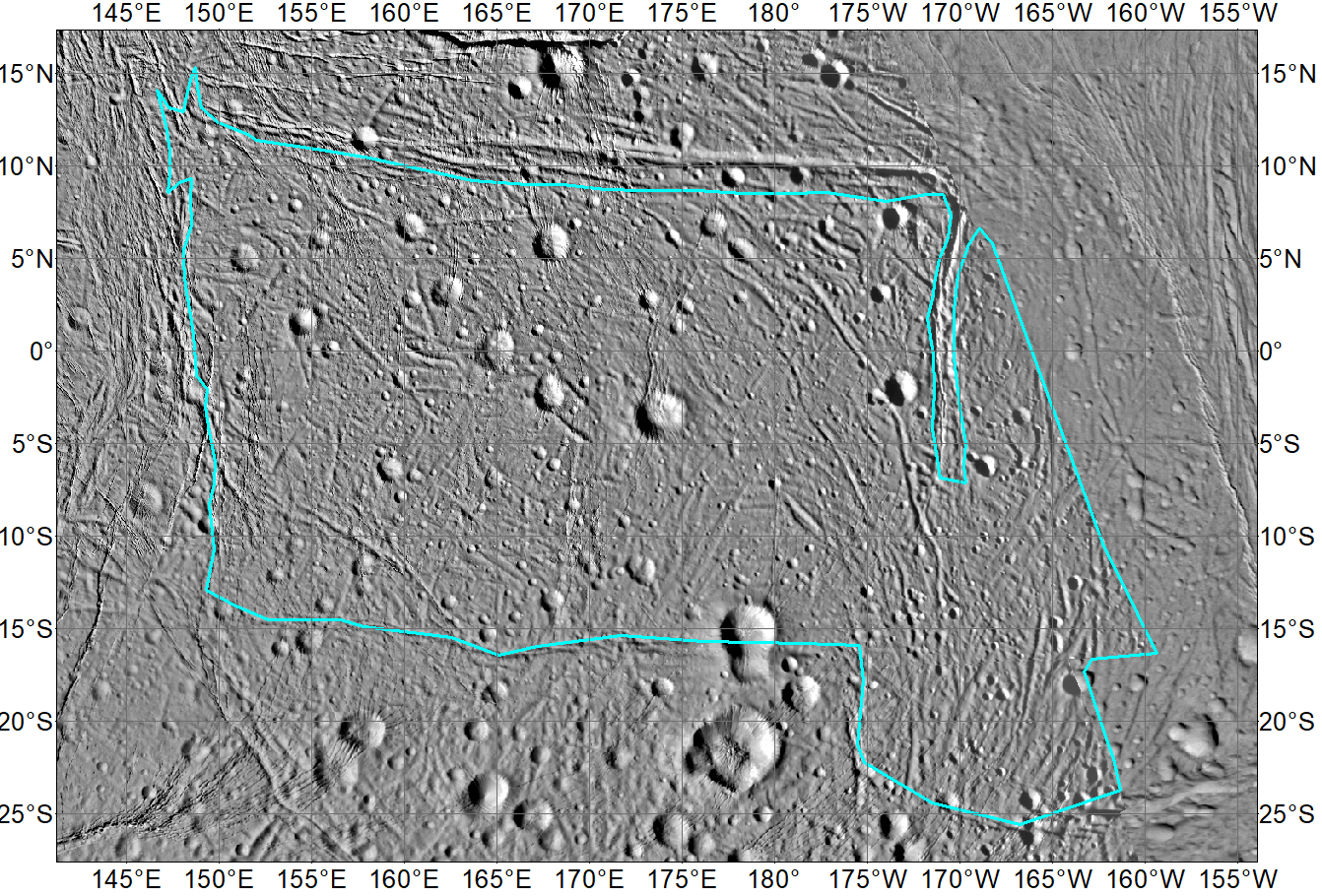}
    \caption{}
    \label{fig:far-eq-cp4}
\end{figure}
This equatorial cratered plain, originally part of the northern equatorial region defined by \citet{Kirchoff2009} (Enc-cp-eq), is mapped separately to avoid areas with prominent surface features and to ensure a representative crater count. The unit is notable for its lack of large craters, with the largest reaching 13~km near the southern boundary (178$^\circ$E, 15$^\circ$S). It is bounded by east–west-trending troughs to the north and narrower, shallower troughs to the south.\\

A distinctive boundary feature is the abrupt shift in trough orientation (at 170$^\circ$W, 8$^\circ$N) from east–west to north–south at 170$^\circ$W, 8$^\circ$N. Poor image resolution east of this point makes it unclear whether this bend reflects tectonic activity from the adjacent, younger leading hemisphere terrain or simply the intersection of two separated troughs. This distinctive feature is excluded from the unit mapping, ensuring that areas where troughs could obscure craters are not included in the crater count.\\

The tectonic history comprises: (1) ancient NE–SW-trending troughs and ridges (1.5 to 2.5~km in width), now smoothed and overprinted by craters; (2) scattered NW–SE-trending troughs ($\sim$3~km wide), extends longer in the east; and (3) pervasive north–south-trending pit chains that overlay most craters. All craters postdate the main trough sequences, indicating a geologically homogeneous surface and minimising the risk of mixed-age populations.\\

Although pit chains obscure some small craters—particularly near 5$^\circ$N, 157$^\circ$E—this effect is limited to diameters below 1.2 km. The crater size–frequency distribution is otherwise smooth from 1.2 km up to the largest observed diameters, providing a continuous and representative record of recent impacts. Consequently, crater counts for diameters $\ge$1.2~km from this unit are well suited for constructing the CPF, as they reflect a well-preserved and undisturbed impact chronology.

\subsection{Far-side equatorial cratered plain 6 (Far-Eq-CP6)}
\begin{figure}[H]
    \includegraphics[width=1.0\linewidth]{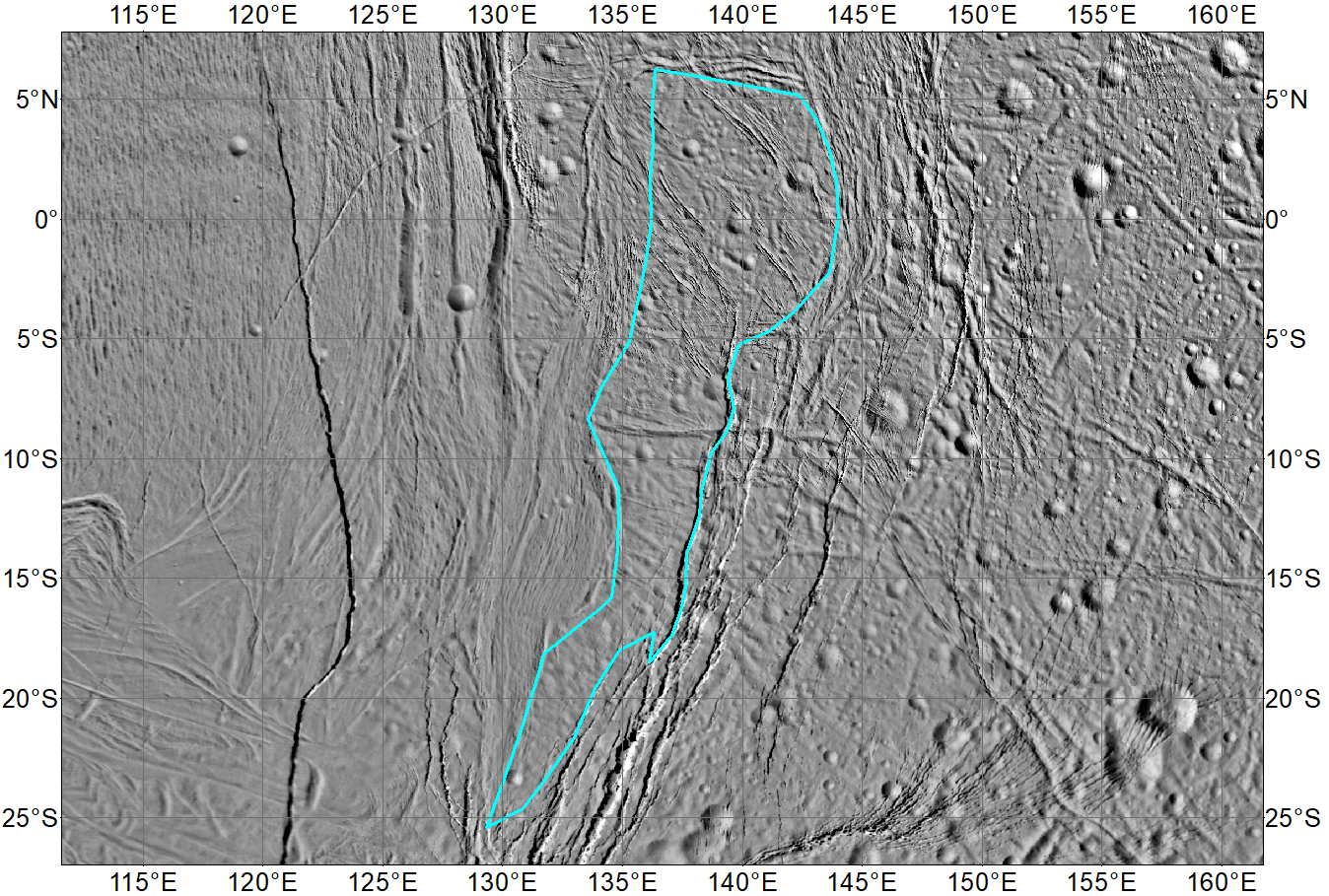}
    \caption{}
    \label{fig:far-eq-cp6}
\end{figure}
A narrow cratered plain unit lies between the trailing hemisphere curvilinear plains (CS-T-curvilinear-right) and the far-side crater plains. Originally part of a larger cratered plain mapped by \citet{Kirchoff2009}, it is now delineated separately to avoid adjacent tectonic features and improve crater count accuracy. The eastern boundary is marked by a prominent corrugated trough or chasmata, with intersecting X- and Y-shaped structures extending from 5$^\circ$S to 40$^\circ$S, nearly spanning the southern hemisphere before ending at the CW-S-curvilinear boundary. In the north, a NW–SE-trending pit chain cuts across both craters and troughs, representing the youngest geological features in the unit \citep{Martin2014}. Aside from the pit chain, the unit is morphologically homogeneous, likely formed in a single event and subsequently accumulated craters, some of which are cross-cut by the pit chain. The absence of large intrusive features minimises contamination and obscuration, making this unit ideal for crater analysis. However, as a cratered plain surrounded by tectonic units and with pit chains potentially obscuring small craters, only craters 2–5 km in diameter are used for analysis. This range avoids undercounting at smaller diameters and provides a smooth, reliable segment of the crater size–frequency distribution for fitting the CPF.

\subsection{Far-side equatorial cratered plain 8 (Far-Eq-CP8)}
\begin{figure}[H]
    \includegraphics[width=1.0\linewidth]{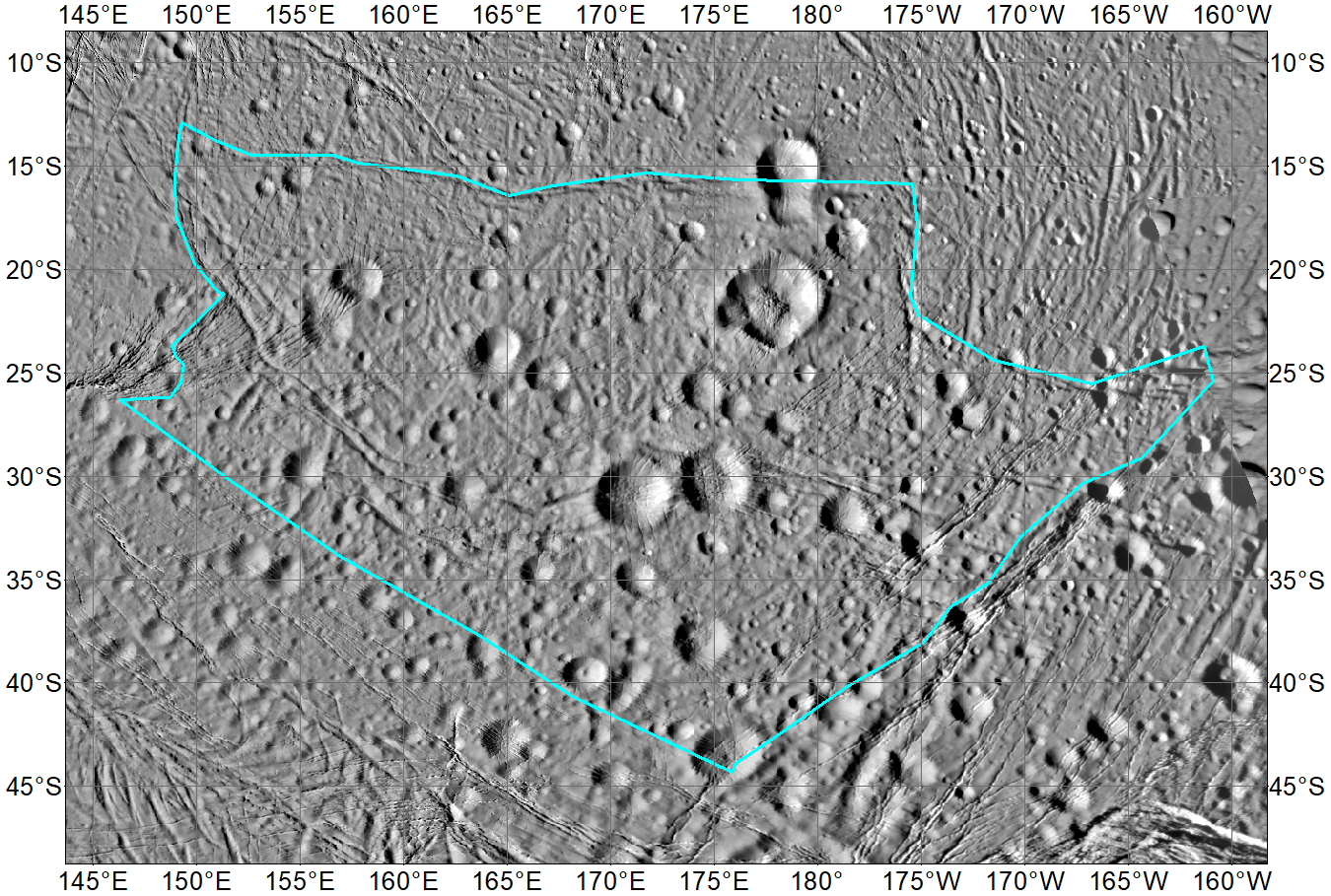}
    \caption{}
    \label{fig:far-eq-cp8}
\end{figure}
Originally part of the equatorial crater plains defined by \citet{Kirchoff2009} (Enc-cp-eq), this unit is mapped separately to better capture its higher crater density and distinct size–frequency profile, particularly the prevalence of larger craters. The southern boundary is drawn to exclude areas influenced by recent south polar resurfacing, ensuring the crater population reflects a stable, undisturbed history.\\

The surface features subdued troughs of varying orientations, often intersected by craters. A prominent ancient double trough crosses the unit, interrupted by numerous craters and pit chains. Three sets of pit chains, formed at different times, are present: (1) a north–south chain at 32°S, 172°E crossing two ~15 km craters; (2) a northwest–southeast chain near 31°S, 177°E, likely contemporaneous with those in the southern Far-Mid-CP8; and (3) a more recent northeast–southwest chain in the east, likely linked to the eastern Far-Mid-CP9 \citep{Martin2014}.\\

Although classified as equatorial cratered plain (in \citealt{Kirchoff2009}), the geomorphology and crater size–frequency distribution closely resemble those of mid-latitude plains. Compared to the northern Far-Eq-CP4, this unit likely experienced less subsurface heating and burial by E-ring material, preserving more large craters (>10 km). Geological evolution has been influenced by neighbouring regions, especially activity from the south polar terrain.\\

Crater diameters in this unit range from 800 m to over 20 km, with a smooth size–frequency distribution and a gradual roll-over below 2 km, and minor flattening near 6 km. This broad, well-preserved diameter range provides a robust dataset for modelling the CPF, capturing both small and large crater populations and offering key constraints on the impactor flux and surface age.

\subsection{Far-side mid-latitude cratered plain 5 (Far-Mid-CP5)}
\begin{figure}[H]
    \includegraphics[width=1.0\linewidth]{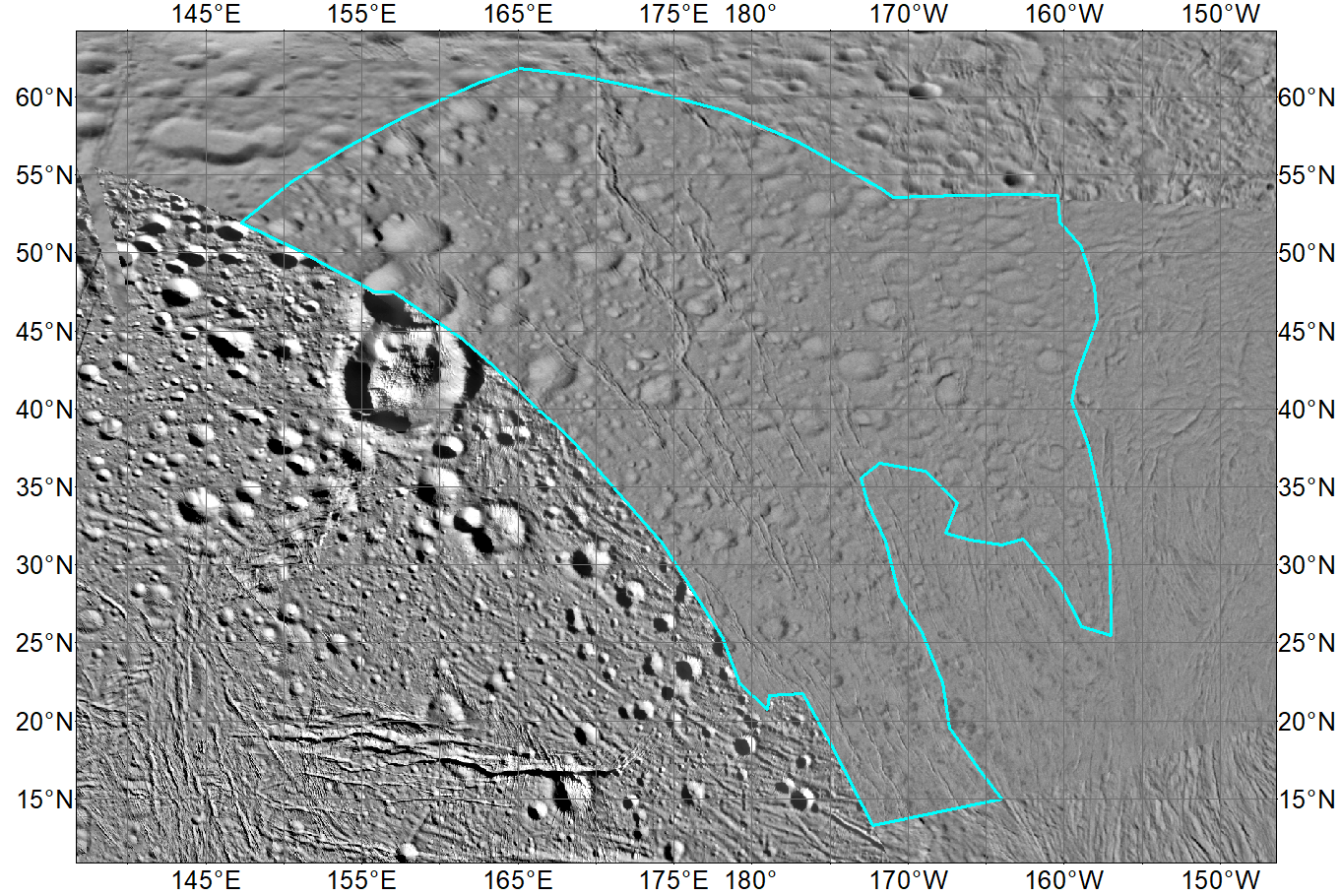}
    \caption{}
    \label{fig:far-mid-cp5}
\end{figure}
This heavily cratered region, previously uncounted due to lower image resolution and unfavourable viewing geometry, is now mapped using higher-resolution data. Limited resolution in past observations likely caused an artificial drop-off in crater detection below ~3.5 km diameter. For Enceladian CPF analysis, we therefore focus on craters $\ge$4.0 km, where detection is complete and reliable.

The terrain features NW–SE-trending corrugated troughs (e.g., from 58$^{\circ}$N, 171$^{\circ}$E to 30$^{\circ}$N, 175$^{\circ}$W) and prominent pit chains, including one intersecting the Al-Haddar, Shahrazad, and Dunyazad craters (15–30.8 km diameter). Despite these structures, craters $\ge$4 km are not significantly obscured, and the size–frequency distribution remains smooth up to 10 km. The triple craters are included in the unit count (also shared with adjacent Far-Mid-CP3) but excluded from CPF fitting due to low-number statistics.

\subsection{Combined near-side mid-latitude cratered plain 1 and 2 (Near-Mid-CP1 \& -CP2)}
The two mid-latitude crater plain units on the near-side can be combined for analysis, despite differences in image resolution, because they are used to construct the CPF at larger crater diameters. Large craters (>10 km) are rare, and craters smaller than 4 km are excluded from the CPF due to potential resolution effects. Thus, variations in image quality do not significantly affect the combined crater statistics.\\

Crater size–frequency measurements show that resolution roll-off begins at larger diameters (~7 km) for the lower-resolution Near-Mid-CP1, and below 4 km for the higher-resolution Near-Mid-CP2. Above 7 km, both units display smooth, consistent size–frequency profiles. Additionally, the similar crater densities and surface morphologies suggest both units experienced the same resurfacing event and subsequent cratering history. Therefore, combining these units improves statistical robustness for the analysis of rare, large craters, providing a reliable basis for modelling the CPF.\\

Individual geomorphological and crater distribution descriptions for the two units are provided below.

\subsubsection{Near-Mid-CP1}
\begin{figure}[H]
    \includegraphics[width=1.0\linewidth]{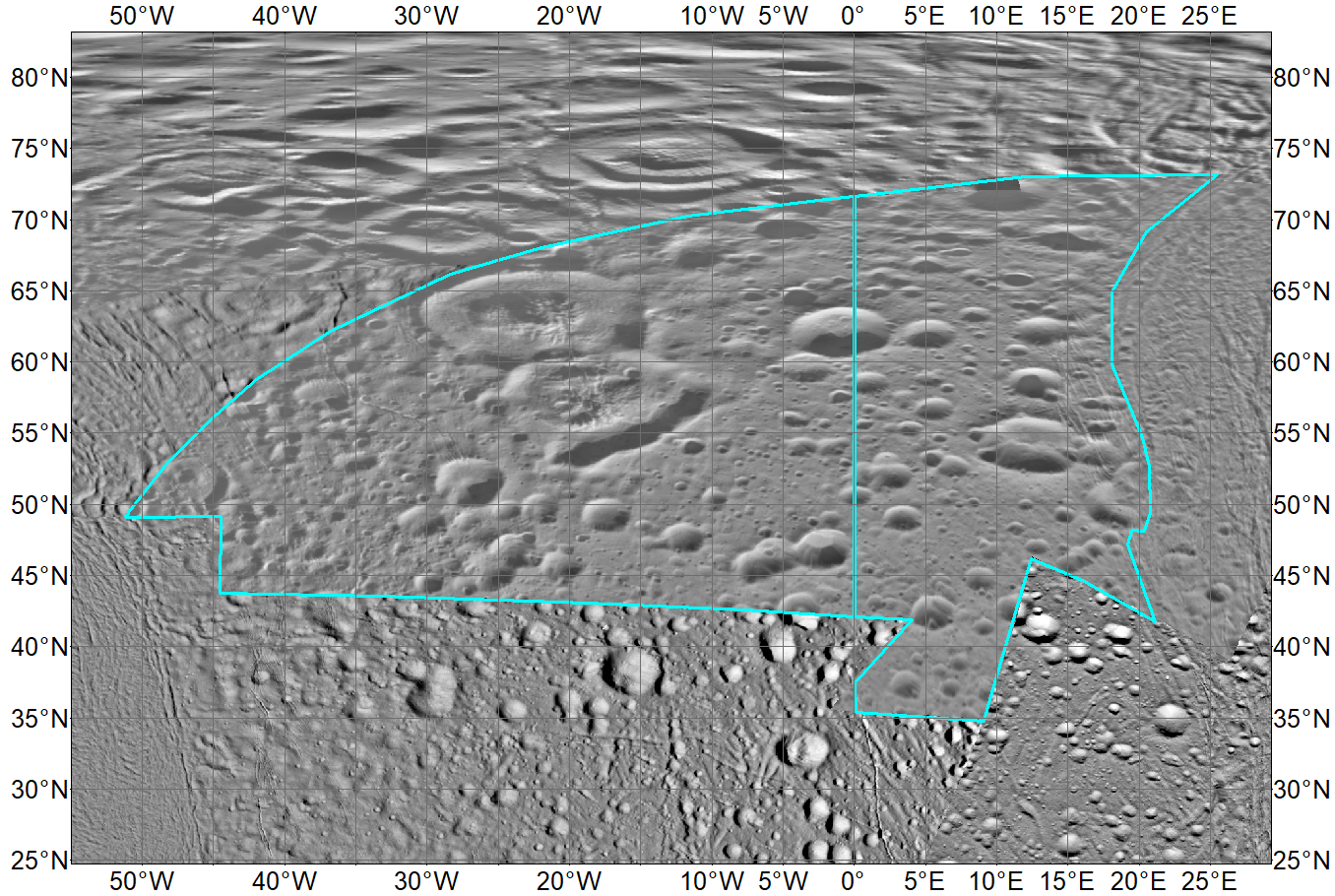}
    \caption{}
    \label{fig:near-mid-cp1}
\end{figure}

This unit contains two of Enceladus’s largest craters: Ali Baba (34 km, 57°N, 18°W) and Aladdin (30 km, 62°N, 22°W), with Ali Baba being the largest on the moon. Both exhibit prominent interior domes, indicative of viscous relaxation. Much of the unit (26°W to 15°E) appears pristine, with minimal tectonic disruption—likely due to limited image resolution (200 m/pixel), which is twice as coarse as the typical 100 m/pixel. In this region, tectonic features are barely visible.\\

Between 45°W and 30°W, thin, roughly north–south ridges and lineations emerge along the unit’s western boundary, adjacent to CW-L-curvilinear. Similar structures extend from the eastern CW-T-north-lineated unit along 16°E.

\subsubsection{Near-Mid-CP2}
\begin{figure}[H]
    \includegraphics[width=1.0\linewidth]{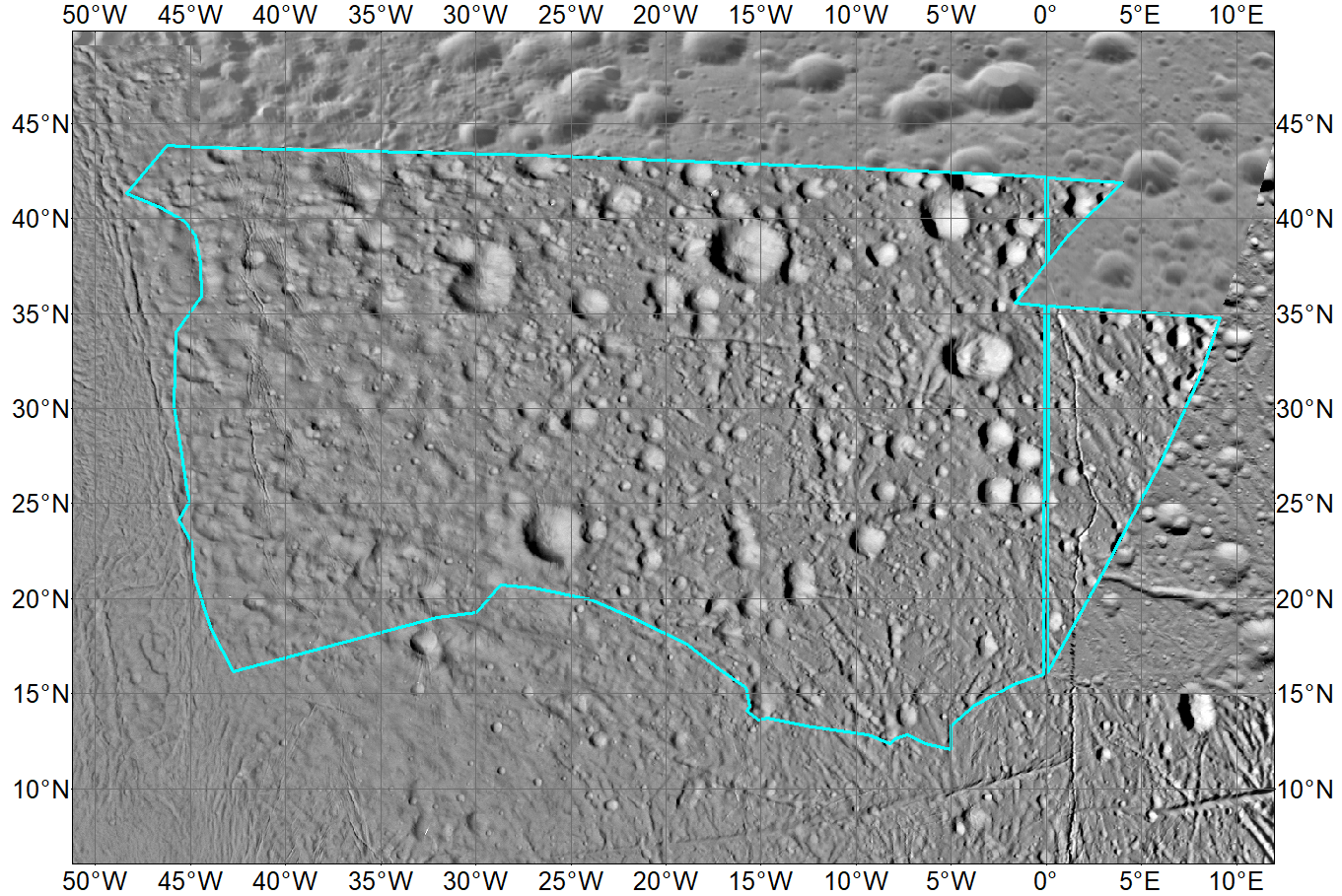}
    \caption{}
    \label{fig:near-mid-cp2}
\end{figure}

Higher-resolution images (110 m/pixel) of this unit reveal three prominent tectonic features:
(1) isolated, fragmented single ridges up to 2 km wide, dispersed from 30°W to 5°E with varied orientations, some truncated by adjacent craters;
(2) a smoother western boundary, adjacent to CW-L-curvilinear, features pit chains that intersect partially eroded craters. These pit chains may extend northward to Near-Mid-CP1, where image resolution is sufficient to clearly distinguish them from thin ridges;
and (3) a deep, narrow trough (~1 km wide, 345 km long) cuts across smoothed ridges and craters, extending from 40°N to 37°S. While it may continue northward, its narrow width makes it undetectable in lower-resolution images. Part of this unit overlaps with regions 3 and 4 of \citet{Kinczyk2024}, where our crater counts yield similar densities and size–frequency profiles within Poisson uncertainties.

\subsection{Combined far-side mid-latitude cratered plain 1, 2, and 3 (Far-Mid-CP1, -CP2, \& -CP3)}

These three units are subdivisions of the mid-latitude cratered plain (270$^\circ$W to 180$^\circ$W) defined by \citet{Kirchoff2009}. Boundaries are mapped along some troughs that transect the units, with the southern edge drawn to exclude areas where craters appear partially eroded by resurfacing or influenced by features extending from the trailing hemisphere curvilinear plain (CW-T-curvilinear (CW15); ridged plain 6 \citep{Kirchoff2009}).\\

All three units display pit chains that cross some craters and smooth background troughs overlain by craters. The crater size–frequency distribution of each subunits rolls off near 3 km, with increased fluctuations above 15 km due to low-number statistics for large craters. Because the units share similar surface and crater morphologies, we combine them to improve statistical reliability at larger diameters for constructing the CPF.\\

Individual geomorphological and crater distribution descriptions for the three units are provided below.

\subsubsection{Far-Mid-CP1}
\begin{figure}[H]
    \includegraphics[width=1.0\linewidth]{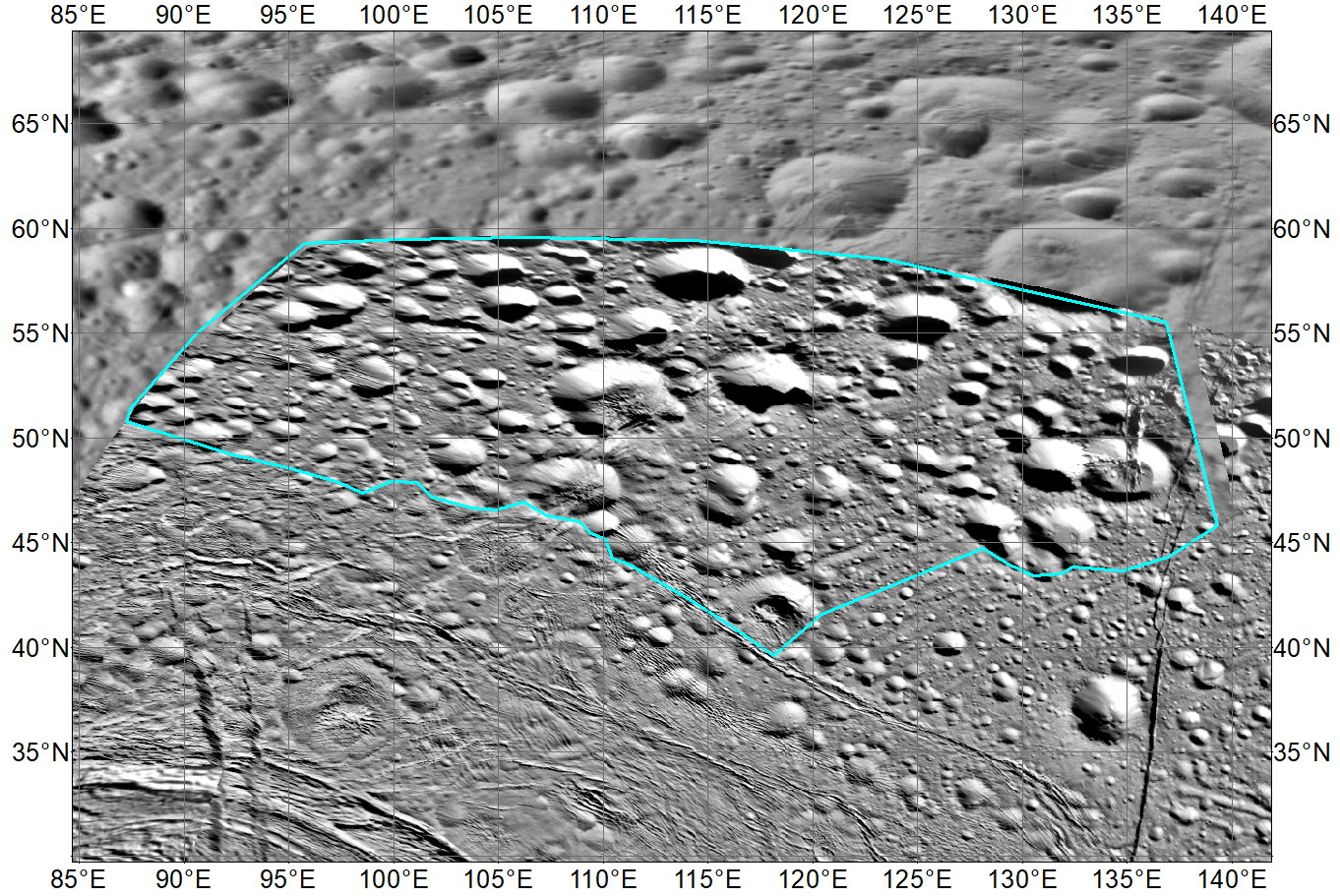}
    \caption{}
    \label{fig:far-mid-cp1}
\end{figure}
This heavily cratered terrain is characterised by bowl-shaped craters, with those larger than 10 km displaying flat floors or central uplifts. Three main tectonic features define the landscape: (1) In the southeast, older NE–SW-trending ridges and bands, formed during terrain solidification, are disrupted by later craters; (2) In the northwest, a younger NW–SE-trending pit chain extends ~60 km, intersecting several small to mid-sized craters; (3) Toward the south, the surface becomes smoother with fewer craters near the boundaries.\\

The sequence of events began with the formation of smooth, subdued ridges, followed by crater accumulation. Later resurfacing and tectonic activity from the south (CW-T-curvilinear-right) removed or smoothed some smaller craters. To avoid areas affected by resurfacing, the unit boundaries are drawn more conservatively than in the original mid-latitude cratered plains defined by \citet{Kirchoff2009}, shifted northward and away from the southern margin. This ensures that included craters are less likely to be eroded or overprinted by activity or features extending from the south.\\

The better craters, particularly those unaffected by recent resurfacing, provide a robust dataset for modelling the CPF.

\subsubsection{Far-Mid-CP2}
\begin{figure}[H]
    \includegraphics[width=1.0\linewidth]{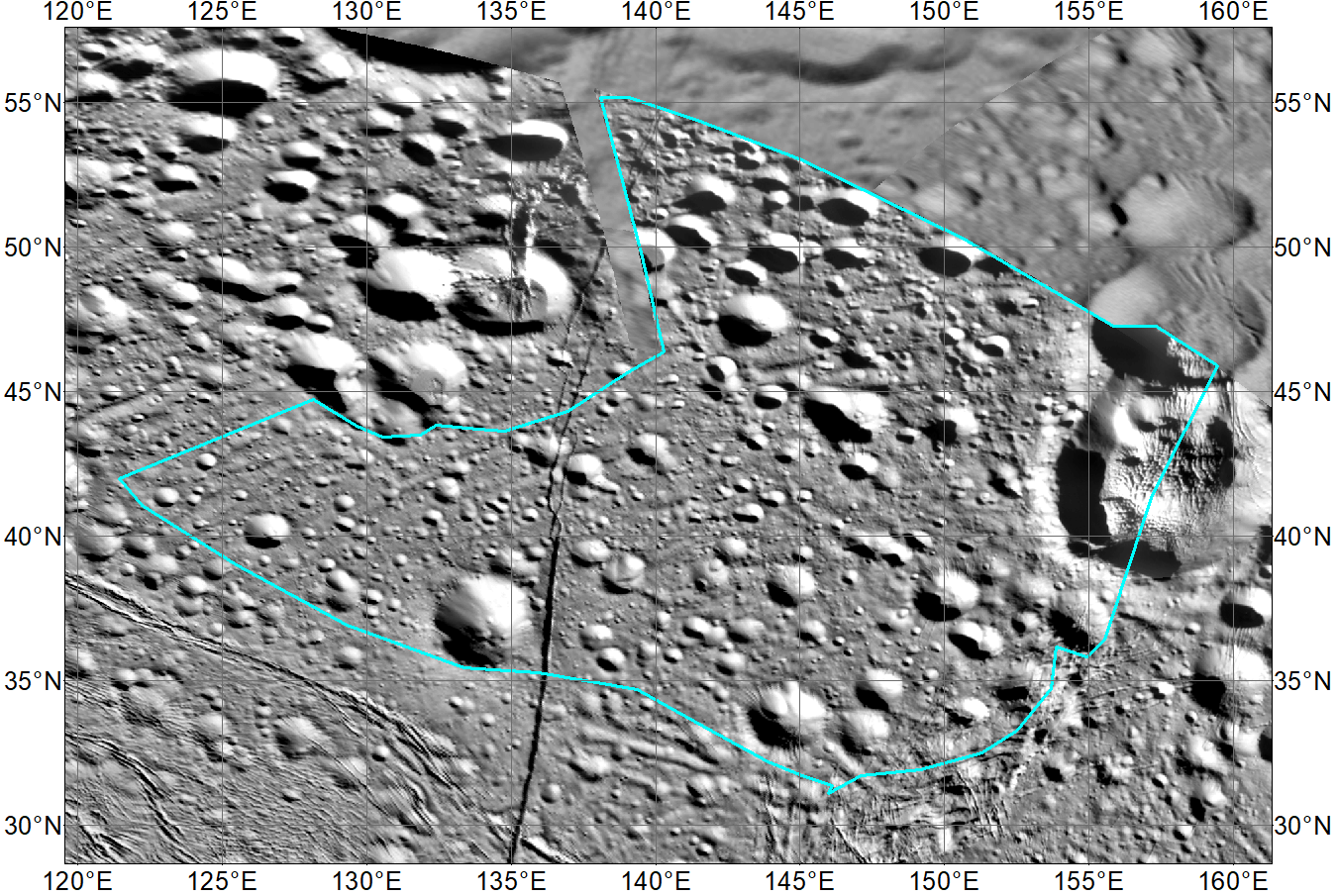}
    \caption{}
    \label{fig:far-mid-cp2}
\end{figure}
The unit displays two key tectonic features: (1) broad, subdued NE–SW-trending ridges and bands, extended from Far-Mid-CP1, likely formed during plains solidification and subsequently disrupted by subsequent crater impacts; and (2) north–south-trending pit chains in the southeast, which cross craters larger than ~7 km. Notably, a prominent pit chain traverses the floors of two large eastern boundary craters, Dunyazad (31 km) and Shahrazad (20 km), both part of a crater triplet in the northern hemisphere.

\subsubsection{Far-Mid-CP3}
\begin{figure}[H]
    \includegraphics[width=1.0\linewidth]{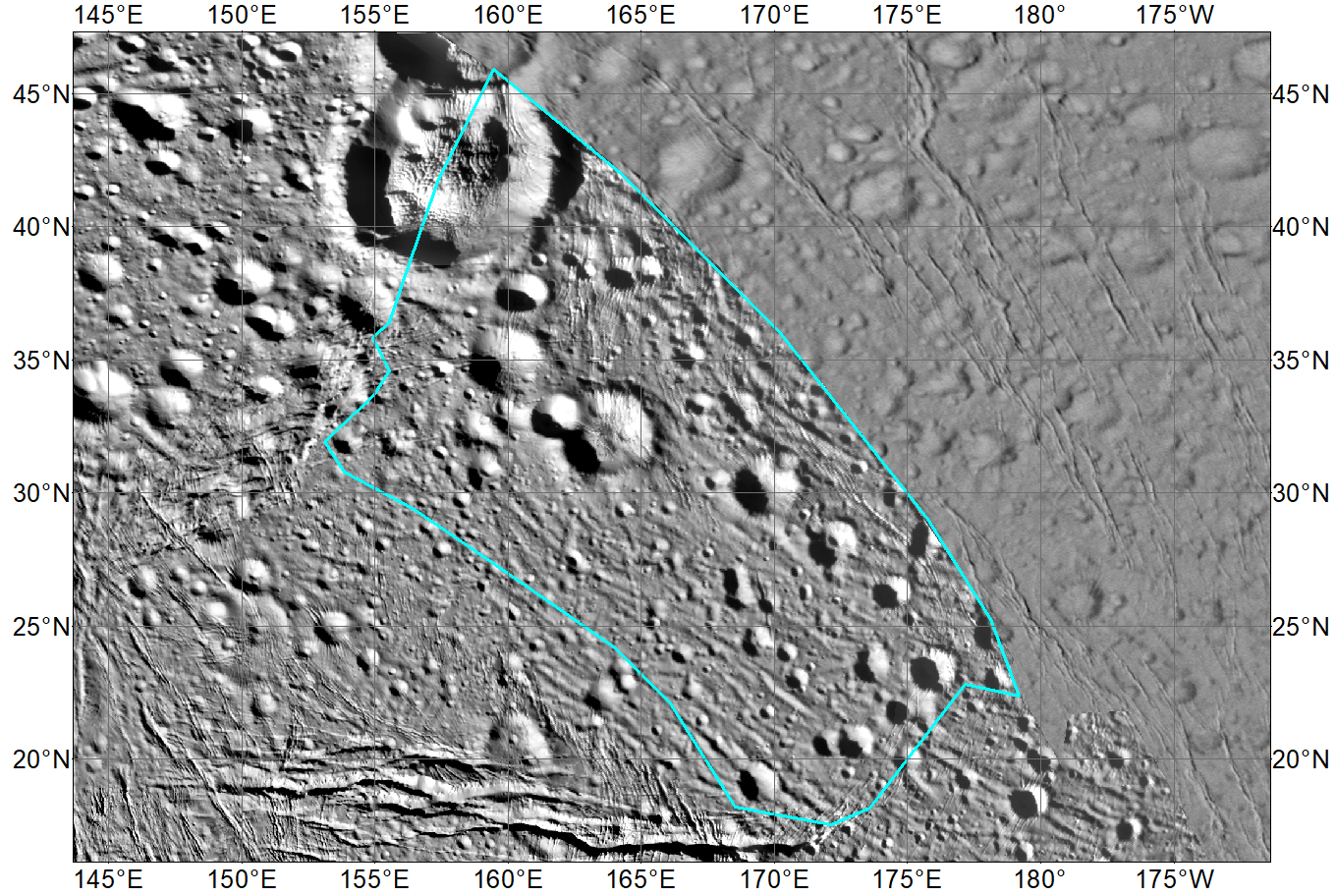}
    \caption{}
    \label{fig:far-mid-cp3}
\end{figure}
This unit, which shares large craters with Far-Mid-CP2, records two main tectonic events:
(1) unit-wide, smooth, parallel NW–SE-trending ridges and bands formed during plains solidification—these are the oldest features and are interrupted by later crater impacts; and
(2) a north–south-trending pit chain that cross-cuts craters in the northern part of the unit. The preservation of these features, along with large craters, allows for reliable crater counts suitable for modelling the CPF.

\clearpage

\setcounter{figure}{0}

\section{Demonstration of crater erosion}
\label{subsec:suppA}
\renewcommand{\thefigure}{S.2.\arabic{figure}}

\begin{figure}[H]
    \includegraphics[width=1.0\linewidth]{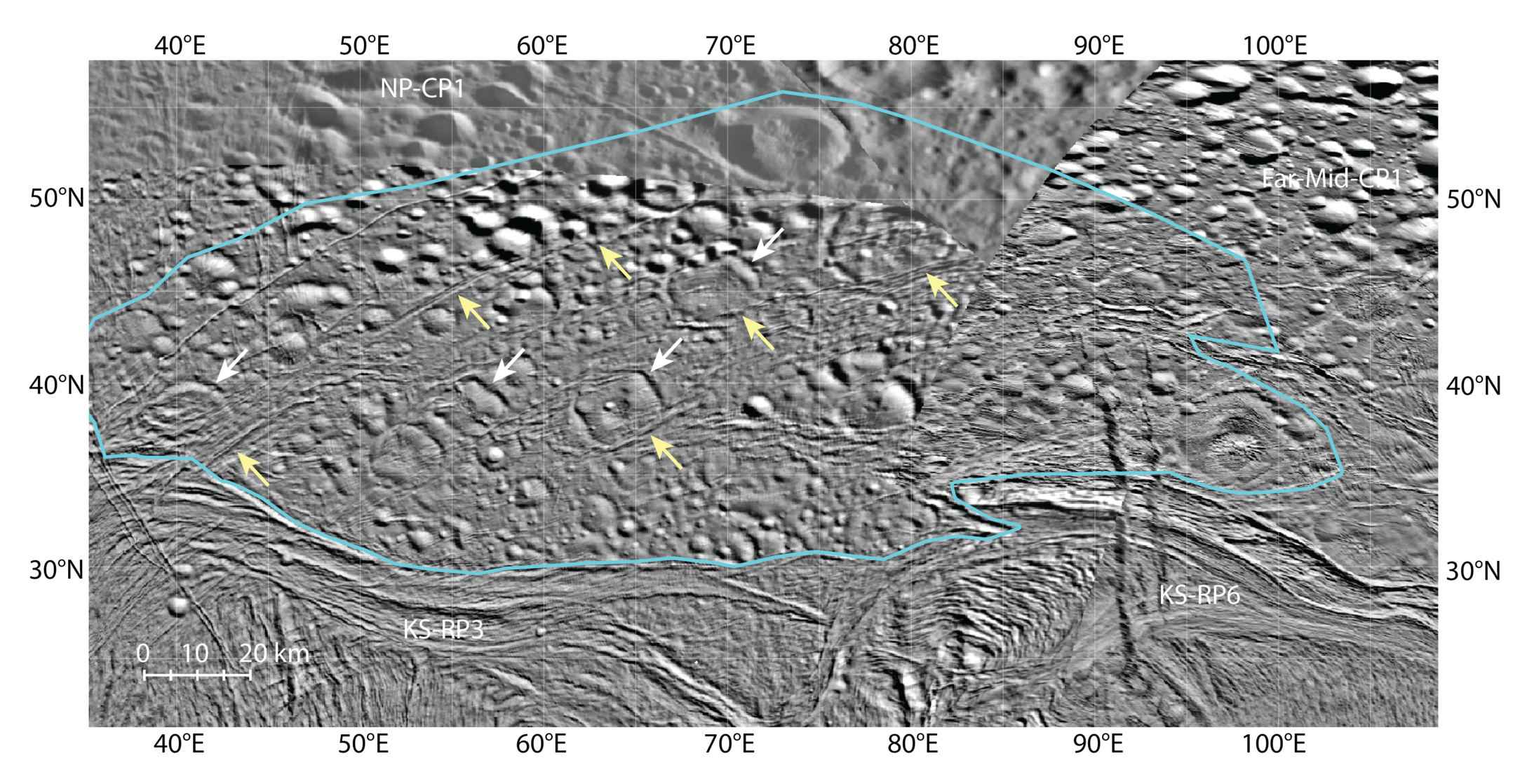}
    \caption{CW-T-square (70°E, 40°N), also known as the Subdued Cratered Plains \citep{CrowWillard2015}, is one of Enceladus' most heavily cratered regions. The craters here have undergone significant modification. Several large $\sim$20 km craters (marked by white arrows) eroded to be shallower and more angular, likely due to relaxation from elevated heat flow \citep{Bland2012} or tectonic deformation extending from the south-adjacent KS-RP3. The squarish craters, influenced by pre-existing faults/troughs (e.g., Meteor Crater in Arizona) or subsequent tectonics, are further evidence of deformation along NE-SW troughs (yellow arrows). As a result of extensive resurfacing, the crater size-frequency measurement rolls off at $\sim$10 km, in contrast to units with comparable crater densities, where it begins at $\le$4 km.}
    \label{fig:app_erosion1}
\end{figure}

\begin{figure}[H]
    \includegraphics[width=1.0\linewidth]{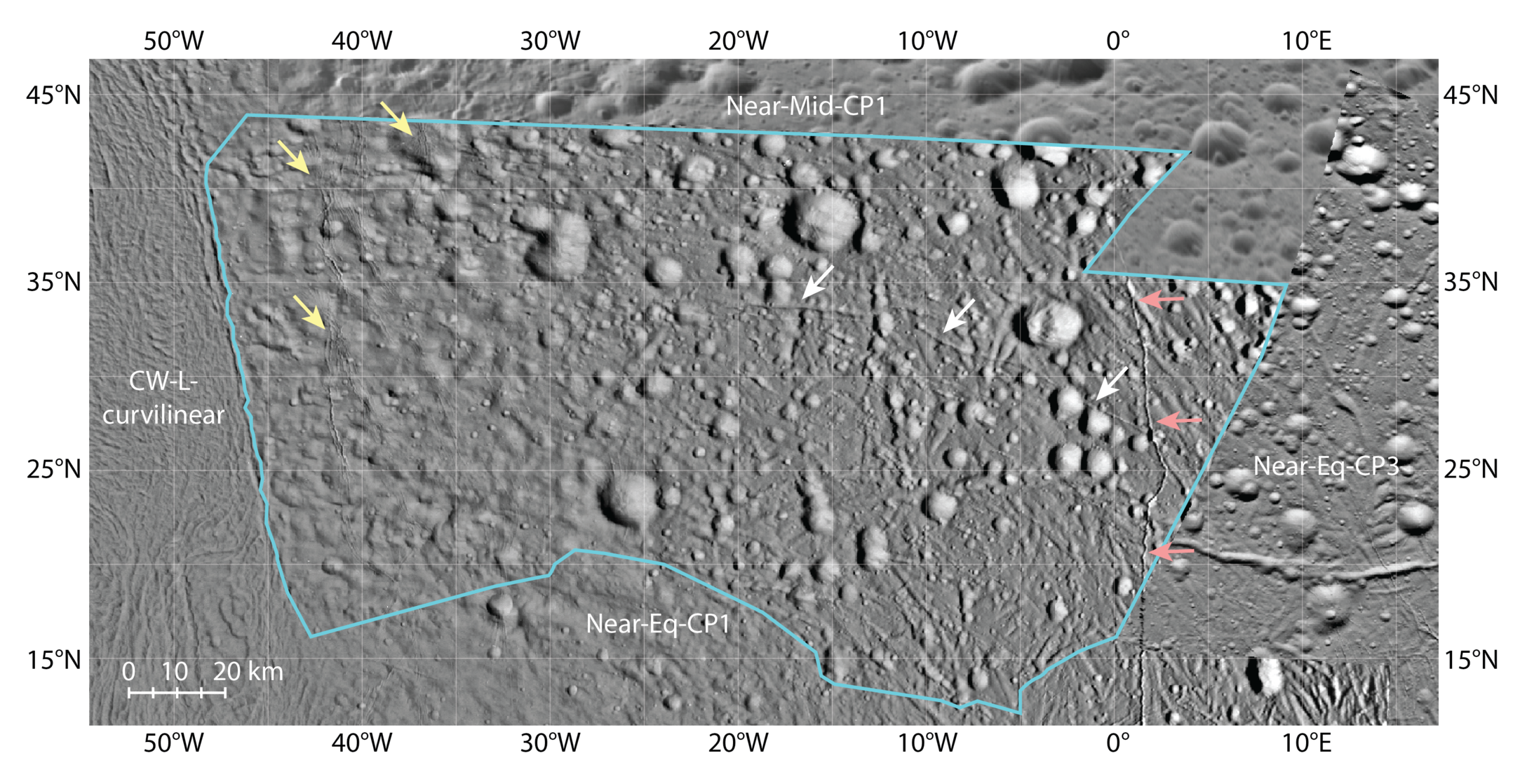}
    \caption{Near-Mid-CP2 (20°W, 30°N), several tectonic features are evident. (i) Isolated ridges, up to 2 km wide, are scattered across the right region with varying orientations, including horizontal and NW-SE. Some ridges appear truncated by craters (yellow arrow), indicating the craters are younger, though higher-resolution images are needed to confirm this. (ii) In the western region, tectonic activity from CW-L-Curvilinear unit has transitioned into this more heavily cratered Near-Mid-CP2, where craters gradually smooth out and lose their rim and bowl shapes. (iii) The western boundary also shows pit chains cutting through these partially eroded, smooth craters (white lines). (iv) A 1 km-wide trough, extending $\sim$340 km from 40°N to 37°S, cuts through ridges and craters (yellow arrows).}
    \label{fig:app_erosion2}
\end{figure}

\begin{figure}[H]
    \includegraphics[width=1.0\linewidth]{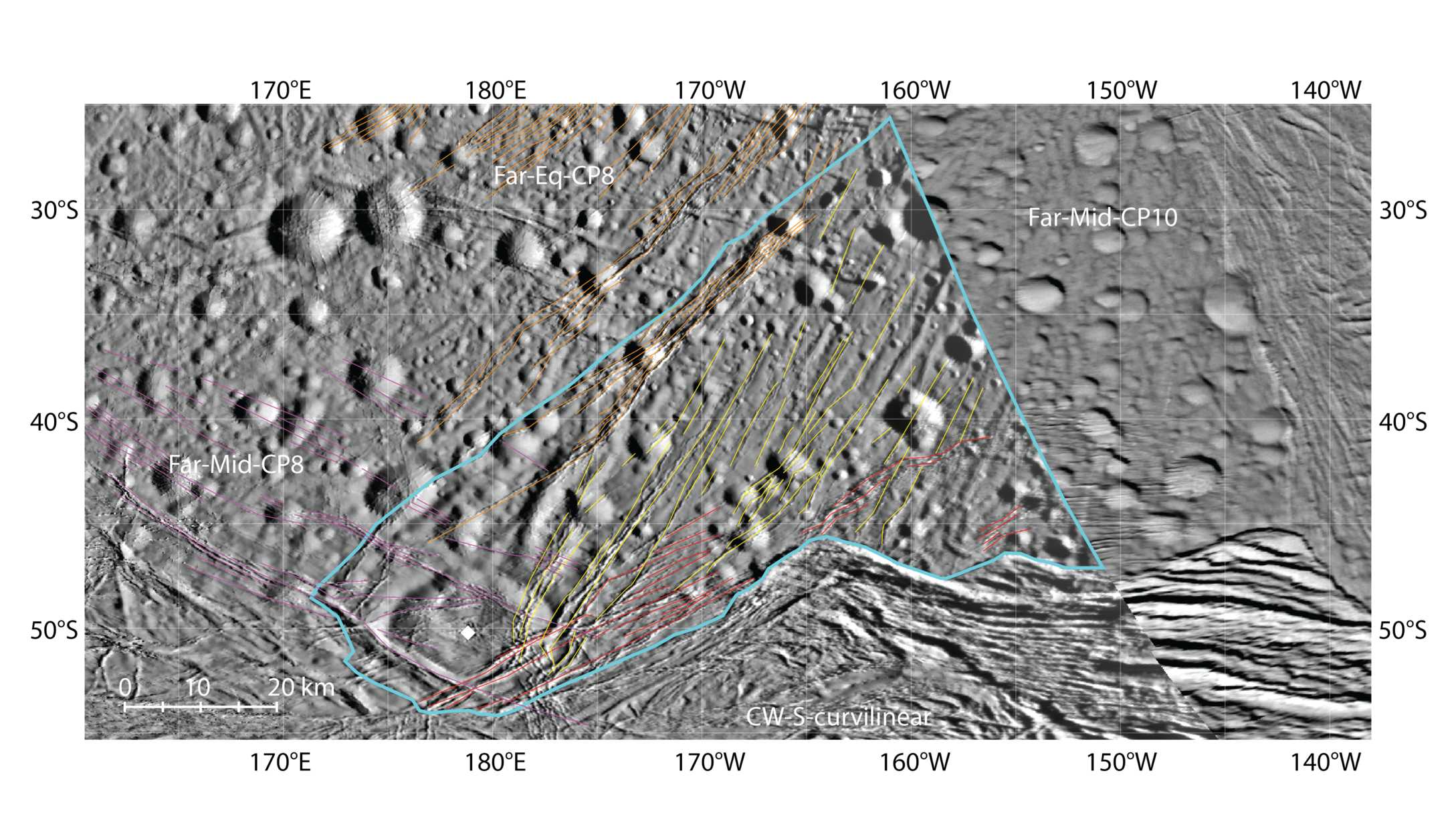}
    \caption{Far-Mid-CP9 (165°W, 40°S) feature three sets of pit chains formed at different times \citep{Martin2014}. The oldest NW-SE-trending chains (labelled as magenta lines) cross from the west, sharing an origin with those from Far-Mid-CP8. Two NE-SW trending chains lie along the southeast boundary adjacent to CW-S-curvilinear (red lines) and the northwest boundary adjacent to Far-Eq-CP8 (orange lines). The newest, also NE-SW chains dominate the terrain (yellow lines), intersecting most craters. The pit chain colours follow those used by \citet{Martin2014}. Due to images’ resolution limits, they indicate the general direction and rough location of each branch, rather than the precise position of individual pit chains. A most obvious feature is a 22 km crater at the southwest tip (white diamond), where pit chains and ridged belts have eroded at least one-third of its rim. Craters appear progressively more eroded and smoother toward the south.}
    \label{fig:app_erosion3}
\end{figure}

\begin{figure}[H]
    \includegraphics[width=1.0\linewidth]{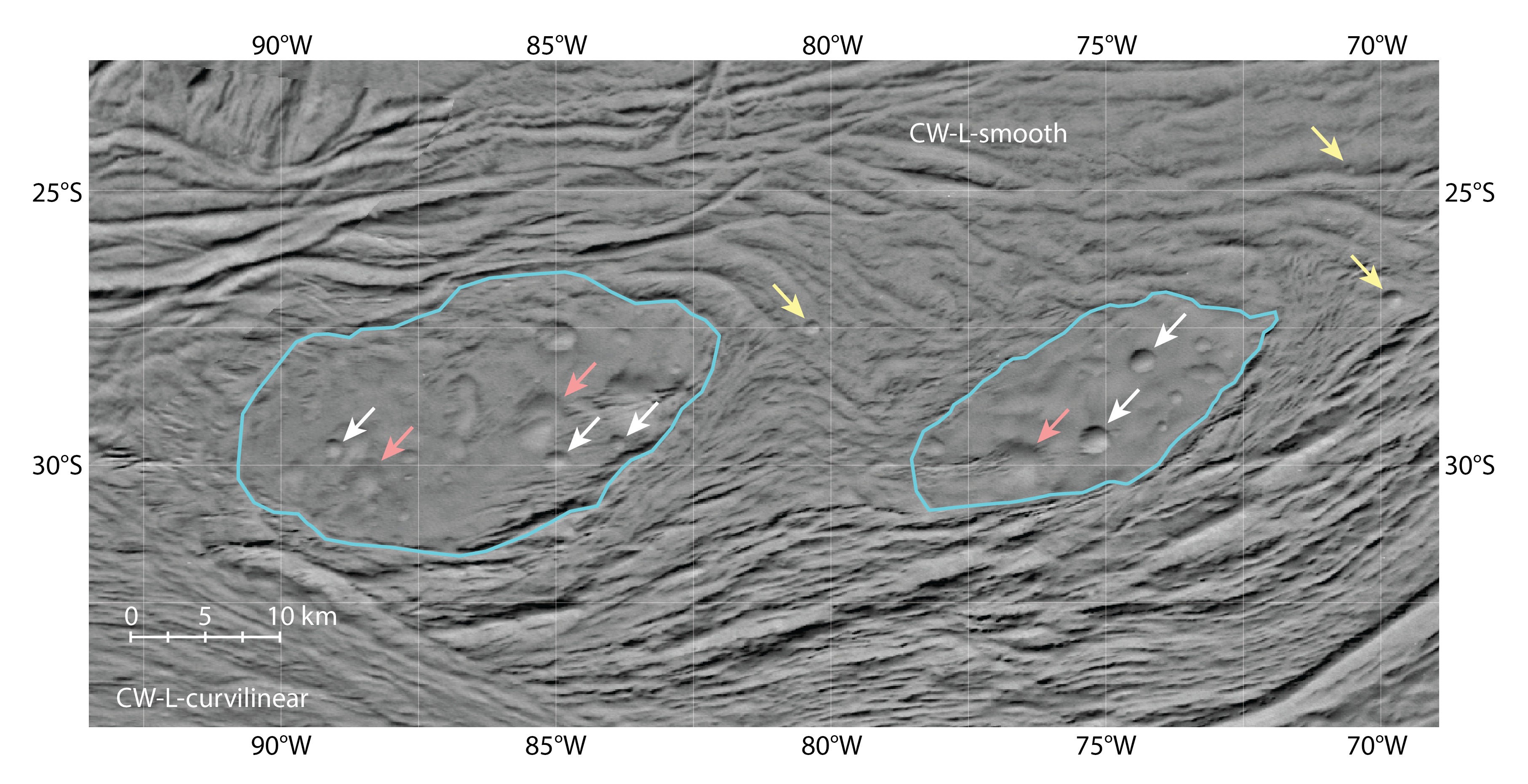}
    \caption{CW-L-CP-island (86°W, 29°S and 75°W, 28°S) consists of two small and heavily cratered ``islands'' surrounded by smoother terrain (CW-L-central). Most craters (up to $\sim$4 km) exhibit significant erosion, with distorted rims and shallow floors. In contrast, a few fresher craters (1.5–-2 km) with clear circular rims stand out, matching the morphology of craters on the younger CW-L-central. These fresher craters likely formed after the resurfacing event that shaped CW-L-central, underscoring the erosion of pre-existing craters. The crater density here aligns with other cratered plains at similar altitudes on both the anti-Saturn side and sub-Saturn side, suggesting a once higher global crater density on Enceladus before resurfacing events like the formation of CW-L-curvilinear and CW-L-central. These relict ``islands'' survived the resurfacing events and preserved pre-existing craters, which can still be distinguishable.}
    \label{fig:app_erosion4}
\end{figure}


\section{Demonstration of crater erosion}
\label{subsec:suppA}
\renewcommand{\thefigure}{S.3}

\begin{figure}[H]
    \includegraphics[width=1.0\linewidth]{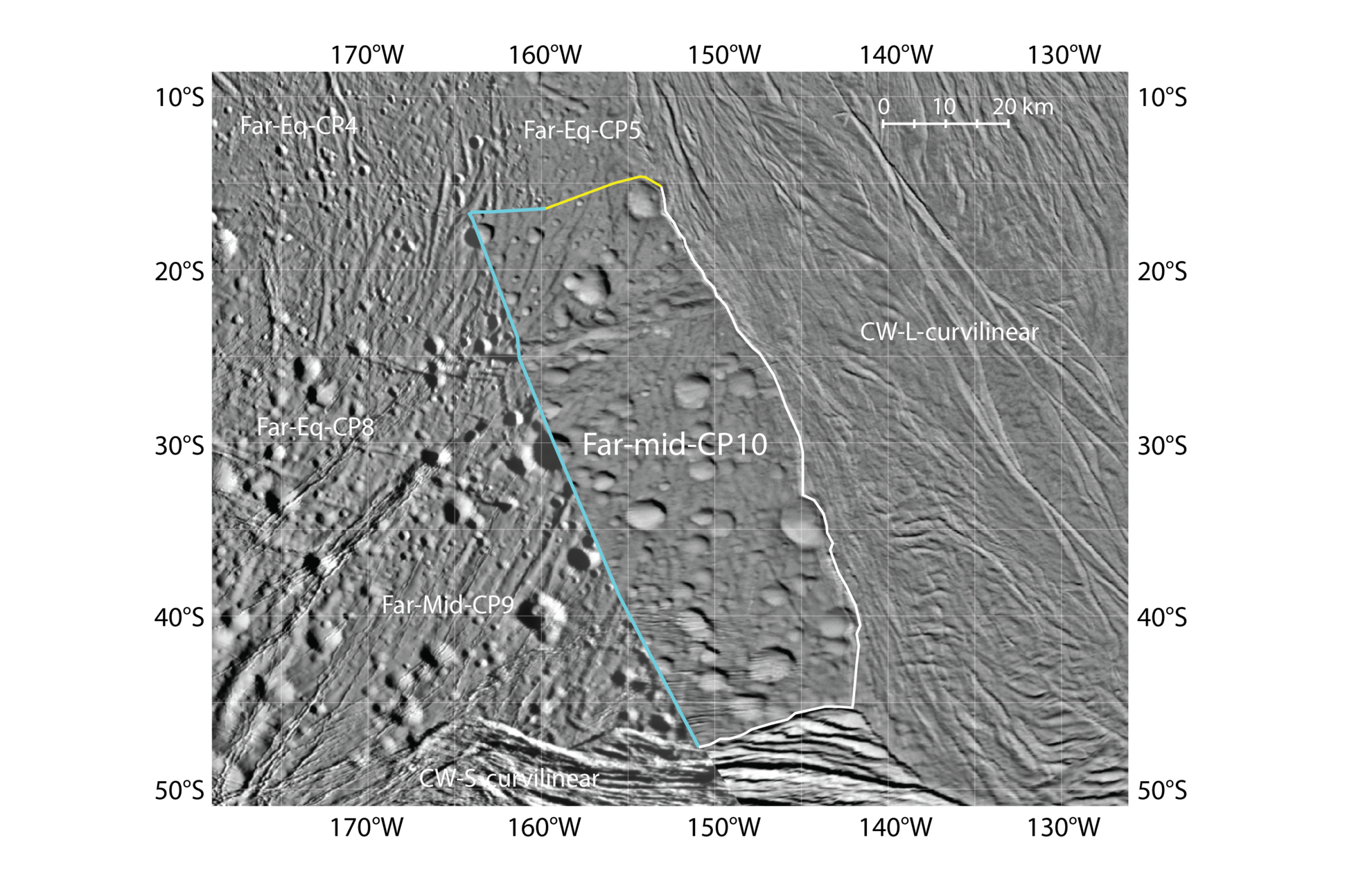}
    \caption{Far-Mid-CP10 (152°W, 30°S) helps illustrating the three criteria applied simultaneously when drawing unit boundaries. The eastern boundary of Far-Mid-CP10, marked in white, is defined by a prominent ridge system: lineated band of closely-spaced scarps or corrugated ridges belt, in the CW-L-curvilinear (criterion in Sec.~\ref{subsec:geomorphology}). A gradual southward increase in crater density marks the transition from Far-Eq-CP5 to Far-Mid-CP10, indicated by the yellow boundary (criterion in Sec.~\ref{subsec:crater_distribution}). A visible discontinuity in image resolution and shadow length and orientations distinguishes Far-Mid-CP9 from Far-Mid-CP10, shown by cyan boundary; the two units are otherwise of similar geological character (criterion in Sec.~\ref{subsec:instrumental}). 
    }
    \label{fig:unit_subdivide_example}
\end{figure}

\end{document}